\documentclass[twocolumn,trackchanges]{aastex701}
\usepackage{fix-cm}

\usepackage{float} 

\usepackage[utf8]{inputenc}
\usepackage{graphicx}
\usepackage{natbib}
\usepackage{amsmath}

\usepackage{url}
\usepackage{booktabs}
\usepackage{siunitx}
\usepackage{ragged2e}
\DeclareSIUnit{\parsec}{pc}
\DeclareSIUnit{\kpc}{kpc}

\usepackage{newtxtext,newtxmath}
\usepackage[italic]{hepnames}

\usepackage{caption}
\usepackage{subcaption}
\usepackage{hyperref}
\begin{document}
\title{Expanding Asteroseismic Studies in Star Clusters Using NASA's TESS and ESA's Gaia Missions}

\author[0009-0003-8168-4474,gname=Carli,sname=Mankowski]{Carli Mankowski}
\affiliation{Department of Astronomy, University of Florida, Gainesville, FL 32611, USA}
\email[show]{carli.mankowski@ufl.edu}

\author[0000-0002-4818-7885,gname=Jamie,sname=Tayar]{Jamie Tayar}
\affiliation{Department of Astronomy, University of Florida, Gainesville, FL 32611, USA}
\email{jtayar@ufl.edu}

\author[0009-0007-3222-0752,gname=Cassidy,sname=Martin]{Cassidy Martin}
\affiliation{Department of Astronomy, University of Florida, Gainesville, FL 32611, USA}
\email{cassidymartin@ufl.edu}


\begin{abstract}
Star clusters have long been central to the study of stellar evolution due to their chemically and chronologically homogeneous populations. Asteroseismology, the analysis of stellar oscillations and pulsations, provides precise information about properties such as masses, radii, and ages of stars in the field. However, these stars lack calibration to an absolute scale, and so this project seeks to utilize the data from NASA's \textit{TESS} mission and ESA's \textit{Gaia} mission to identify additional cluster stars suitable for asteroseismic analysis and calibration. In this work we analyze 14 stars belonging to 3 well-populated clusters, 5 additional stars that are the only detected oscillators in their respective clusters, and 3 detected oscillators of unknown cluster membership. By significantly expanding the number of clusters with measured oscillating giants, this project increases the opportunity for cross-validation between classical stellar models and asteroseismic methods, allowing for improvements in both calibration techniques and age estimations across the galaxy.

\end{abstract}

\section{Introduction}
The determination of stellar age is foundational in our understanding of galactic formation. In order to accurately infer stellar ages, we need to be able to calibrate models using well-studied examples and observed data. Increasing our understanding of the relationship between stellar properties and age enables connections across diverse areas of astronomy. Ages can be combined with chemical diagnostics \citep{Tucci2016}, nucleosynthetic pathway studies \citep{Casali2022} and large-scale kinematic maps \citep{Buder2021} to help reconstruct the Milky Way’s history. They also enhance chemical evolution models, improve stellar mass estimates, and inform other observational analyses \citep{Sharma2023}. Its large-scale applicability and long-standing academic scrutiny secure stellar age as a cornerstone diagnostic element of galactic study. Examples of the contextual applicability of stellar age include the topics of planetary system evolution \citep{Yang2023}, galaxy mergers \citep{Horta2024}, galactic chemical enrichment \citep{Feuillet2018}, and epochs of star formation \citep{Thomas2005}. Our very understanding of the Milky Way is a prime example of the integration of age estimations and observational constraints (i.e., photometric, spectroscopic, and kinematic diagnostics). Using the ages of galactic globular clusters allows us to study early formation; using the ages of non-primordial clusters allows us to study our galaxy's accretion events from long ago \citep{YingChaboyer2025_GCages}. In this way, stellar age serves as a unifying metric for individual stellar histories and the evolution of stellar, galactic, and cosmological studies.

Star clusters have long been central to the study of stellar evolution because their chemically and chronologically homogeneous populations make them ideal laboratories for testing and calibrating age estimates and other theoretical models. This remains true across a diverse sample of masses and evolutionary stages \citep{Miglio2021}. Using observational diagnostics such as color–magnitude diagrams (CMDs; \citet{CantatGaudin2020}) and theoretical isochrones \citep{Reyes2024} can also aid in the determination of stellar properties across this same wide‑scale sample. Although star clusters have long served as key laboratories for testing and calibrating stellar-evolution and age-dating models, the majority of observable stars reside outside of bound clusters. Calibrating these models on cluster members then enables their application to field stars whose ages cannot be independently confirmed. Recent advances in asteroseismology have demonstrated the potential to infer ages for hundreds of thousands of field stars if their accuracy can be anchored to coeval benchmarks such as clusters \citep{Hon2024, theodoridis2025_asteroseismically}.

Asteroseismology is the study of stellar oscillation for the purpose of gaining insight into internal stellar structure \citep{Brown1989, KjeldsenBedding1995}.  Characterizing the frequencies of these resonant oscillations in solar-like oscillators provides insight into various properties such as stellar evolution \citep{Chaplin2013}, stellar masses, radii, and luminosities \citep{Hekker2011}. Global asteroseismic observables include $\Delta\nu$, the large frequency separation between overtone modes of the same angular degree, and $\nu_{\text{max}}$, the frequency at which the oscillation power envelope reaches its maximum \citep{Hekker2011}. scillatory behavior is closely correlated with stellar mass and radius \citep{Ash2024}. Using asteroseismology with scaling relations \citep{Hon2024} allows for accurate age estimation and is thus a gateway for quantitative progress in understanding stellar history and evolutionary timescales \citep{Lebreton2008}. 

At this point, a variety of space-based photometry missions have provided the necessary data to characterize the asteroseismic oscillations of thousands of stars, and almost all of them have included subsets of stars in clusters that can be used for calibration. CoRoT (COnvection, ROtation and planetary Transits) (2006--2013) \citep{Baglin2006} pioneered space-based asteroseismology by detecting thousands of red giant oscillations among solar-like stars \citep{Mosser2018}, and the \textit{Kepler} mission (2009--2013) \citep{Borucki2010} subsequently revealed tens of thousands more oscillators in a similar stellar sample \citep{Yu2018}. The K2 campaign (2014--2018) \citep{Howell2014} included cluster members, giants, and a few dwarf stars through thousands of detections \citep{Zinn_2022}, while \textit{TESS} (2018--present) \citep{Ricker2015} has surveyed the bright all-sky field and identified over 158,000 oscillators, predominantly red giants \citep{Hon2021}.

Combining asteroseismology with supplemental information such as \textit{Gaia} photometry \citep{Risbud2025, Myers2022} and cluster membership \citep{Hunt2023, Tayar2025} can expand the impact of asteroseismology and constrain the chemical and dynamical evolution of stars; integration of these large, consistent datasets allows for deeper comparative analysis. However, their age estimations are still at odds with those derived from isochrone fitting \citep{Tayar2025, Palakkatharappil2023}.

In this context, NASA’s \textit{TESS} mission, launched on 2018 April 18, provides nearly all-sky high-precision photometry. \textit{TESS} operates in a 27.4-day highly elliptical Earth orbit, observing consecutive ($24^\circ \times 96^\circ$) sectors \citep{Ricker2015} and delivering both light curves and full-frame images (FFIs) \citep{Stassun2019TICv8}, ultimately achieving continuous coverage of more than 95\% of the sky \citep{Jenkins2016}. We can now also advance asteroseismic studies using \textit{TESS} data due to its all-sky coverage of many targets with high precision. Since \textit{TESS} provides asteroseismic data for solar-type stars \citep{Ricker2015} all over the sky, it opens the possibility of greatly increasing the number of clusters with asteroseismic data that could then be used to calibrate the ages of hundreds of thousands of oscillators across the galaxy. Prior to this work, asteroseismic detections in giant stars had been reported in only seven open clusters and six globular clusters ( \citet{Edmonds1996, Stello2009, Howell2025}; see also \citet{Miglio2021} for a review). Our analysis of \textit{TESS} data adds eight open clusters, three well-populated and five with single detections, more than doubling the known sample of open clusters with seismically characterized giants.

However, the fundamental operation of \textit{TESS} can impose certain limitations that can impair the study of asteroseismology. Since \textit{TESS} has a large pixel size, light from nearby sources can contaminate that of target stars, especially in crowded fields, complicating the extraction of clean light curves for individual stars \citep{Hatt2023}. Additionally, upon its initial launch, \textit{TESS}'s 30-minute cadence imposed a Nyquist frequency limit of \SI{278}{\micro\hertz}: the highest frequency that can be accurately sampled without aliasing, which may affect the asteroseismology of stars with oscillation frequencies at or around that Nyquist limit \citep{Murphy2015,Barclay2020}.  \textit{TESS} also faces analytical limitations, including lost detections caused by data gaps and short observing sectors \citep{Guerrero2021_TOI_Prime} and challenges arising from its month-to-month discontinuous observing pattern \citep{Lu2020}.

Therefore, to promote the efficient analysis of \textit{TESS} asteroseismology in star clusters, we aim to establish a database of cluster stars that are suitable for seismic analysis, even with the limitations of the \textit{TESS} mission as described above. For example, we are interested in whether the ages derived from global asteroseismic parameters are consistent with cluster ages obtained from isochrone fitting. Previous work has suggested that seismically inferred ages may differ systematically from those obtained using other techniques, often yielding larger values \citep{Tayar2025}. However, the small number of clusters analyzed to date, together with their limited coverage in metallicity and age, makes it difficult to assess the true accuracy and precision of ages inferred from asteroseismology.

In this study we combine precise astrometry, membership information, and isochrone-based ages for nearby open clusters from the \textit{Gaia} mission with all-sky photometric data from \textit{TESS}. We demonstrate that, despite the relatively short time series and large pixel scale of \textit{TESS}, its data are sufficient in many cases to characterize the oscillations of red giant cluster members. We then compare ages inferred from asteroseismic scaling relations and stellar age modeling to literature cluster parameters in order to map the reliability of seismic ages, enabling their broader application to infer the ages of large samples of field stars across the Milky Way \citep{theodoridis2025_asteroseismically}. To facilitate future work, we  provide a catalog of cluster stars that may be suitable for seismic analysis with \textit{TESS}.

\section{Methods}
\subsection{Catalog Selection}
The foundation of this work relies on careful characterization and association of cluster stars. We use \textit{Gaia} DR3, whose all-sky parallaxes, proper motions, magnitudes, BP-RP colors, and other photometry allow for accuracy in determining open-cluster membership, precise distances, and clean HR-diagram placement \citep{Gaia2023}. This photometric data is pulled from cluster membership studies such as \citet{Hunt2023}, \citet{Schiavon2022}, and \citet{CantatGaudin2020}, which have already grouped members of the same clusters together with reasonable confidence. 

Using the \citet{Hunt2023}, \citet{Schiavon2022}, and \citet{CantatGaudin2020} catalogs, we queried the \textit{TESS} Input Catalog (TIC v8; \citealt{Stassun2019TICv8}) via the S3 STScI public dataset using \texttt{astroquery.mast.cloud} to isolate targets with unique TIC IDs. To account for systematic errors such as field variation, cosmic noise, and electro-optical response differences \citep{Gai2022_TESS_astrometry}, the right ascension and declination of each target were given a radial error window of 0.1 degrees. After confirming that each target was present in the \textit{TESS} Input Catalog, we performed a column-wise merge to match \citet{Hunt2023}, \citet{Schiavon2022}, and \citet{CantatGaudin2020} members with a CSV file containing TIC IDs from the 2019 TIC-7 Input Catalog \citep{TICv7}. These two datasets were merged on a shared identifier column (\textit{Gaia} DR3), producing a functional catalog of stars with secure membership classifications and relevant astrometry. This catalog contains 38311 members for further consideration, all within the \textit{TESS} input catalog.

\subsection{Distance Cuts}
We filter our catalog to eliminate distant targets, following the findings of \citet{CantatGaudin2020}, which concluded that \textit{Gaia} DR3 proper motion uncertainties contribute at $\sim$\SI{500}{\parsec} and dominate at distances beyond \SI{1000}{\parsec}. Using the identity $d = \frac{1000}{\varpi}$, where $d$ is the distance in parsecs and $\varpi$ is the parallax in milliarcseconds, we use catalog-provided parallax data to remove any stars at distances greater than \SI{1}{\kilo\parsec}. This reduces the systematic noise and measurement uncertainty from the dataset. Filtering out distant stars leaves a nearby sample with more reliable absolute magnitudes, because parallax-based distance uncertainties and absolute-magnitude uncertainties are typically a factor of $\sim5$–10 smaller for stars within a few kiloparsecs than for more distant objects \citep{Lindegren2018}.

\subsection{Magnitude Filtering}
We take into consideration the intense “bleeding" effect of bright stars on \textit{TESS}'s camera in the creation of our catalog. Stars that are exceptionally bright will not yield any photometric data, as the \textit{TESS} pipeline does not process anything past its brightness threshold \citep{TESS_DRN115_2025}. For targets under that threshold, we cannot extract reliable brightness data from stars whose light bleeds into multiple pixels, introducing the need for a “cap'' on how bright a star can be. This is explored further in section \S\ref{subsec:crowding}. We use the \textit{TESS} QLP Aperture Lightcurve Products, which are generated only for stars brighter $T \approx 13.5$ and can exclude the very brightest targets when bleed-trail pixel stamps preclude standard processing \citep{MIT_QLP,DRN_S59}.

We also account for the increased contamination and background noise affecting fainter \textit{TESS} targets, which becomes a significant issue beyond $G \sim 15$ and renders stars fainter than $G \gtrsim 16$ largely unsuitable for precise asteroseismology \citep{Riello2021,Boyle2025}. Following previous asteroseismic work and \citet{Stassun2019_TIC_CTL}, we restrict our sample to stars with $G \le 13$~mag so solar-like oscillations can be reliably detected at high signal-to-noise and to prioritize high signal-to-noise targets for analysis \citep{Hon2021,Boyle2025,Zhou2024,Grusnis2025_CVZ}.

\subsection{Color Magnitude Diagrams and Red Giant Selection}

We require all stars in our sample to have reported \textit{Gaia} photometry (G and $G_{\rm BP}-G_{\rm RP}$). To obtain these values, we merge the cluster member list with the \textit{Gaia} DR3 photometric catalog of \citet{Andrae2023}.
 This addition allows us to construct a color--magnitude diagram (CMD) that traces the evolutionary state of each star \citep{GaiaCollaboration2018_HRD}. Following the methods of \citet{Konchady2020_AASNova_HR_Gaia}, we create a CMD from the mean G-band magnitude (G) and color index $(G_{\rm BP}-G_{\rm RP})$ provided by the XGBoost-based \textit{Gaia} DR3 catalog. In our analysis, this XGBoost catalog is used for the photometry and atmospheric parameters ($T_{\mathrm{eff}}$, [Fe/H], and $\log g$). Where these values are missing or physically unrealistic, we use atmospheric parameters adopted from the \texttt{gspec} spectroscopic catalog. By applying the distance modulus,
\begin{equation}
M_G = G - 5 \log_{10}(d) + 5,
\end{equation}
we convert all of the Gmag values in the catalog ($G$) into the absolute magnitude $M_G$.

\begin{figure}[htpb]
\centering
\includegraphics[width=\columnwidth,height=0.35\textheight,keepaspectratio]{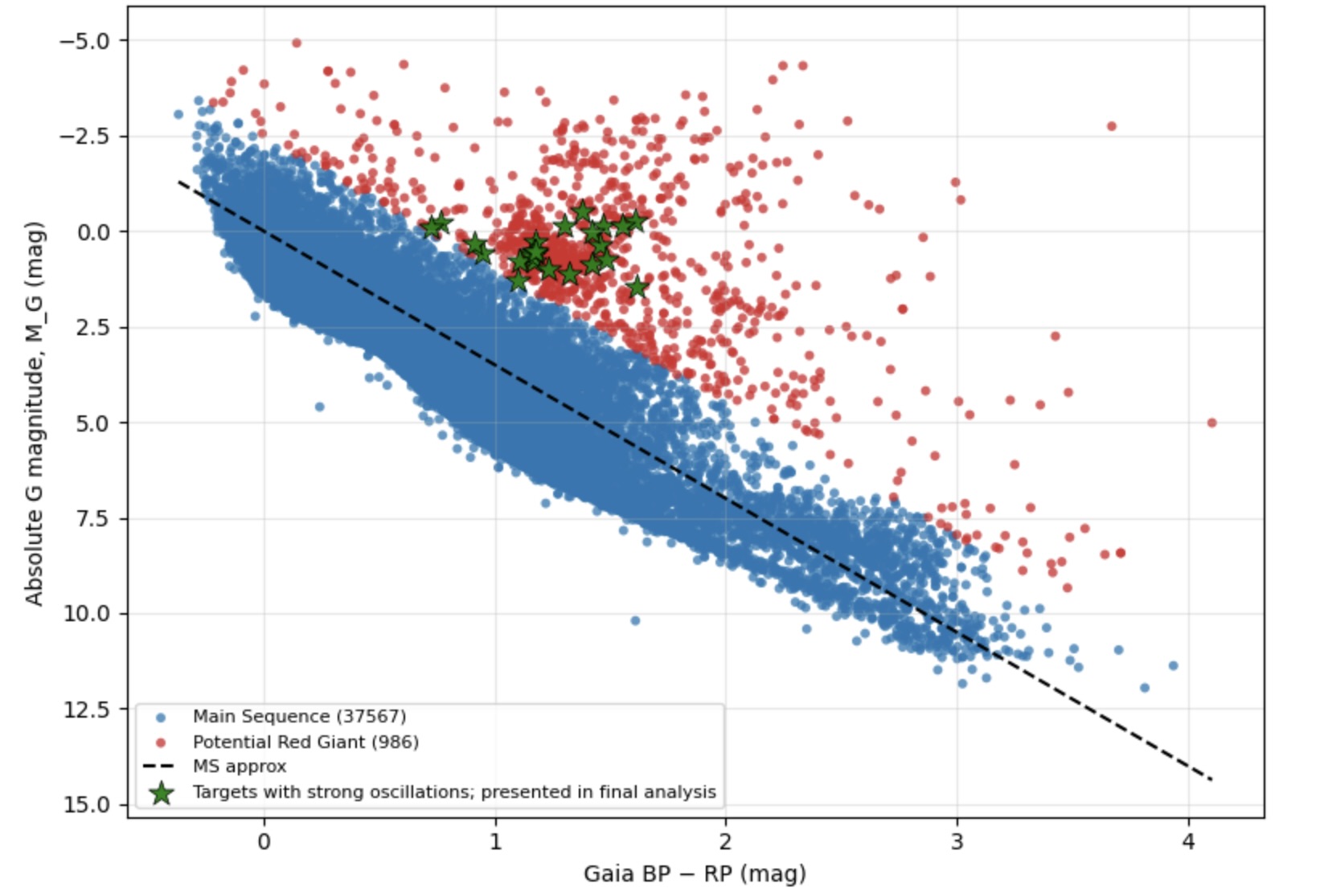}
\caption{Global HR diagram using (\textit{Gaia}) absolute magnitude vs BP-RP color. Main sequence stars are shown in blue, potential red-giants in red, seismic detections marked as green stars, and the main sequence approximated by the black dashed line.}
\label{fig:hr_all}
\end{figure}

\begin{figure}[htpb]
\centering
\begin{subfigure}[t]{0.49\columnwidth}
\centering
\includegraphics[width=\linewidth,height=0.23\textheight,keepaspectratio]{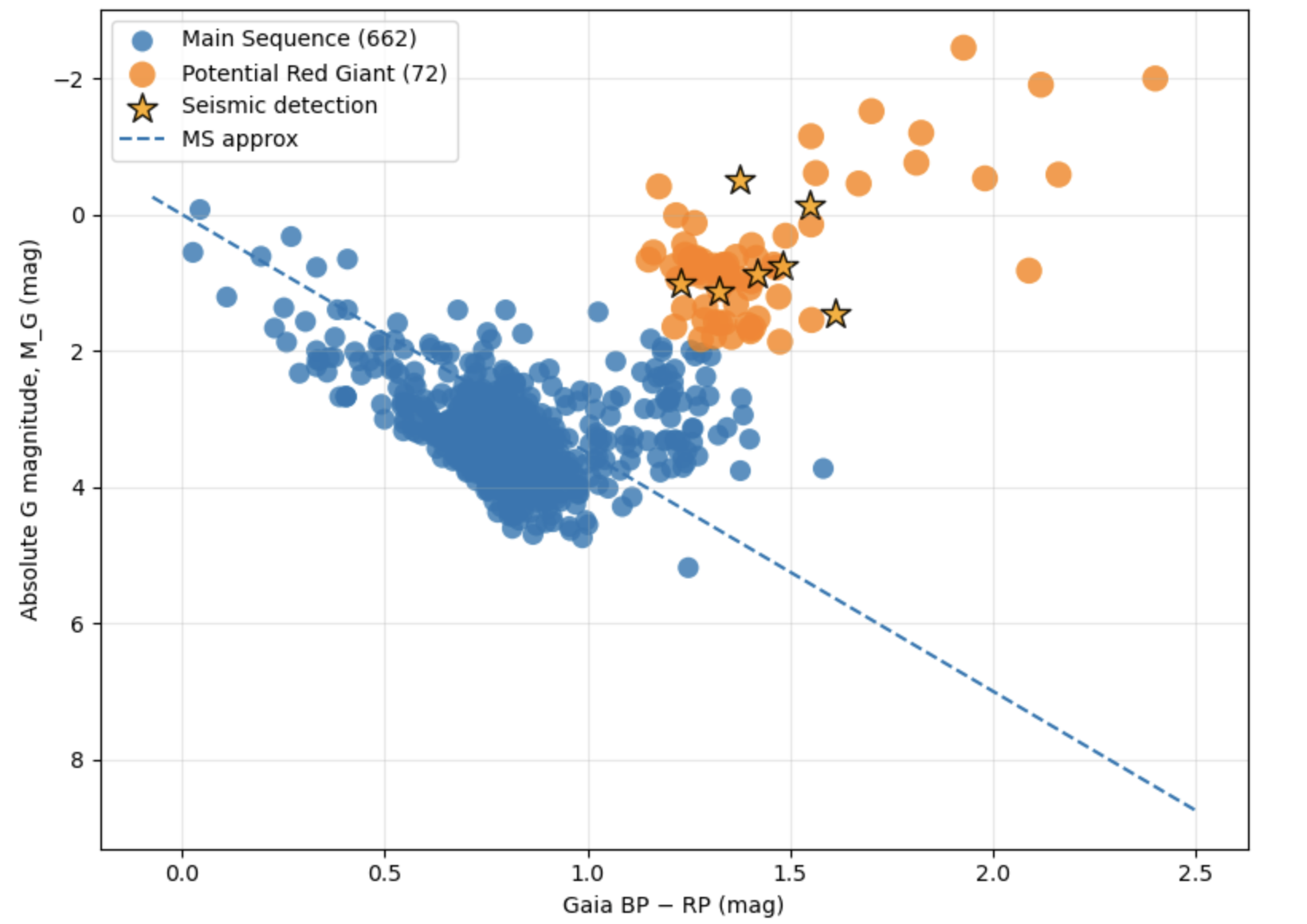}
\caption{Theia 6046}
\end{subfigure}\hfill
\begin{subfigure}[t]{0.49\columnwidth}
\centering
\includegraphics[width=\linewidth,height=0.23\textheight,keepaspectratio]{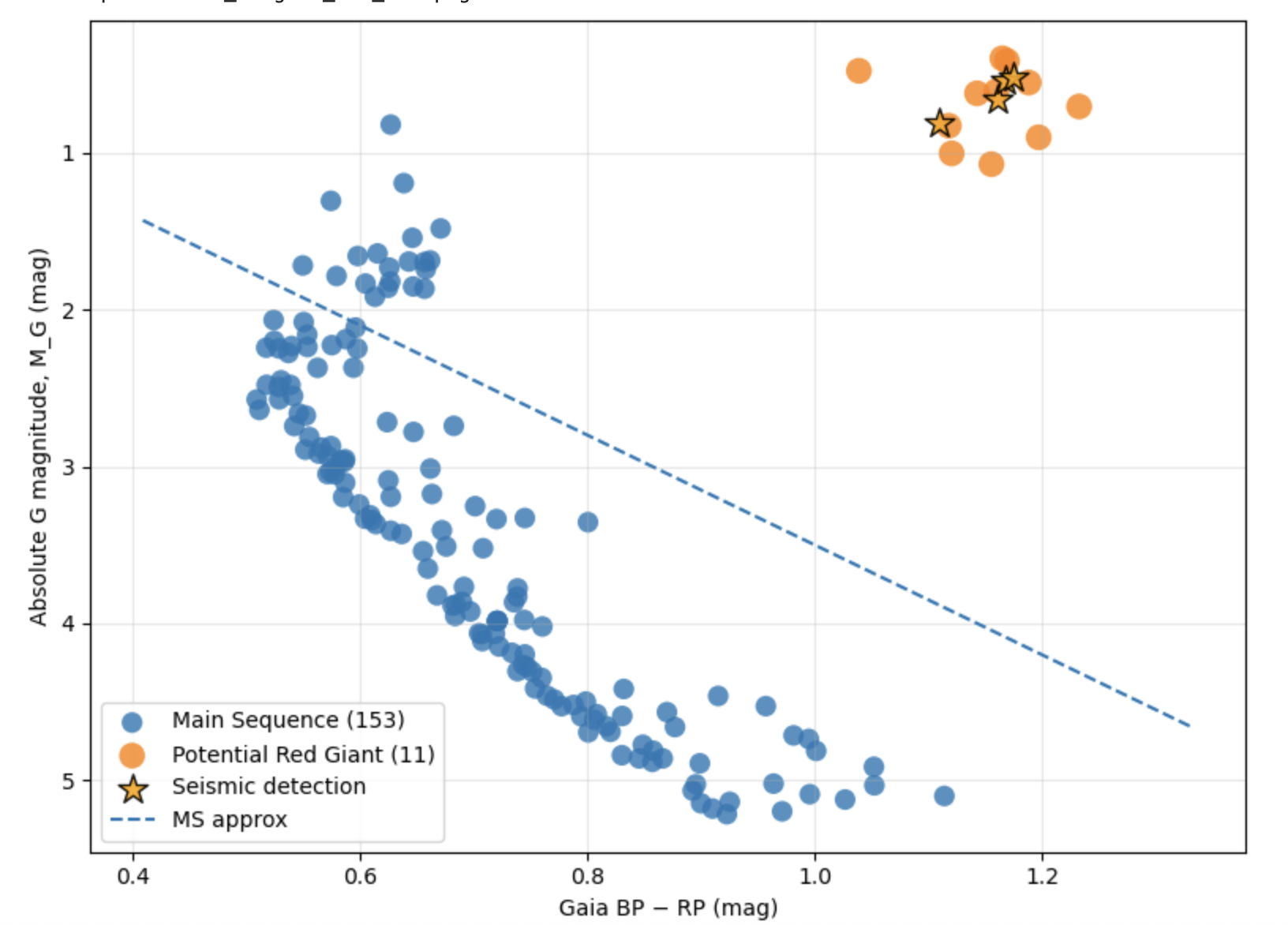}
\caption{NGC 752}
\end{subfigure}
\vspace{0.4em}
\begin{subfigure}[t]{0.49\columnwidth}
\centering
\includegraphics[width=\linewidth,height=0.23\textheight,keepaspectratio]{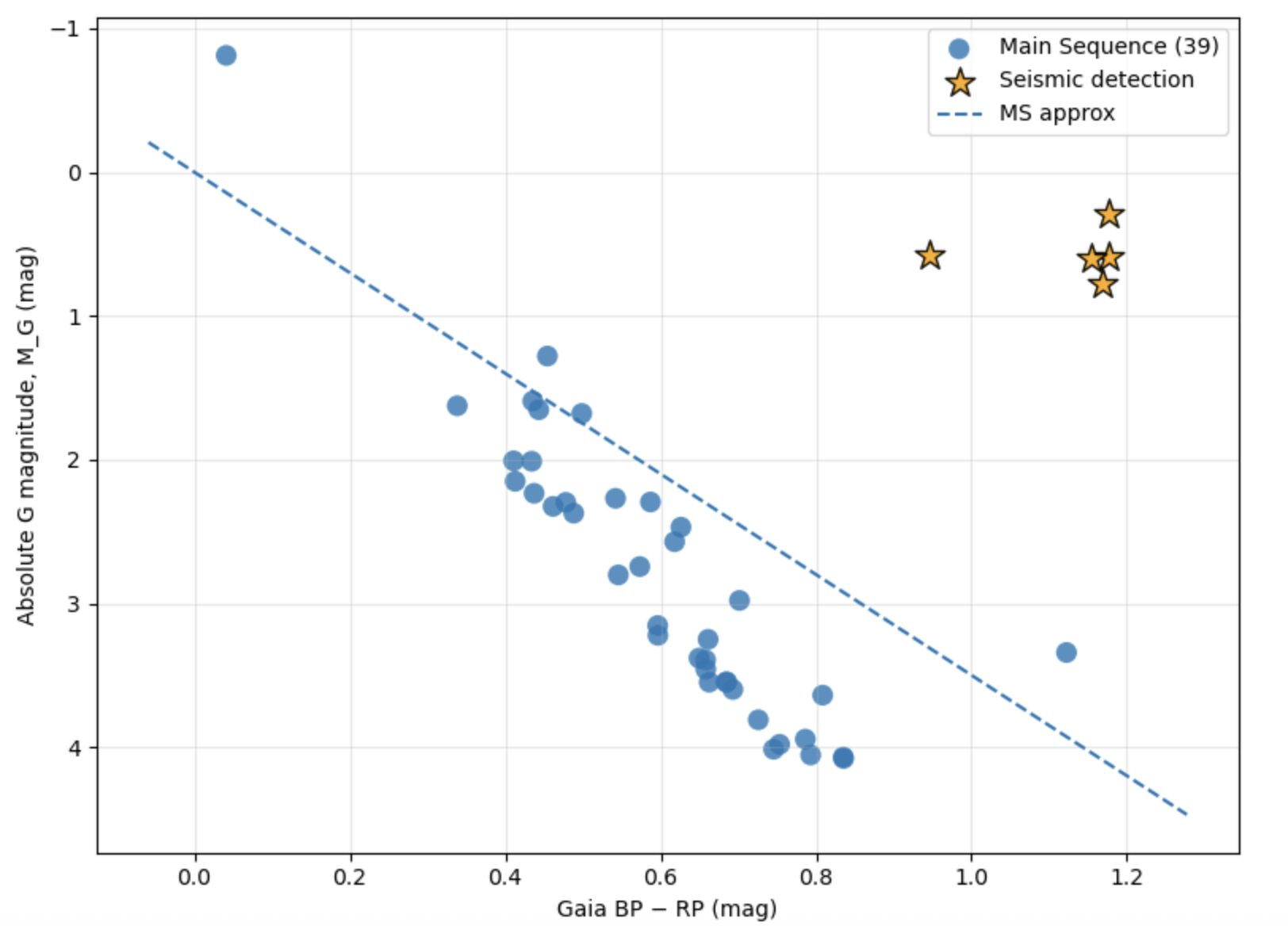}
\caption{Casado Alessi-1}
\end{subfigure}\hfill
\begin{subfigure}[t]{0.49\columnwidth}
\centering
\includegraphics[width=\linewidth,height=0.23\textheight,keepaspectratio]{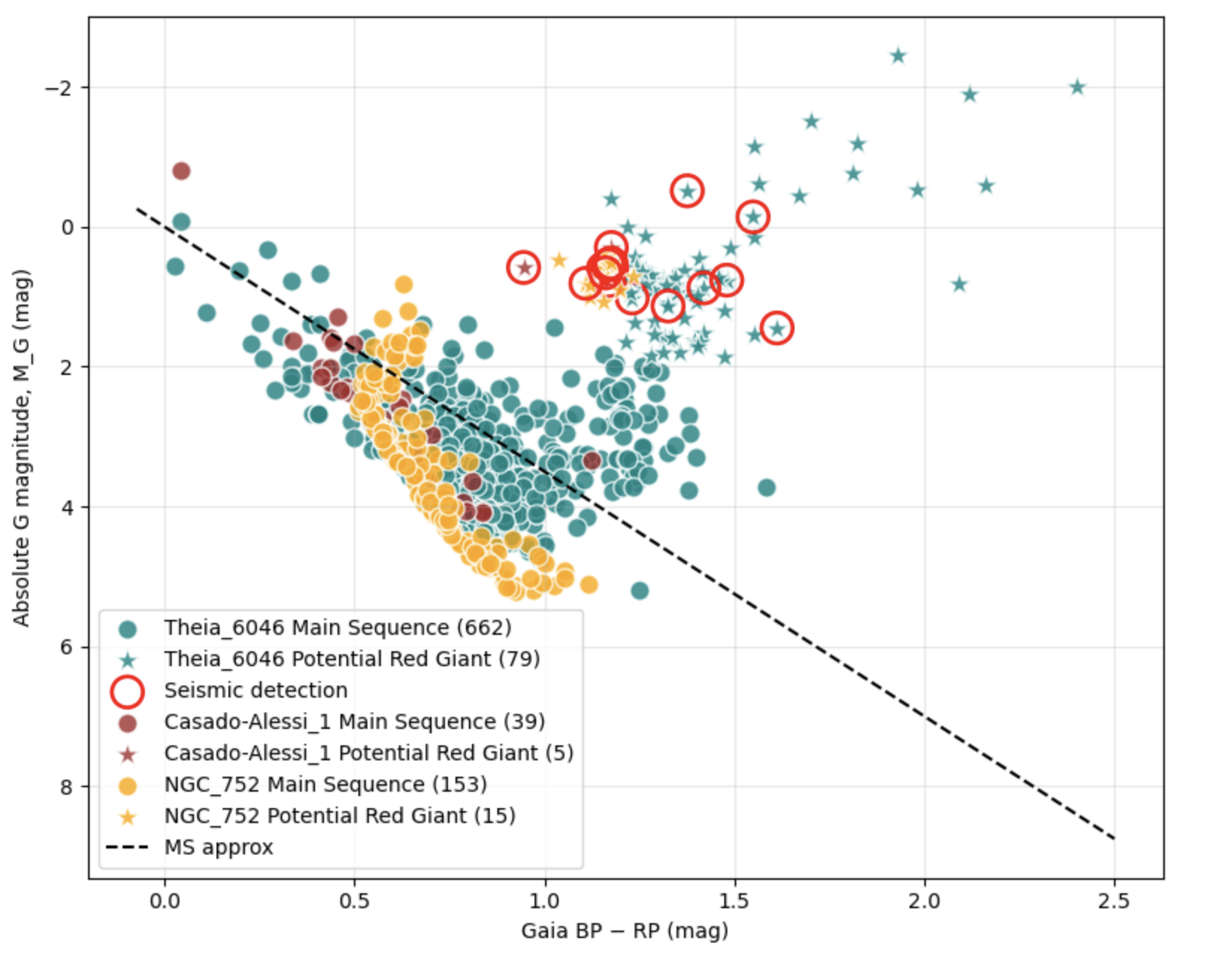}
\caption{Three-cluster overlay}
\end{subfigure}
\caption{Cluster HR diagrams for the three clusters in our study with more than one seismic detection in a giant. Panels (a)-(c) show individual clusters with main sequence stars in blue, potential red-giants in orange, and seismic detections marked as orange stars. Panel (d) overlays all three clusters: Theia 6046 in teal, NGC 752 in orange, Casado Alessi-1 in magenta, with potential red-giants shown as stars, seismic detections circled in red, and the main sequence approximated by the black dashed line.}
\label{fig:hr_clusters}
\end{figure}

This catalog follows the selection principles of \citet{Hon2021} and \citet{Garcia2022}, prioritizing red giants because their solar-like oscillations have larger amplitudes and are detected at higher rates in \textit{TESS} data. Since red giants span a broad range of stellar ages \citep{Xiang2022}, this should allow us to test the reliability of seismically inferred ages more generally. We visualize the full sample on a \textit{Gaia}-based CMD (Figure~\ref{fig:hr_all}) and we also show the three most populated clusters with seismic detections in our sample, Casado Alessi-1, Theia 6046, and NGC 752 (Figure~\ref{fig:hr_clusters}). We approximate the location of the main sequence in this diagram in order to exclude stars that are unlikely to be red giants. To do this, we define an empirical main-sequence locus in color–magnitude space and remove stars that fall close to this sequence rather than on the red giant branch. The main sequence line used in our CMD approximation is given by \( M_G = 3.5 \times (BP - RP) \) where \( M_G \) is the absolute \textit{Gaia} G-band magnitude and \( BP - RP \) is the \textit{Gaia} color index. In order to be fairly confident that our sample only contains red giants, we filter this CMD so that each target greater than 3.5 units of magnitude away from the main sequence line in the $+y$ direction is deemed a potential red giant. Out of the initial sample of 38,311 distance and magnitude corrected stars, 986 are estimated to be potential red giants and the other 37,667 are estimated to be potential main sequence.


\subsection{Crowding, Bleeding and Duration Requirements} \label{subsec:crowding}

\textit{TESS}’s observing pattern, large pixels, and susceptibility to saturation around the brightest stars can result in a degradation of data quality that makes seismic analysis challenging or impossible. This suggests that not every cluster red giant in our sample will have the quality of data needed to perform asteroseismology. For this analysis, we want to focus on stars with clean point-spread functions and sufficient data. To address this, we individually inspect each potential star to ensure the \textit{TESS} data exist. Then, we check if nearby stars allow for the extraction of a clean light curve, or if local saturation and bleeding would result in the need for more advanced light curve techniques \citep{Pope2016}. Specifically, we used each star’s right ascension (RA), declination (Dec), and \textit{Gaia} DR3 name to create a 10x10 pixel cutout around each target star using NASA's Timeseries Integrated Knowledge Engine (TIKE). Nearby bright neighbors present the possibility for contaminated light-curves; we identified extraneous stars within both 3 magnitudes and 2 arcminutes of our target. We marked those nontarget stars to evaluate potential contamination of seismic signals.

\begin{figure*}[htbp]
    \centering

    \begin{subfigure}[t]{0.32\textwidth}
        \centering
        \includegraphics[width=\linewidth]{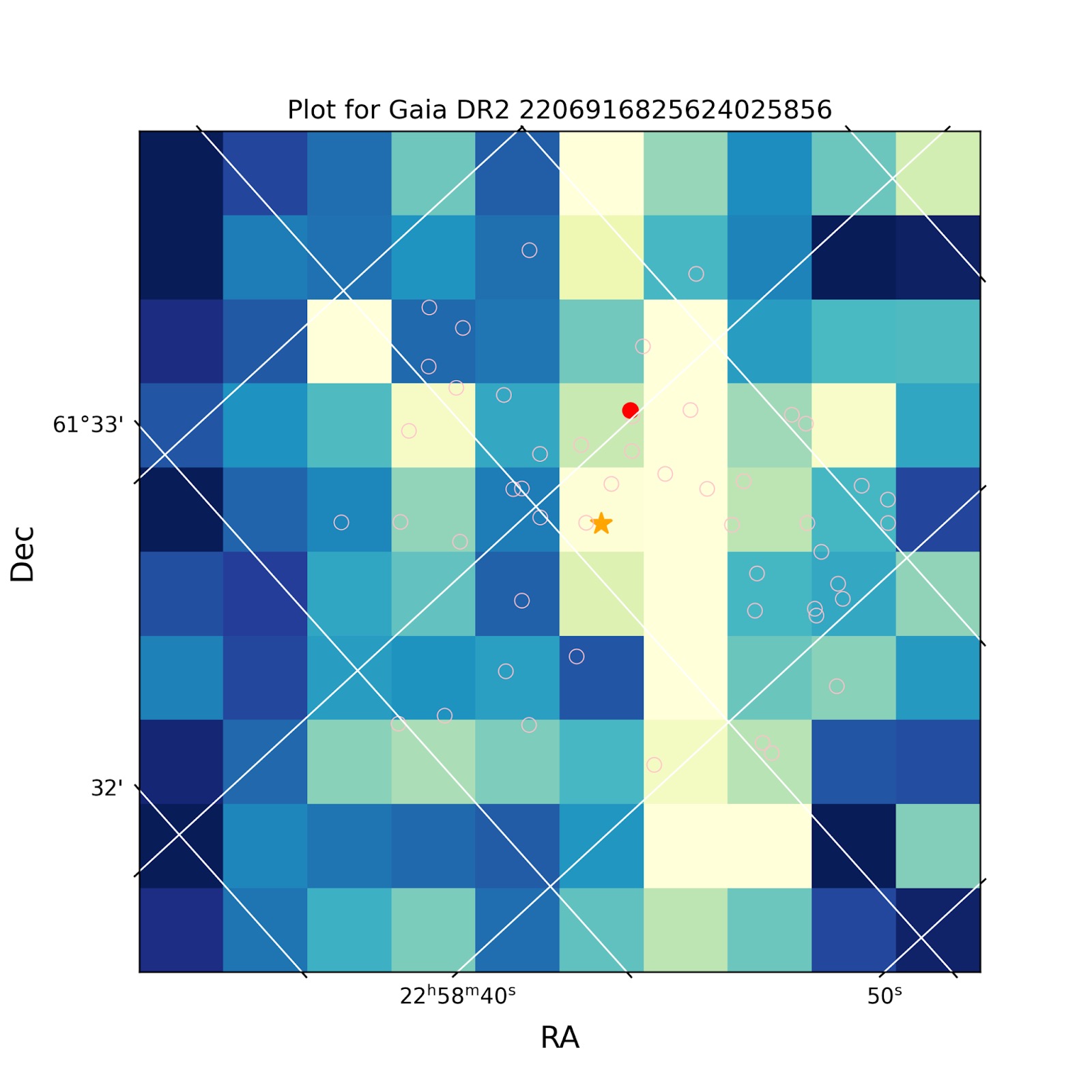}
        \caption{Bleeding = 1, Crowding = 1}
    \end{subfigure}\hfill
    \begin{subfigure}[t]{0.32\textwidth}
        \centering
        \includegraphics[width=\linewidth]{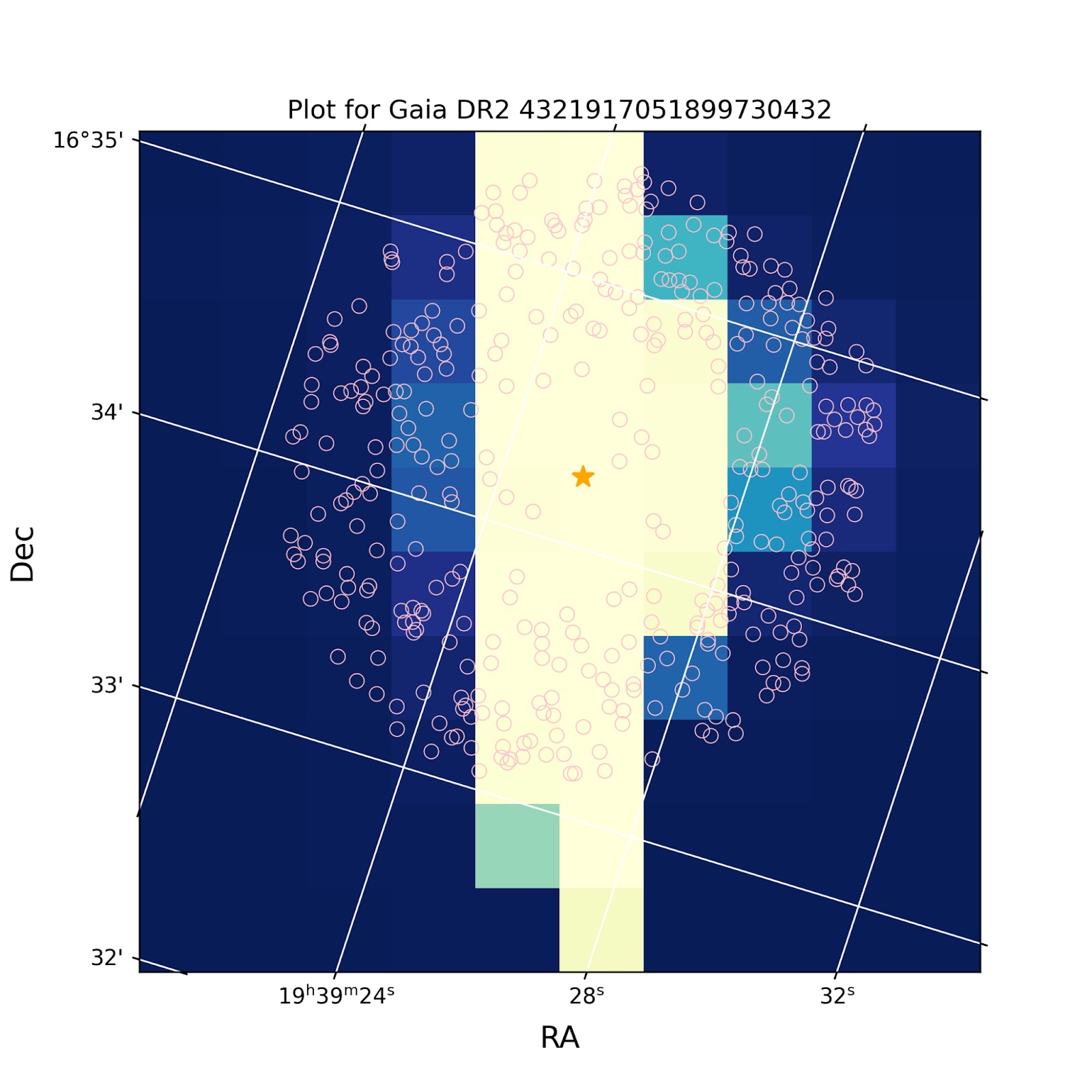}
        \caption{Bleeding = 1, Crowding = 4.5}
    \end{subfigure}\hfill
    \begin{subfigure}[t]{0.32\textwidth}
        \centering
        \includegraphics[width=\linewidth]{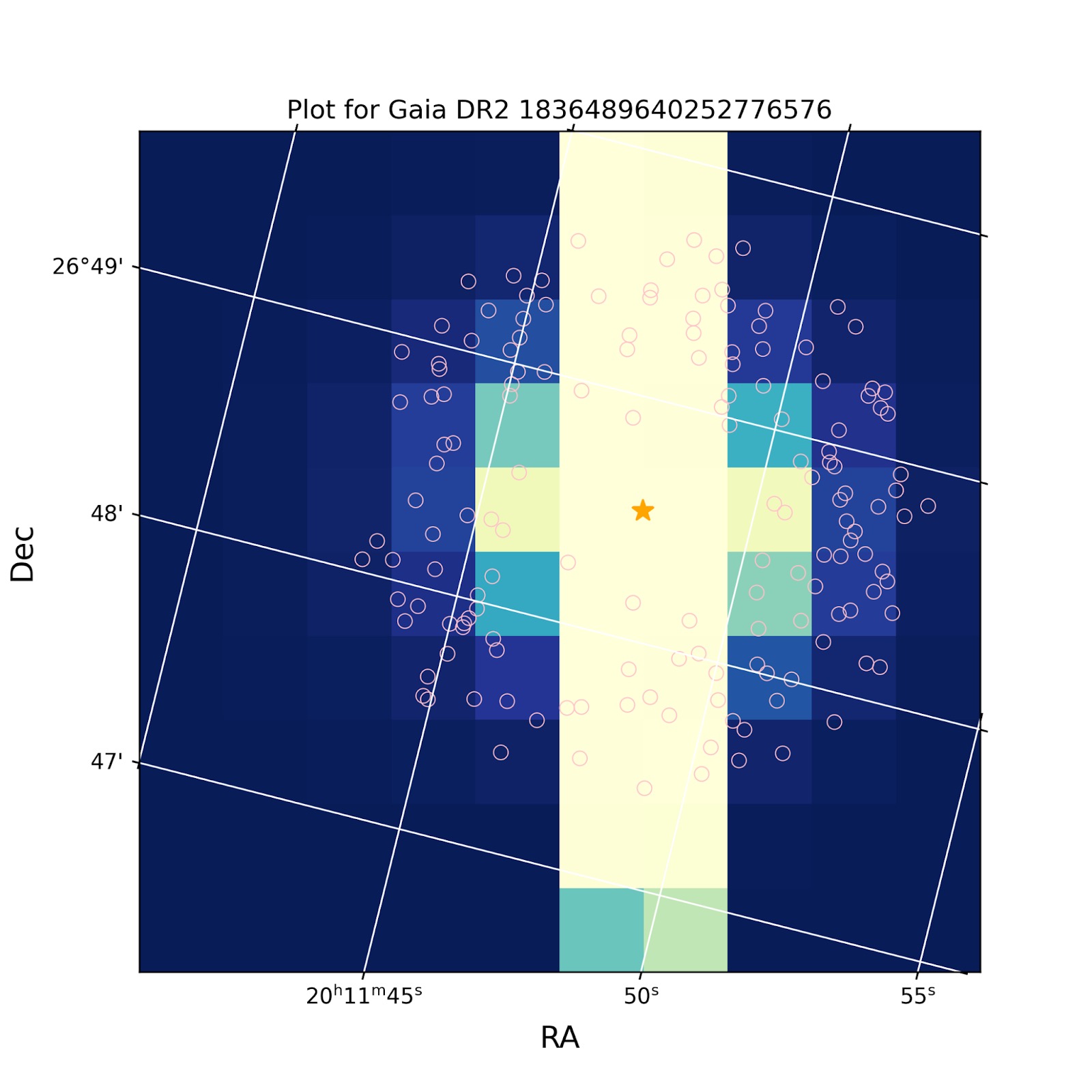}
        \caption{Bleeding = 1, Crowding = 5}
    \end{subfigure}

    \vspace{0.6em}

    \begin{subfigure}[t]{0.32\textwidth}
        \centering
        \includegraphics[width=\linewidth]{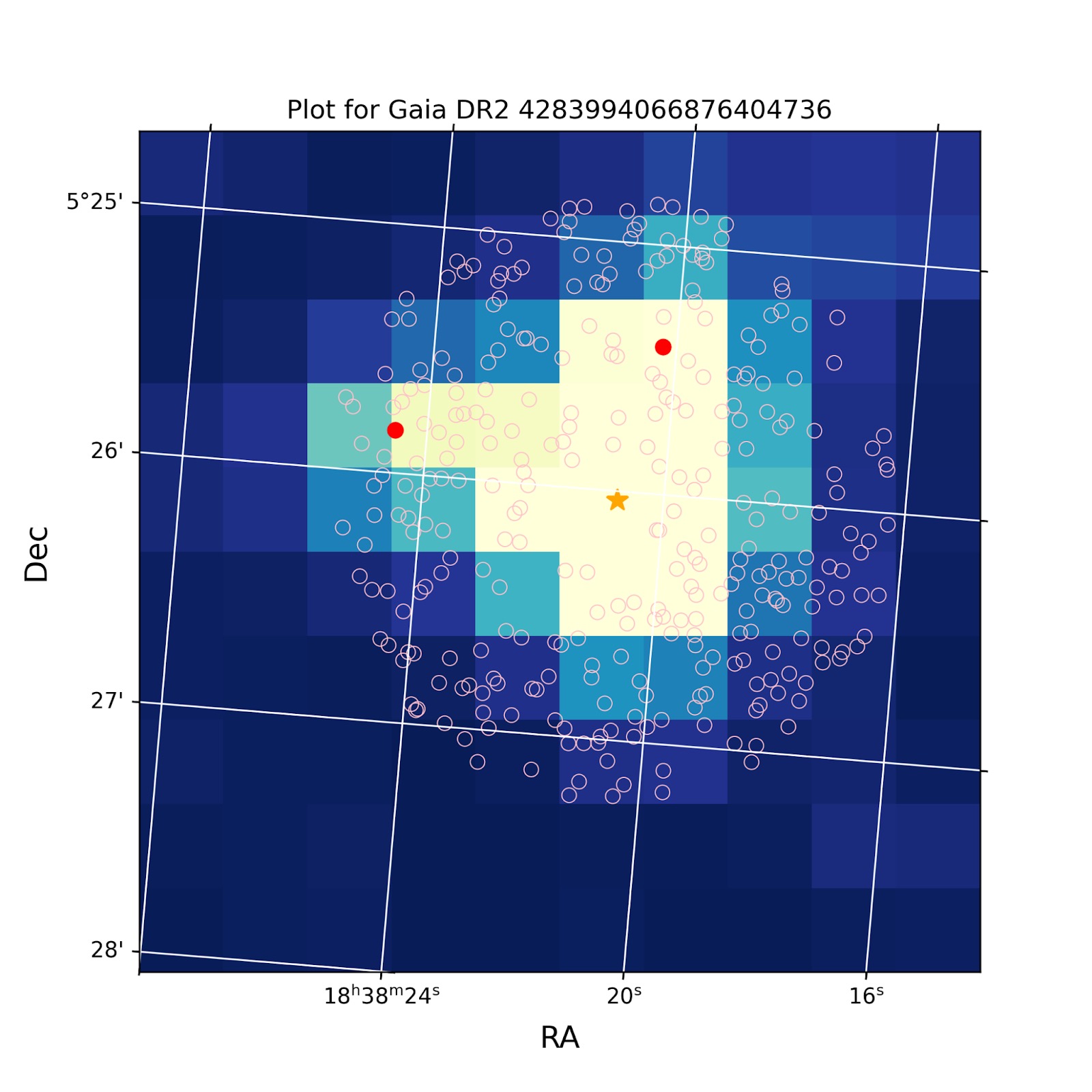}
        \caption{Bleeding = 3, Crowding = 1}
    \end{subfigure}\hfill
    \begin{subfigure}[t]{0.32\textwidth}
        \centering
        \includegraphics[width=\linewidth]{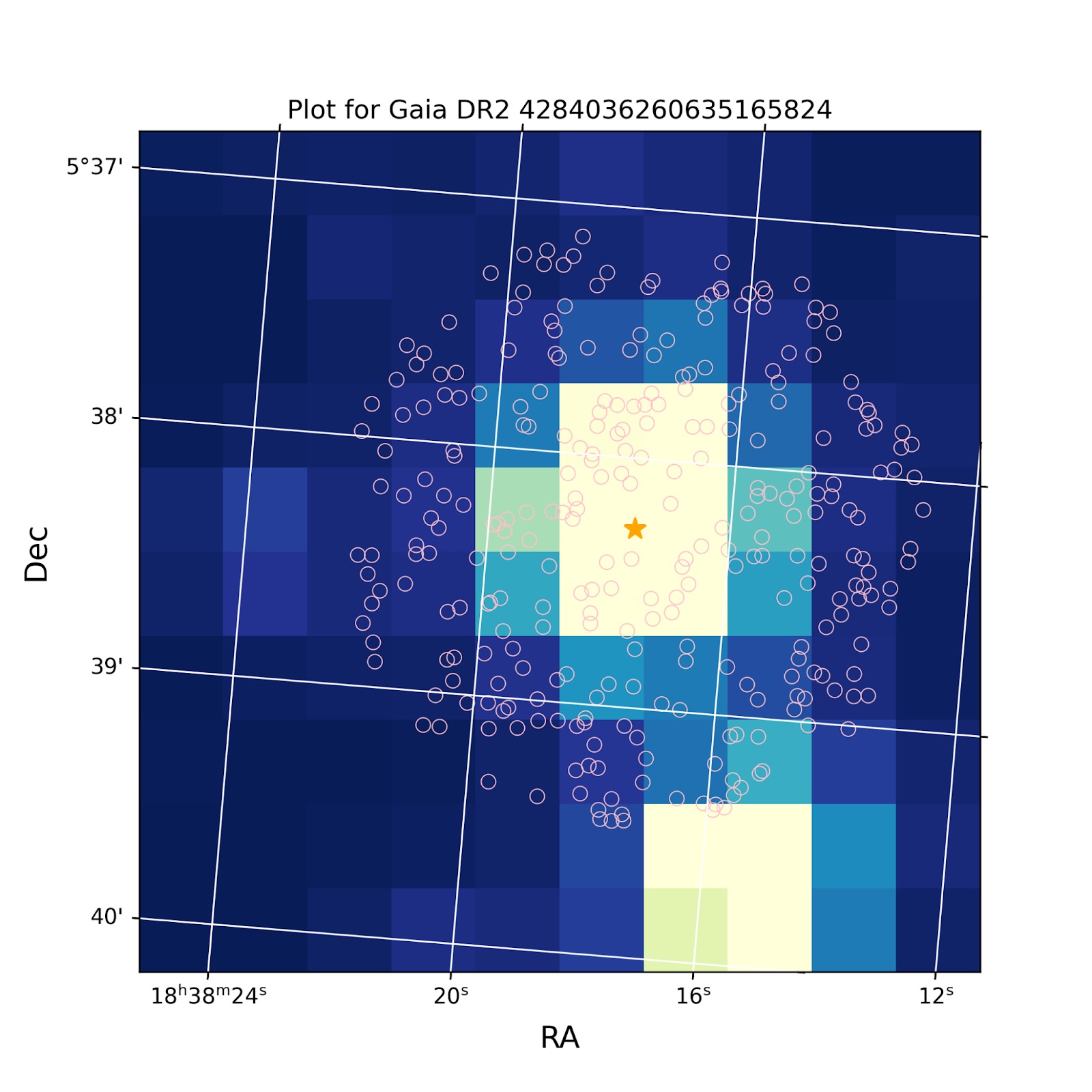}
        \caption{Bleeding = 3, Crowding = 3}
    \end{subfigure}\hfill
    \begin{subfigure}[t]{0.32\textwidth}
        \centering
        \includegraphics[width=\linewidth]{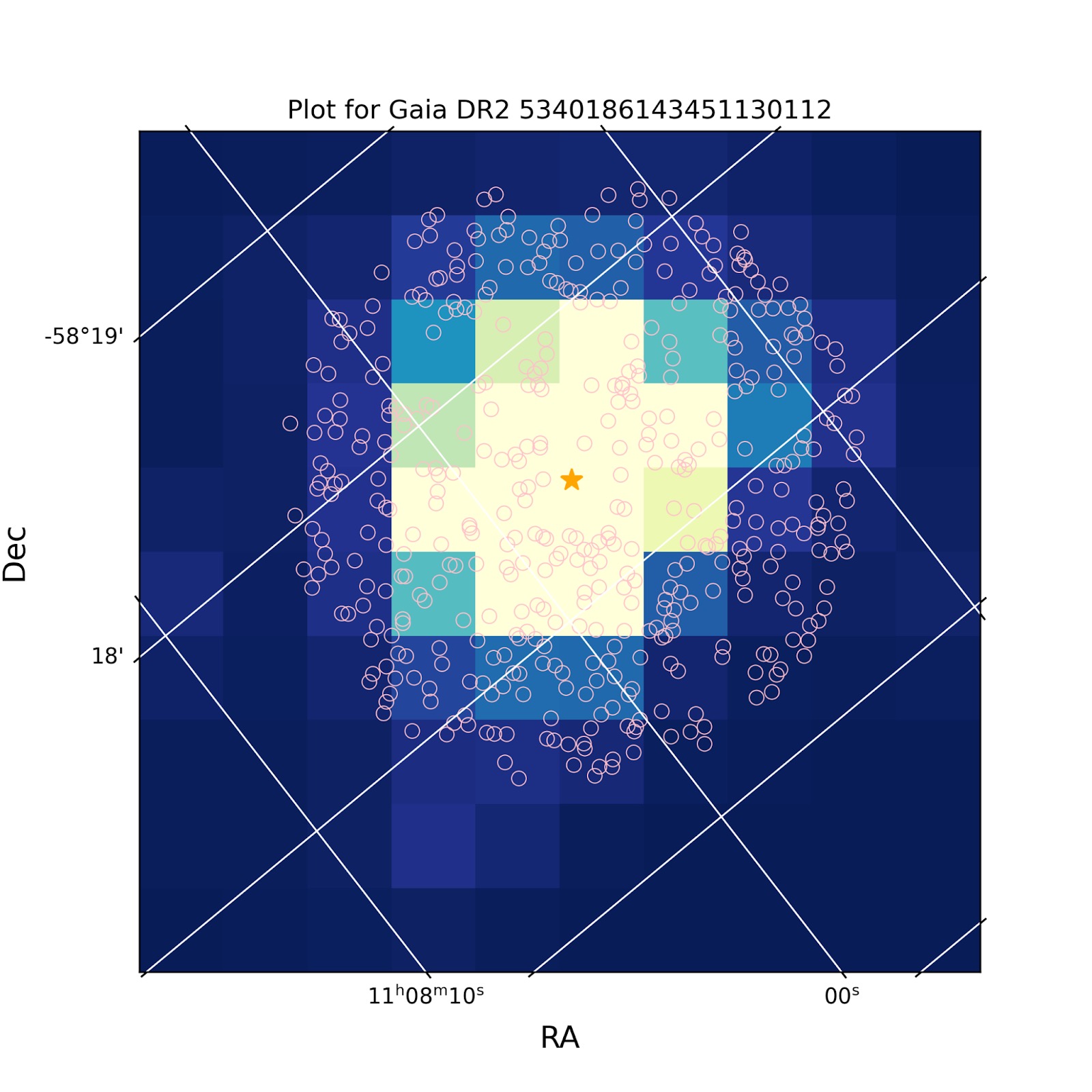}
        \caption{Bleeding = 3, Crowding = 5}
    \end{subfigure}

    \vspace{0.6em}

    \begin{subfigure}[t]{0.32\textwidth}
        \centering
        \includegraphics[width=\linewidth]{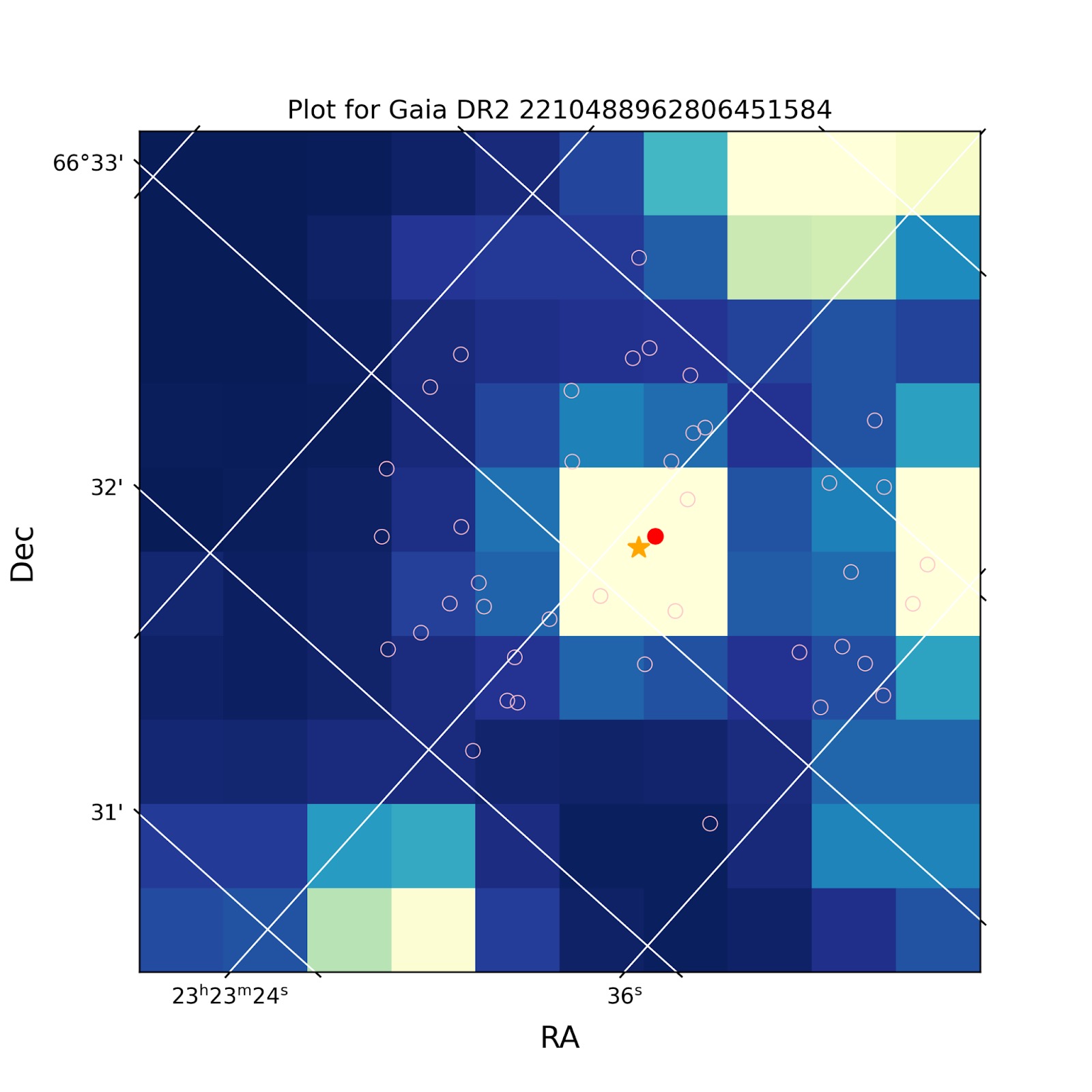}
        \caption{Bleeding = 5, Crowding = 1}
    \end{subfigure}\hfill
    \begin{subfigure}[t]{0.32\textwidth}
        \centering
        \includegraphics[width=\linewidth]{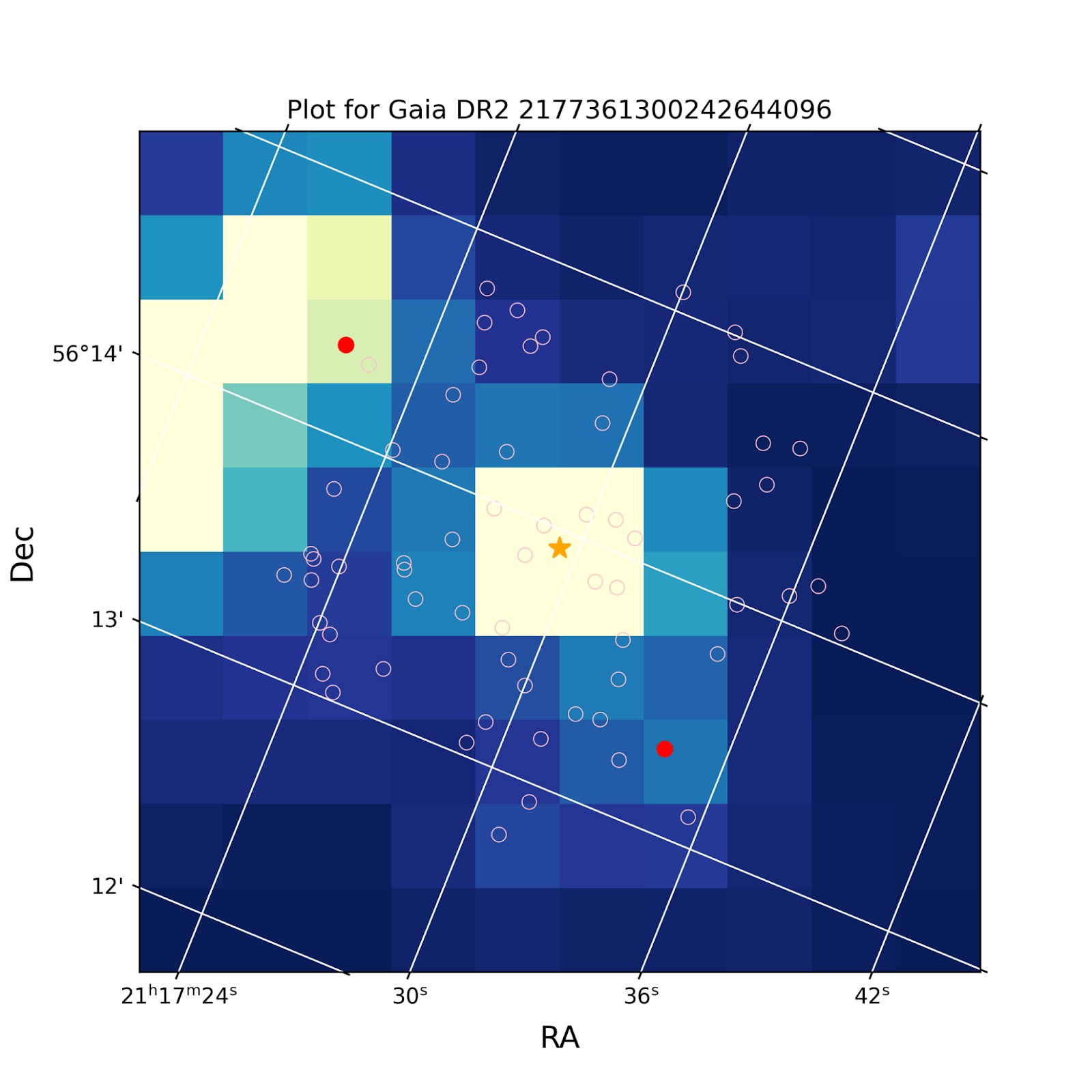}
        \caption{Bleeding = 5, Crowding = 3}
    \end{subfigure}\hfill
    \begin{subfigure}[t]{0.32\textwidth}
        \centering
        \includegraphics[width=\linewidth]{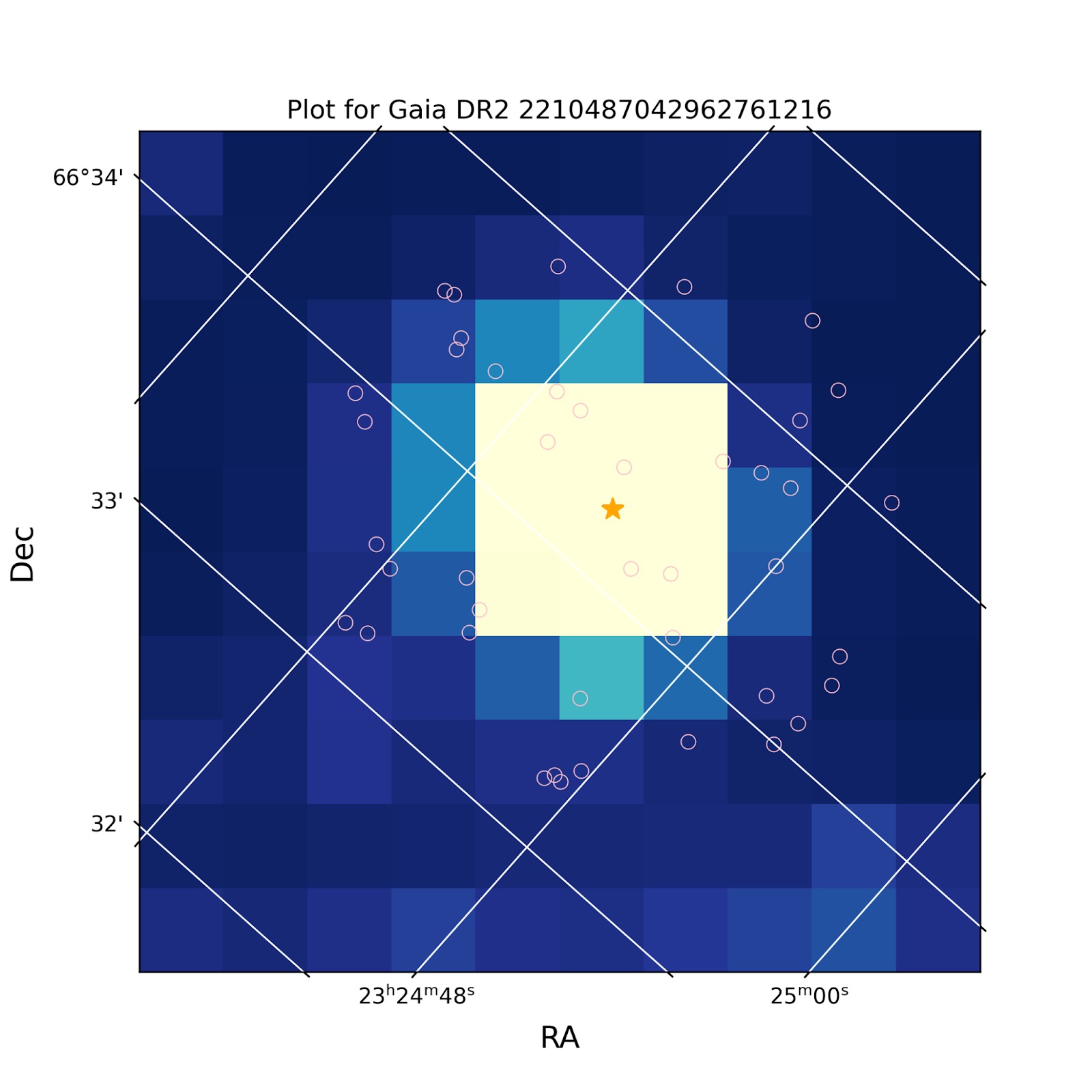}
        \caption{Bleeding = 5, Crowding = 5}
    \end{subfigure}

    \caption{Nine-panel grid of \textit{TESS} targets across various bleeding (grouped horizontally) and crowding (grouped vertically) ratings.}
    \label{fig:bleed_crowd_grid}
\end{figure*}

For each star in our sample, we compiled individual identification information as well as visual inspection ratings into a table. Ratings were allocated to each target using a 1 to 5 scale with 5 representing the cleanest data. Several stars had no available sectors, these are marked with “-1” in the table and are not considered for further analysis. To create these visualizations, the first available sector was used. Exceptions were made for scenarios in which the target was either excessively faint or on the edge of the detector, and the next suitable sector was used. In Figure \ref{fig:bleed_crowd_grid}, we show representative cutouts with fixed bleeding and varying levels of crowding, and the impact of varying bleeding values at fixed crowding levels. Our full table of results is available on 
\href{https://zenodo.org/records/16884612}{Zenodo}, and the code to generate these cutouts can be found \href{https://zenodo.org/records/16938930}{here}.

We restricted our sample to stars with bleeding and crowding values of at least 2 and at least five available sectors. Only this subset of red giants was used in the subsequent analysis, though results for all targets are provided for reference and to enable future work. All quality cuts up to this point are replicable and the process is outlined on \href{https://github.com/carlimankowski/mankowski2025-open-cluster-asteroseismology}{GitHub}.

\subsection{TESS Light Curve Retrieval and Pre-Processing}
We then retrieved the raw light curves for the targets that had adequate bleeding, crowding, and sector metrics. Using the \texttt{lk.search\_lightkurve} function, we located each \textit{TESS} light curve from the QPL pipeline and downloaded them. For each sector,  we normalized and flattened the data with a 299-cadence window, and removed $2.75\sigma$ outliers. We then stitched all available sectors into a single light curve and sorted the points by time \citep{Jenkins2016}. To avoid large gaps between sectors in the analysis, whenever we encountered a gap larger than 10 days in the time array, we subtracted the size of that gap from all subsequent time stamps to form a continuous time series for Fourier analysis, though we note potential biases for larger gaps as discussed in \citet{BeddingKjeldsen2023}. This pre-processing is an essential prerequisite to \textit{TESS} data analysis, since the high values of systematic noise can damage results. The light curves were converted into power spectra, oversampled by a factor of five to achieve a frequency resolution of $\sim0.1\,\mu$Hz. The spectra were binned, using five-point binning for $\nu_{\max}$ estimation and 25-point binning for background fitting, following methods presented in \citet{Chontos2022_pySYD}.

\subsection{PySYD Seismic Analysis}
Using PySYD, an open-sourced Python adaptation of the SYD method, we aimed to have code automatically detect the asteroseismic parameters $\nu_{\max}$ and $\Delta\nu$. As noted in \citet{Huber2009}, the underlying SYD methodology was commonly used to measure oscillatory parameters in \textit{Kepler} stars. Extending this to \textit{TESS}, we needed to neutralize the noisy data with Gaussian smoothing of the oscillation power excess, an approach motivated by \citet{Huber2009} and justified by \citet{Hon2024}. We set 50 Monte Carlo iterations, restricted $\nu_{\max}$ to 1--100\,$\mu$Hz initially, refined the power spectrum to 10--60\,$\mu$Hz, and fitted the background between 5--525\,$\mu$Hz. Running PySYD produced $\nu_{\max}$ and $\Delta\nu$ estimates, background model fits, uncertainties, and diagnostic plots. 
Window restrictions were adjusted appropriately for each target, and uncertainties for $\nu_{\max}$ were calculated considering both PySYD values and visual inspection, referring to how far in each direction can be feasibly interpreted as $\nu_{\max}$. Our typical fractional uncertainty on $\nu_{\max}$ is 3.3$\%$, comparable to the APOKASC-3 values reported by \citet{Pinsonneault2025}. The Jupyter notebook used to run this PySYD analysis is openly available on \href{https://github.com/carlimankowski/mankowski2025-open-cluster-asteroseismology}{GitHub}.

\begin{figure}[htbp]
\centering
\begin{subfigure}[t]{0.48\columnwidth}
  \centering
  \includegraphics[width=\linewidth]{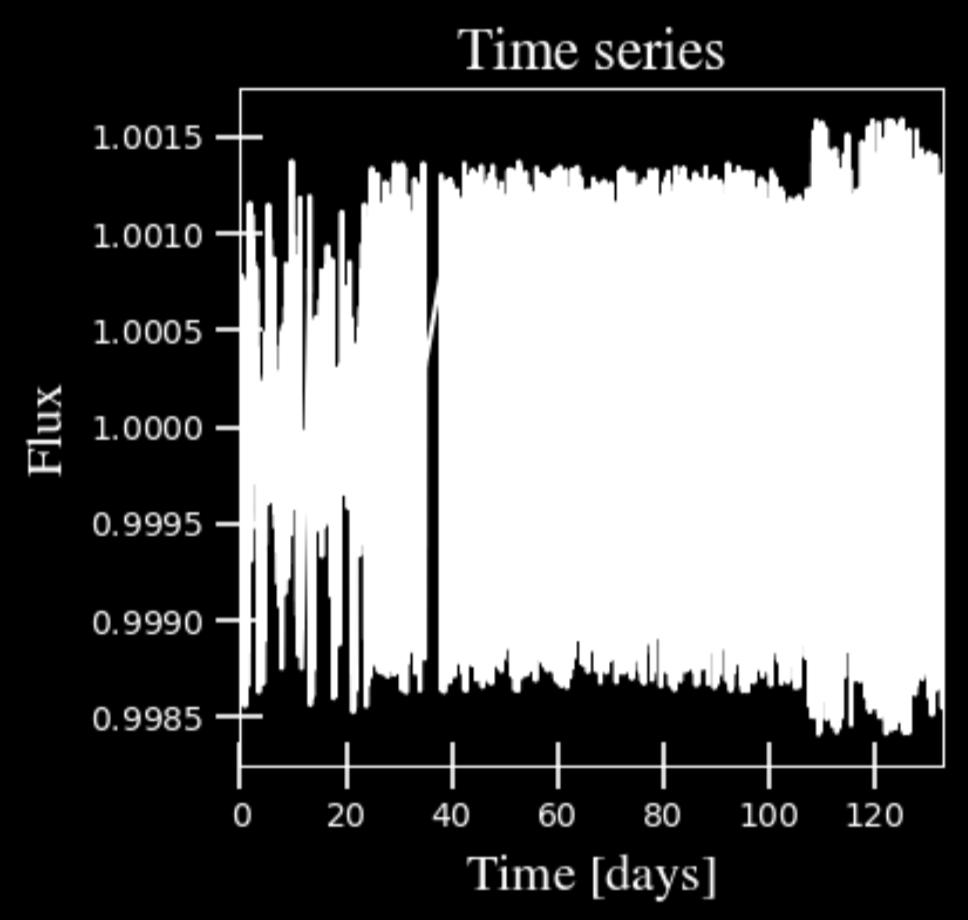}
  \caption{Detrended light curve showing brightness variations.}
  \label{fig:numax_306955660_a}
\end{subfigure}%
\hfill
\begin{subfigure}[t]{0.48\columnwidth}
  \centering
  \includegraphics[width=\linewidth]{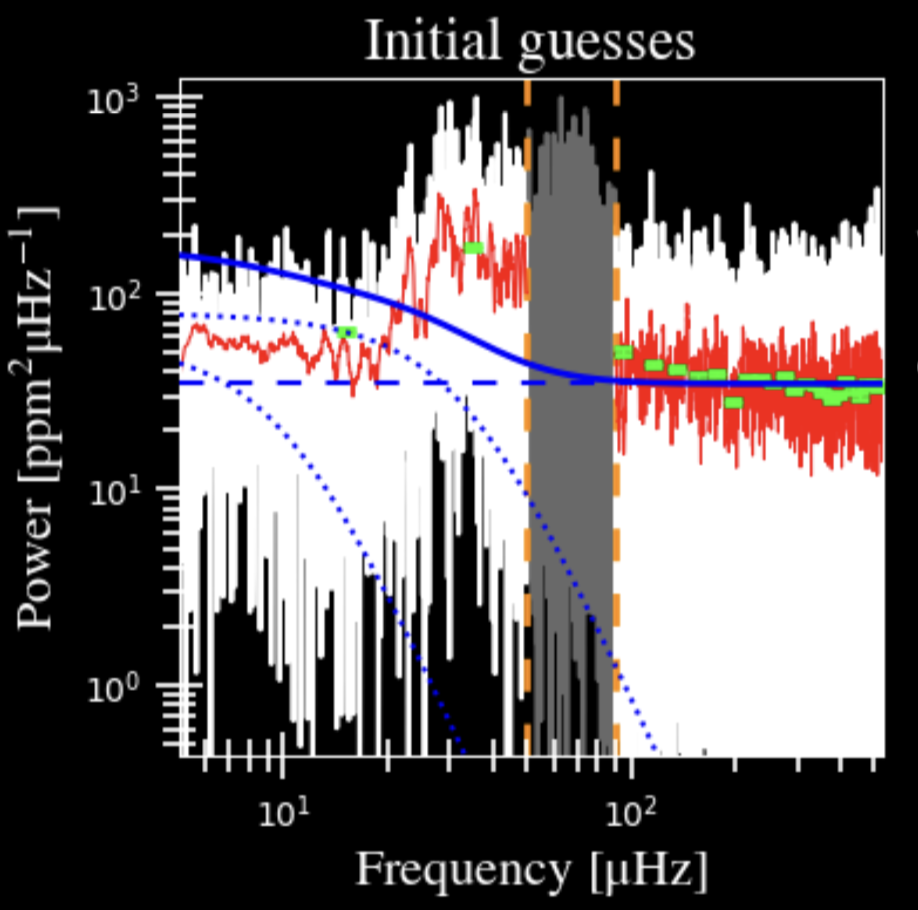}
  \caption{Power spectrum (black) with initial background model fit (red) separating granulation, activity, and white noise.}
  \label{fig:numax_306955660_b}
\end{subfigure}

\begin{subfigure}[t]{0.48\columnwidth}
  \centering
  \includegraphics[width=\linewidth]{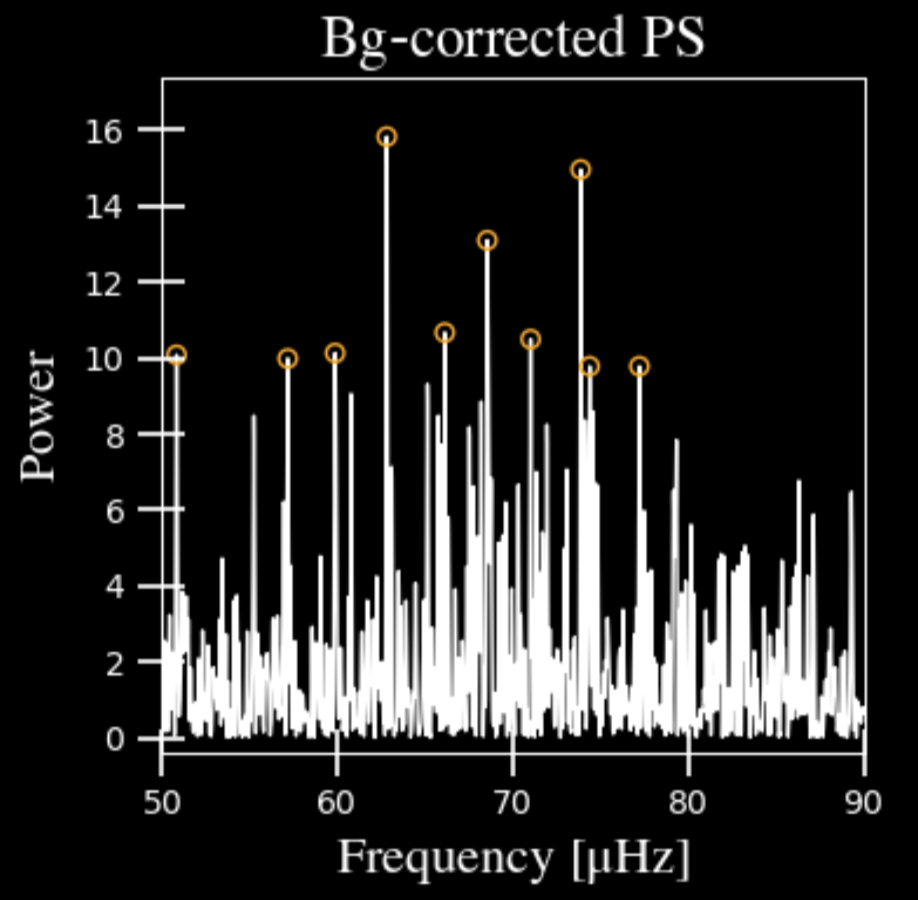}
  \caption{Background-corrected spectrum; the oscillation power excess becomes visible.}
  \label{fig:numax_306955660_c}
\end{subfigure}%
\hfill
\begin{subfigure}[t]{0.48\columnwidth}
  \centering
  \includegraphics[width=\linewidth]{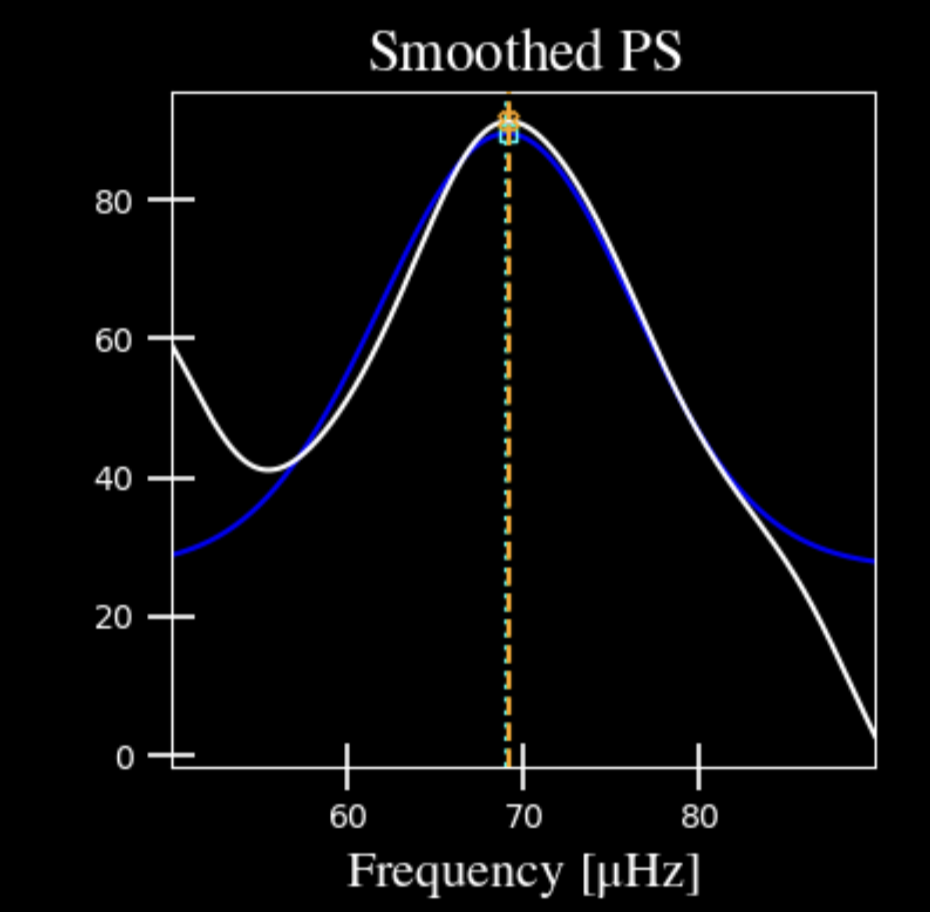}
  \caption{Smoothed spectrum (black) with Gaussian fit (red); dashed line marks $\nu_{\max}$.}
  \label{fig:numax_306955660_d}
\end{subfigure}

\caption{Example of the background modeling and $\nu_{\max}$ measurement process for TIC~306955660. Panel~(a) shows the detrended time series. Panel~(b) displays the initial background fit to the power spectrum. Panel~(c) presents the background-corrected power spectrum, where the oscillation signal emerges. Panel~(d) shows the smoothed spectrum with a Gaussian fit to the oscillation envelope; the vertical dashed line indicates the measured $\nu_{\max}$. This workflow shows how we extract global asteroseismic parameters from \textit{TESS} data for our cluster sample.}
\label{fig:numax_306955660}
\end{figure}

Figure~\ref{fig:numax_306955660} is a visual demonstration of the workflow we follow to extract $\nu_{\max}$ from the \textit{TESS} data. Starting from the detrended light curve (panel~a), we compute the power spectrum and fit a background model that accounts for granulation, stellar activity, and white noise (panel~b). After subtracting this background (panel~c), the oscillation power excess becomes clearly visible. We then smooth the background‑corrected spectrum and fit a Gaussian to the oscillation envelope to determine $\nu_{\max}$ (panel~d, dashed line). This same procedure is applied consistently to all targets.

\subsection{\'{E}chelle
 Diagram Construction and Validation}
\begin{figure*}
\gridline{
  \fig{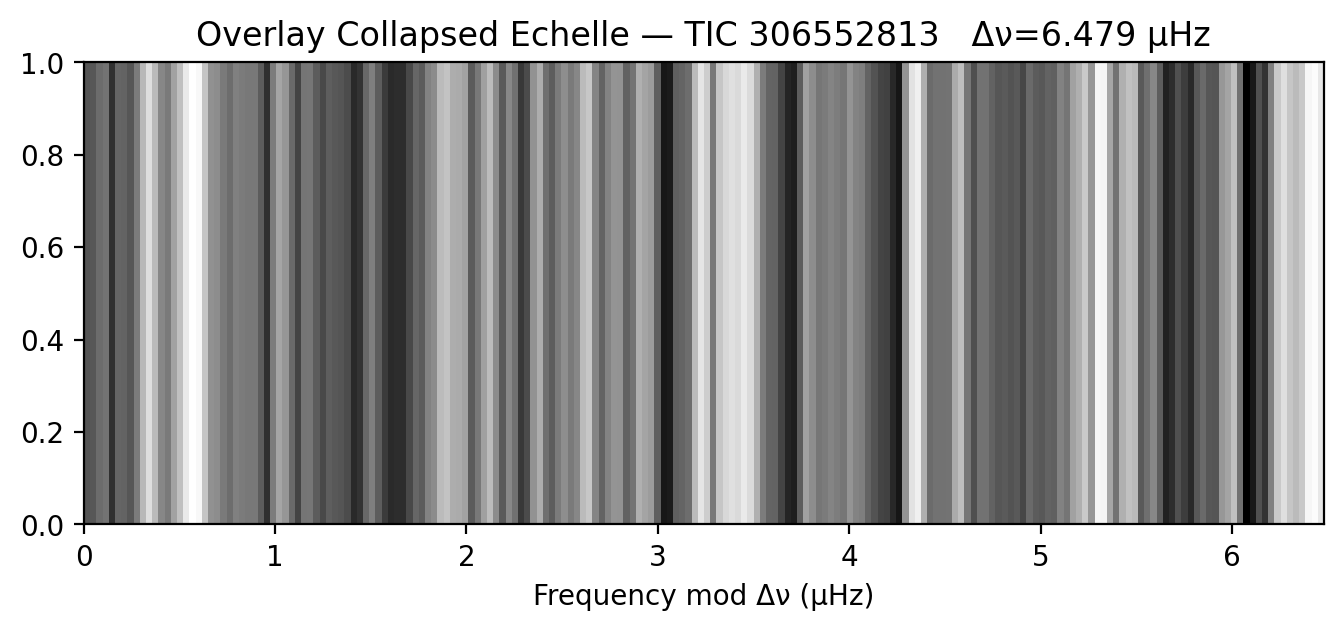}{0.45\textwidth}{(a) Collapsed/overlay \'{e}chelle (1D summary)}
  \fig{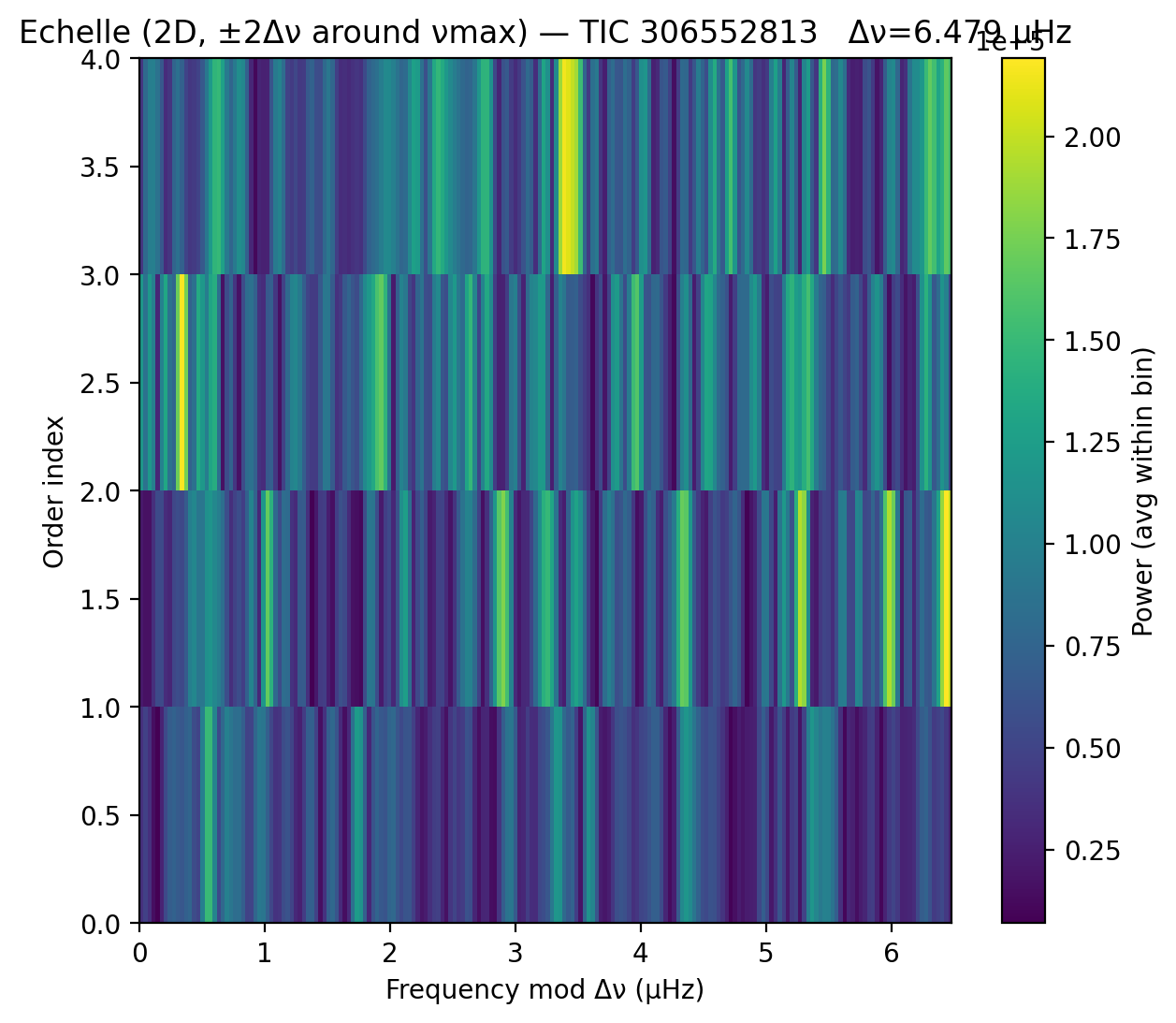}{0.45\textwidth}{(b) 2D \'{e}chelle ($\pm2\Delta\nu$ around \texorpdfstring{$\nu_{\max}$}{nu_max})}
}
\caption{\'{E}chelle diagnostics for TIC~306552813 with adopted large separation $\Delta\nu=6.479~\mu\mathrm{Hz}$. Panel (a) shows the collapsed/overlay representation: the power spectrum is folded by $\Delta\nu$ (x-axis: frequency modulo $\Delta\nu$) and collapsed across the displayed $\pm2\Delta\nu$ window; the collapse is replicated in y for visualization. Panel (b) shows the 2D \'{e}chelle: x is frequency modulo $\Delta\nu$ and y is an order index spanning the $\pm2\Delta\nu$ window around \texorpdfstring{$\nu_{\max}$}{nu_max}. Mode ridges appear as near-vertical bands of enhanced power in the 2D \'{e}chelle; for this target the most prominent vertical bands are concentrated near $\nu \bmod \Delta\nu \approx 0.5~\mu\mathrm{Hz}$, $\nu \bmod \approx 3.2$--$3.6~\mu\mathrm{Hz}$,  and $\nu \bmod \approx 5.3$--$5.4~\mu\mathrm{Hz}$. These same x-locations correspond to the strongest columns in the collapsed/overlay panel.}
\label{fig:echelle_triptych_306552813}
\end{figure*}

To ensure that we report the most accurate $\Delta\nu$ values, we created two different interactive \'{E}chelle diagrams: a 2D heatmap and a collapsed ridge alignment strip (See Figure \ref{fig:echelle_triptych_306552813}), using the \citet{daniel_hey_2020} code as a base and starting point. These all operate within a $\pm2\Delta\nu$ window around $\nu_{\max}$ and are adjustable using a Bokeh package Custom JavaScript (JS) slider which instantaneously populates the graph with $\Delta\nu$ values as the slider is manipulated. Each target was passed through both plots, inspected, and the best $\Delta\nu$ was recorded along with more accurate uncertainties, referring to how far in each direction can be feasibly interpreted as $\Delta\nu$. Our typical fractional uncertainty on $\Delta\nu$ is 2.26$\%$ , comparable to the APOKASC-3 values reported by \citet{Pinsonneault2025}. 

In Figure~\ref{fig:echelle_triptych_306552813}, the diagnostic purpose is ridge coherence. With the adopted $\Delta\nu$, the oscillation power repeats at consistent $\nu \bmod \Delta\nu$ from one radial order to the next, yielding clearly defined, near-vertical mode ridges in the 2D heatmap and corresponding concentrated columns in the collapsed strip. Although the spectrum is noisy, the ridge alignment remains stable across the $\pm2\Delta\nu$ window around $\nu_{\max}$, providing a visual confirmation that the measured $\Delta\nu$ captures the underlying comb-like spacing; perturbing $\Delta\nu$ visibly degrades this coherence by introducing ridge drift and broadening the collapsed power. The Jupyter notebook used to produce all \'{E}chelle figures and $\Delta\nu$ estimates is openly available on \href{https://github.com/carlimankowski/mankowski2025-open-cluster-asteroseismology}{GitHub}.

\subsection{Evolutionary State Determination, Corrections, and Scaling Relations}\label{subsec:evstate}

\begin{figure}[htbp]
\centering
\includegraphics[width=\linewidth]{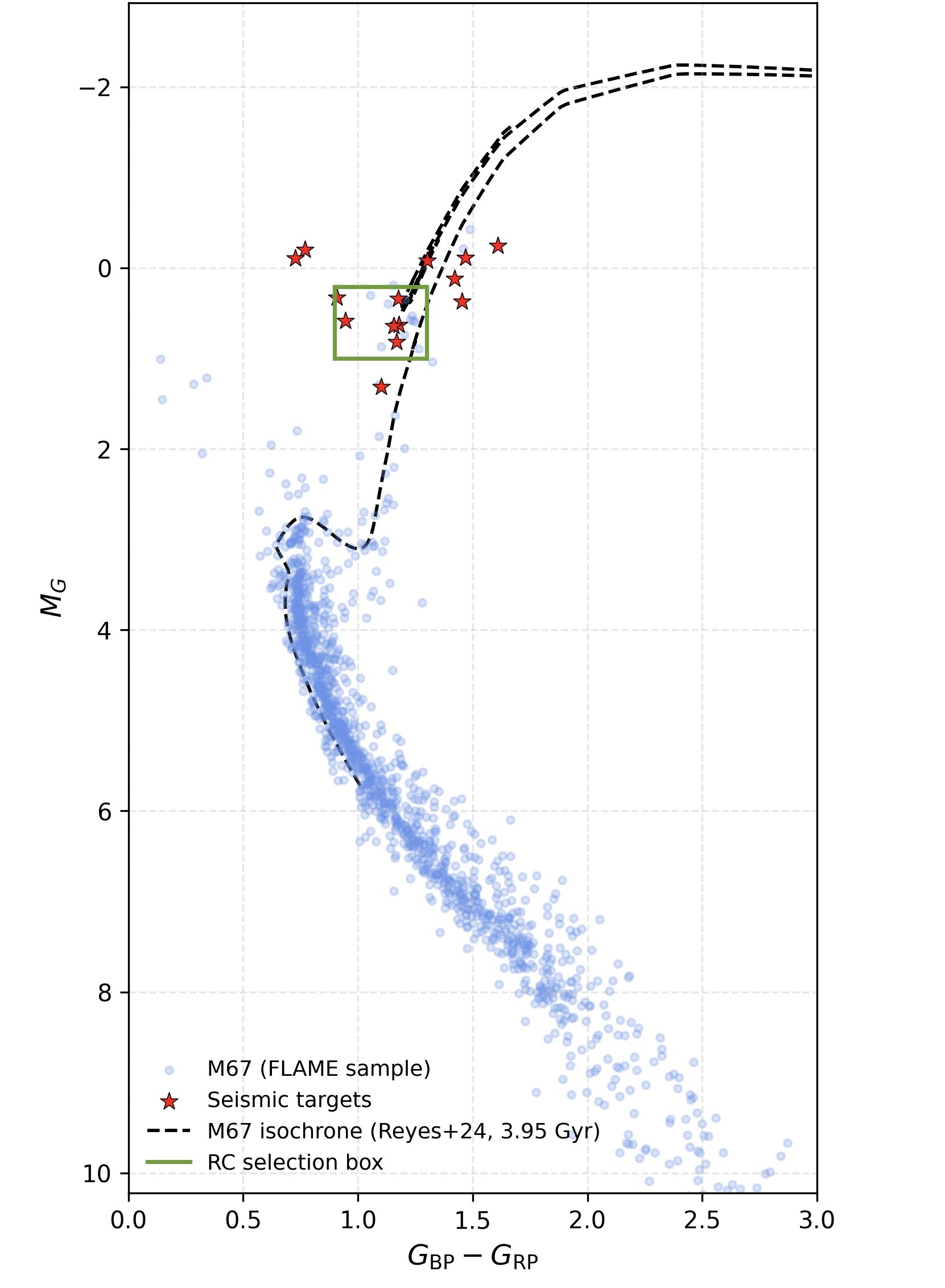}
\caption{
\textit{Gaia} color--magnitude diagram of seismic targets from our sample (red stars) and the 3.95\,Gyr M67 isochrone from \citet{Reyes2024}; black dashed line) with the open cluster M67 overplotted for reference (blue points). The green box marks the adopted red clump (RC) selection region in $(G_{\rm BP} - G_{\rm RP})$ and $M_G$.
}

\label{fig:cmd_rc}
\end{figure}

In order to apply the appropriate corrections to the $\Delta\nu$ estimates, we first identified the evolutionary state of each star. While in some cases it is possible to infer the evolutionary state directly from the oscillations\citep{Mosser2012_powerExcessKeplerRG, Vrard2025_KeplerEvoStatusCatalog, Elsworth2019_APOKASC_EvoState}, this is challenging in low signal to noise cases like our data. Therefore, we infer evolutionary states from the \textit{Gaia} color-magnitude diagram here, although future work may be able to improve upon this looking at each cluster in detail. Using the \textit{Gaia} color--magnitude diagram, we tagged stars as red clump (RC) if they lay within a conservative rectangular region centered on the RC overdensity,  using \texttt{RC\_COLOR\_MIN, RC\_COLOR\_MAX = 0.9, 1.3} in $(BP{-}RP)$ and \texttt{RC\_MG\_MIN, RC\_MG\_MAX = 0.2, 1.0} in $M_G$. \textit{Gaia}’s \textit{FLAME} radii are the stellar radii estimated by the \textit{Gaia} DR3 \textit{FLAME} module, which combines bolometric fluxes, effective temperatures, and parallaxes \citep{GaiaDR3_FLAME}, and has been used for population studies \citep{Andrae2023}. Applying the same selection box to an M67 \textit{Gaia} CMD constructed from \textit{FLAME} radii shows that it also cleanly traces the observed RC overdensity in that open cluster and shows that our cuts are consistent with well-behaved benchmark populations (Figure~\ref{fig:cmd_rc}). We feel this box is an appropriate approximation for the compact RC feature near $(BP{-}RP)\!\approx\!1.1$ and $M_G\!\approx\!0.6$ visible in \textit{Gaia} DR3 HRDs \citep[][their Fig.~10]{GaiaCollaboration2018_HRD}.
The approximation also parallels the RC calibrations in \textit{Gaia} bands that map $M_G$ (Figure \ref{fig:cmd_rc}) as a function of color in \citep{RuizDern2018_GaiaRC} and the \textit{Gaia}Unlimited example that selects RC stars by cutting around a fiducial RC ridge line in the \textit{Gaia} CMD \citep{GaiaCollaboration2018_HRD}. Evolutionary states were utilized in the application of model-based large-separation corrections $f_{\Delta\nu}$, calculated using \texttt{asfgrid} interpolations over stellar-models as a function of evolutionary state, $T_{\rm eff}$, [Fe/H], and the raw $(\Delta\nu,\nu_{\max})$ \citep{sharma2016, Sharma2023, Sharma2022}.

To integrate our PySYD measurements into a community-average scale, we used the mean SYD-to-ensemble ratios $(X_{\Delta\nu},X_{\nu_{\max}})$ from the APOKASC-3 inter-pipeline calibration and the sub-percent PySYD$\leftrightarrow$SYD consistency \citep{Pinsonneault2025, Chontos2022_pySYD, Grusnis2025_CVZ, Marasco2025_MetalPoor_pySYD}. We note that the APOKASC-3 calibration factors are defined as the consensus value divided by the pipeline value, which is the reciprocal of the original definition of the correction factor used in  \citet{White2011} and \citet{sharma2016}. We therefore take the reciprocal of the APOKASC-3 factors to obtain $X_{\Delta\nu}$ and $X_{\nu_{\max}}$, so that they represent the factor by which the pipeline value must be divided to bring it onto the consensus scale. This aligns with the definition of the model-based correction $f_{\Delta\nu}$ from \texttt{asfgrid}.

We computed corrected parameters using
\[
\begin{aligned}
\Delta\nu_{\rm corr} &= \frac{\Delta\nu_{\rm pySYD}}{X_{\Delta\nu}\,f_{\Delta\nu}},\\
\nu_{\max,{\rm corr}} &= \frac{\nu_{\max,{\rm pySYD}}}{X_{\nu_{\max}}\,f_{\nu_{\max}}},
\end{aligned}
\]
where $f_{\nu_{\max}}$ is a small adjustment derived by aligning seismic radii with \textit{Gaia}-based radii after the $\Delta\nu$ scale is fixed. After observation of Figure~\ref{fig:rad}, $f_{\nu_{\max}}$ is given a value of 1 since no meaningful correction was deemed necessary. This is consistent with our expectations for $f_{\nu_{\max}}$ around solar metallicity \citep{Lindsay2025}.

The corrected seismic parameters were then propagated through the standard scaling relations, with evolutionary-state dependent correction factors applied. Stellar radius and mass were computed as

\[
\begin{aligned}
\frac{R}{R_\odot} &= 
\left(\frac{\nu_{\max,{\rm corr}}}{\nu_{\max,\odot}}\right)
\left(\frac{\Delta\nu_{\rm corr}}{\Delta\nu_\odot}\right)^{-2}
\left(\frac{T_{\rm eff}}{T_{\rm eff,\odot}}\right)^{1/2}, \\[6pt]
\frac{M}{M_\odot} &= 
\left(\frac{\nu_{\max,{\rm corr}}}{\nu_{\max,\odot}}\right)^3
\left(\frac{\Delta\nu_{\rm corr}}{\Delta\nu_\odot}\right)^{-4}
\left(\frac{T_{\rm eff}}{T_{\rm eff,\odot}}\right)^{3/2}.
\end{aligned}
\]
Here, $\nu_{\max,{\rm corr}}$ and $\Delta\nu_{\rm corr}$ are the corrected parameters defined above, incorporating both the empirical SYD-to-ensemble calibration factors $(X_{\nu_{\max}},X_{\Delta\nu})$ and the model-based $f_{\Delta\nu}$ correction together with the empirical $f_{\nu_{\max}}$ factor. The solar reference values $(\nu_{\max,\odot},\Delta\nu_\odot,T_{\rm eff,\odot})$ adopted in this work are identical to those used in the APOKASC-3 calibration to maintain consistency across pipelines \citep{Pinsonneault2025}.

\subsection{Seismic Radii} \label{subsec:mrval}
We determine stellar radii from asteroseismic scaling relations \S\ref{subsec:evstate} with uncertainties from \S\ref{subsec:error}. For validation, we report them against external constraints. We use the \textit{Gaia} DR3 \textit{FLAME} catalog radii for red giants in the benchmark open cluster M67; these radii are derived from \textit{Gaia} parallaxes and inferred angular diameters/stellar parameters (i.e., not seismic). The detailed \textit{FLAME} comparison and our empirical re-calibration of radius uncertainties are presented in \S\ref{subsubsec:flame_radius_errors}.

\subsection{Seismic Masses and Cluster-wise Reporting} \label{subsec:mval}
We determine stellar masses from asteroseismic scaling relations \S\ref{subsec:evstate} with uncertainties from \S\ref{subsec:error}. For validation, we report the \emph{seismic} mass distributions of three clusters (Casado Alessi-1, NGC~752, and Theia~6046), each target plotted against the masses of other members of the same cluster  (Figure \ref{fig:mass}). This graphical analysis assesses internal, co-eval consistency rather than comparison to an external reference. In computing the cluster means and medians shown, we exclude the single Theia~6046 outlier (shown in Figure~\ref{fig:mass}).
Stars present in our broader sample but not confirmed as members of these three clusters are not plotted and do not enter the cluster-level statistics, but their seismic masses are included in the catalog.

\subsection{Stellar Age Estimation}\label{subsec:age}

To estimate ages for our red giant cluster members, we use the age modeling tools in \texttt{Kiauhoku} \citep{Claytor2020ASCL, Claytor2020}. The process compares each star to four different sets of stellar-evolution models following the multi-model strategy suggested by \citet{Pinsonneault2025}, with the individual models taken from YREC \citep{tayar2022_uncertainties}, Dartmouth/DSEP \citep{Dotter2008}, GARSTEC \citep{weiss2008, Serenelli2013}, and MIST \citep{Choi2016MIST}. We use EEP-based tracks so that evolutionary stages are defined in a consistent way across all models. These models span the mass and metallicity ranges expected for our open clusters and let us test how different model assumptions affect red giant age estimates \citep{Morales2025}. For each star, we provide \texttt{Kiauhoku} with the observed quantities it can compare directly to the models: asteroseismic mass, $\log g$, and metallicity.  
\texttt{Kiauhoku} then interpolates within each model and computes a $\chi^2$ measure of how well the model matches the observed values.  
These $\chi^2$ values are turned into likelihood weights, and we construct an age distribution for each by marginalizing over the weighted models. We emphasize that, given the specific set of inputs provided to \texttt{Kiauhoku}, this procedure is a deterministic grid lookup and interpolation rather than a Bayesian inference. For the final age estimate of each star, we use MIST as our primary model. The Dartmouth, YREC and GARSTEC models are used as comparison models, and we report their mean ages to allow direct comparison with other work. Since different model families can give ages that differ \citep{theodoridis2025_asteroseismically,tayar2022_uncertainties,Morales2025,Pinsonneault2025}, using multiple models provides an estimate of the model-dependent uncertainty. This systematic age uncertainty is then added in quadrature to the observational age uncertainty, which we infer by perturbing the input mass and metallicity by $\pm1\sigma$ following the procedures of \citet{theodoridis2025_asteroseismically} and \citet{Morales2025}. The Jupyter notebook used to produce all age estimations is openly available on \href{https://github.com/carlimankowski/mankowski2025-open-cluster-asteroseismology/tree/main8l}{GitHub}.

\subsubsection{Cluster-Level Age Estimation}
\label{subsubsec:cluster_age_estimation}
For each star, the MIST stellar-evolution grid in \texttt{Kiauhoku} returns an age estimate $t_i$ in Gyr. We define the cluster age as the arithmetic mean of the member-star ages, \begin{equation}
\bar{t} = \frac{1}{N} \sum_{i=1}^{N} t_i.
\end{equation}

This quantity represents the characteristic red-giant age of the cluster as inferred from the MIST models, and we adopt $\bar{t}_{\mathrm{MIST}}$ as our final seismic cluster age.

\subsection{Error Propagation}\label{subsec:error}

We propagated uncertainties with first-order partial-derivative error propagation, assuming independent inputs. We describe (i) how we defined the measurement uncertainties on $\nu_{\max}$ and $\Delta\nu$ from PySYD and by-eye checks, (ii) how these propagate into radius and mass and are tested against catalog radii, and (iii) how we constructed the final age error bars at both the stellar and cluster levels.
\subsubsection{Seismic Measurement Uncertainties from PySYD and By-Eye Checks}

For each star, we first measured $\nu_{\max}$ and $\Delta\nu$ with \texttt{PySYD} and then inspected the power spectra by eye to confirm that the reported peaks and uncertainties were reasonable. The adopted $\sigma_{\nu_{\max}}$ and $\sigma_{\Delta\nu}$ reflect the formal \texttt{PySYD} uncertainties, moderated by manual checks. Uncertainties in $T_{\mathrm{eff}}$, $\log g$, and $[\mathrm{Fe}/\mathrm{H}]$ were taken directly from the relevant photometric or spectroscopic catalogs.

We treat $\nu_{\max}$, $\Delta\nu$, and $T_{\mathrm{eff}}$ as independent input quantities in the standard asteroseismic scaling relations. Using a first-order (Gaussian) Taylor expansion in $\ln X$. 
Applying standard error propagation yields the expressions below.

\begingroup\small
\vspace{0.5cm}
\noindent\textbf{Seismic mass error propagation:}\;

\(
\left(\tfrac{\sigma_M}{M}\right)^2 =
9\left(\tfrac{\sigma_{\nu_{\max}}}{\nu_{\max}}\right)^2
+ 16\left(\tfrac{\sigma_{\Delta\nu}}{\Delta\nu}\right)^2
+ \tfrac{9}{4}\left(\tfrac{\sigma_{T_{\rm eff}}}{T_{\rm eff}}\right)^2
\)
\vspace{0.5cm}

\textbf{Seismic radius error propagation:}\;

\(
\left(\tfrac{\sigma_R}{R}\right)^2 =
\left(\tfrac{\sigma_{\nu_{\max}}}{\nu_{\max}}\right)^2
+ 4\left(\tfrac{\sigma_{\Delta\nu}}{\Delta\nu}\right)^2
+ \tfrac{1}{4}\left(\tfrac{\sigma_{T_{\rm eff}}}{T_{\rm eff}}\right)^2
\)
\vspace{0.5cm}

\textbf{Seismic surface gravity error propagation:}\;

\(
\left(\tfrac{\sigma_g}{g}\right)^2 =
\left(\tfrac{\sigma_{\nu_{\max}}}{\nu_{\max}}\right)^2
+ \tfrac{1}{4}\left(\tfrac{\sigma_{T_{\rm eff}}}{T_{\rm eff}}\right)^2
\)
\endgroup

\subsubsection{FLAME-Based Radius Error Calibration}
\label{subsubsec:flame_radius_errors}

To validate our seismic radius scale and calibrate realistic radius uncertainties, we compare the scaling–relation radii to \textit{Gaia} DR3 \textit{FLAME} catalog values \citep{GaiaDR3_FLAME}. In the benchmark open cluster M67, the published \textit{FLAME} radii for 18 coeval giants show a residual distribution compared to the radii in \citet{Reyes2024}, where we plot $(R_{\rm seis} - R_{\rm \textit{FLAME}})/\sigma_{\rm \textit{FLAME}}$ for M67 giants. The broad and asymmetric shape of this distribution represents underestimated uncertainties in the \textit{FLAME} radii. To promote realistic radius errors in our analysis, we adopt an uncertainty of $\pm 2.0\,\sigma$ of the \textit{FLAME}-reported uncertainty for red-giant radii. This data-driven treatment of the scatter provides the most straightforward and conservative set of error bars while remaining consistent with our seismically derived radius estimates. Because these systematics are significant, we regard the \citet{Reyes2024} seismic radii as the most accurate and up-to-date radius determinations available for these stars, and we use them as our reference scale.

\begin{figure}[htpb]
  \centering

 \includegraphics[width=0.9\columnwidth]{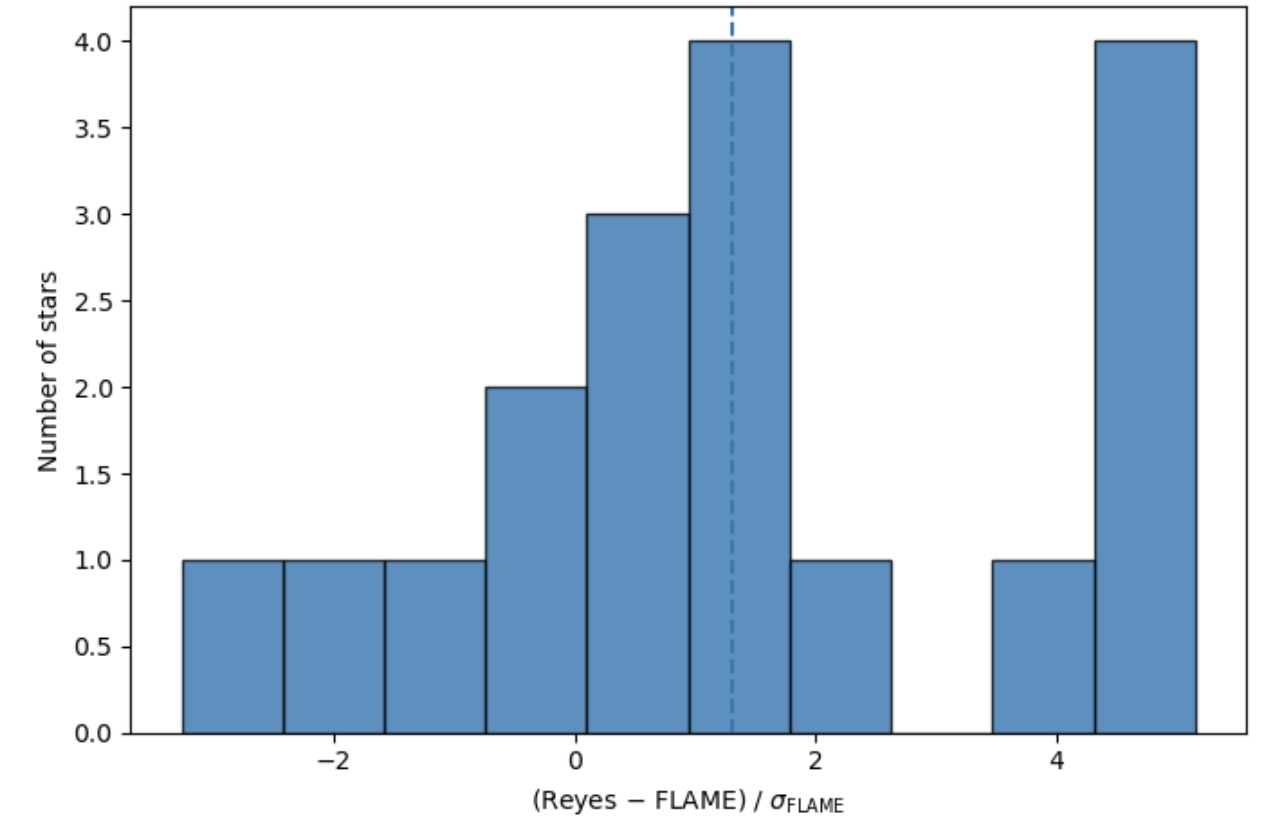}\\[4pt]
\includegraphics[width=0.9\columnwidth]{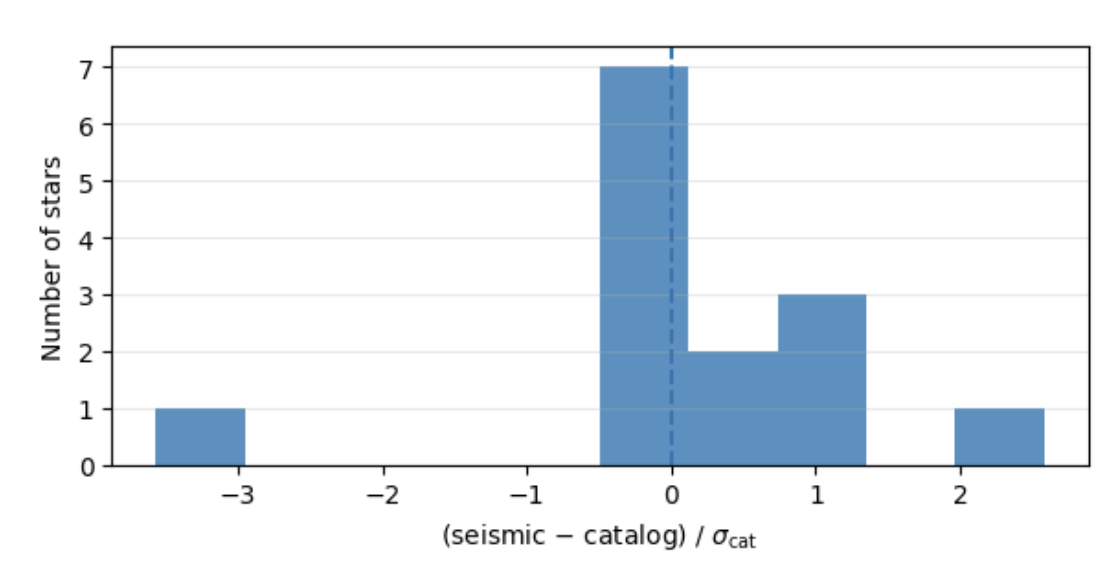}

  \caption{
  Normalized seismic--\textit{FLAME} radius differences for M67.
  The upper panel reproduces the \textit{FLAME}–asteroseismic comparison from \citet{Reyes2024}, while the lower panel shows our own comparison
  using the same \textit{FLAME} radii.
  In both cases, the distribution of
  $(R_{\rm seis}-R_{\rm \textit{FLAME}})/\sigma_{\rm \textit{FLAME}}$ demonstrates the large
  disparity between the \textit{FLAME} radii  and motivates our adoption of a
  conservative $\pm2\sigma$ uncertainty for the radii of the stars studied here.
  }
  \label{fig:flame}
\end{figure}

\subsubsection{Age Uncertainties at the Stellar and Cluster Levels}

For each star, we define age uncertainties by rerunning the \texttt{Kiauhoku} fits after shifting the input
mass and metallicity by their $\pm 1\sigma$ errors. The resulting changes in the recovered ages capture the observational contribution to the age error (through mass and metallicity). We then add in quadrature the spread in age between the different stellar-evolution grids at fixed star to account for model dependence. The stellar age uncertainty can be written as
\begin{equation}
    \sigma_{t,\,\star}^{2} = \sigma_{\rm obs}^{2} + \sigma_{\rm model}^{2},
\end{equation}

where $\sigma_{\rm obs}$ comes from perturbing the input mass and metallicity and $\sigma_{\rm model}$ reflects the grid-to-grid dispersion for that star. These $\sigma_{t,\,\star}$ values set the error bars on all star-by-star age plots.
At the cluster level, the MIST grid provides a single mean age for the member stars (Section~\ref{subsubsec:cluster_age_estimation}) and an associated uncertainty that reflects both the typical stellar age errors and the scatter of ages within the cluster. We therefore adopt the MIST cluster age, $t_{\mathrm{MIST}}$, as our final seismic age and assign an uncertainty
\begin{equation}
    \sigma_{\mathrm{MIST}}^{2}
    = \frac{1}{N^{2}} \sum_{i=1}^{N} \sigma_i^{2}
      + s_{\mathrm{star}}^{2},
\end{equation}
where $\sigma_i$ is the age uncertainty for star $i$, and $s_{\mathrm{star}}$ is the sample-standard deviation of the member-star ages. In practice, the quoted $t_{\rm MIST} \pm \sigma_{\rm MIST}$ captures both uncertainty in the stellar inputs and the intrinsic spread of inferred ages within each cluster.

\section{Analysis}
\subsection{Catalog Presentation}
We begin by presenting the full sample of stars selected for this study, including their basic astrometric and spectroscopic parameters. The catalog is constructed from \textit{Gaia} DR3 and cross-matched with \textit{TESS} Input Catalog identifiers. A sample of the full table is presented as Table \ref{tab:coldefs}. We illustrate the sample in a color--magnitude diagram (CMD) as a consistency check in Figure \ref{fig:hr_all}. This catalog includes columns which indicate whether or not a target has “passed” each cut, marked with "y" for yes and "n" for no. They are organized so that targets with the most “y”s are at the top and the least at the bottom, ensuring simplicity when searching for high quality targets. The full machine-readable catalog is available on \href{https://zenodo.org/records/17932039}{Zenodo}.

\subsubsection{Unknown Clusters}
The clusters marked as "Unknown 1" (TIC306552754) and "Unknown 2" (TIC306552795) in the catalog were both designated as "Casado Alessi-1" in the \citet{Hunt2023} paper, but upon examination of another focused cluster membership study \citep{CantatGaudin2020}, we determined that they are likely non-members. A third source, "Unknown 3" (TIC459055617), originally tagged as an NGC\,2682 (M67) member, but the target is not analyzed by \citet{Reyes2025} and its asteroseismic parameters place it well outside the established locus for the cluster, leading us to classify it as a non-member. However, all three objects have strong asteroseismic signals and are thus reported in the table.

\subsection{\texorpdfstring{$\Delta\nu$–$\nu_{\max}$}{Delta nu--nu max} Relation of the Quality-Selected Sample}

\label{subsec:analysis_dnu_numax}
\begin{figure}[htpb]
  \centering
  \includegraphics[width=\columnwidth]{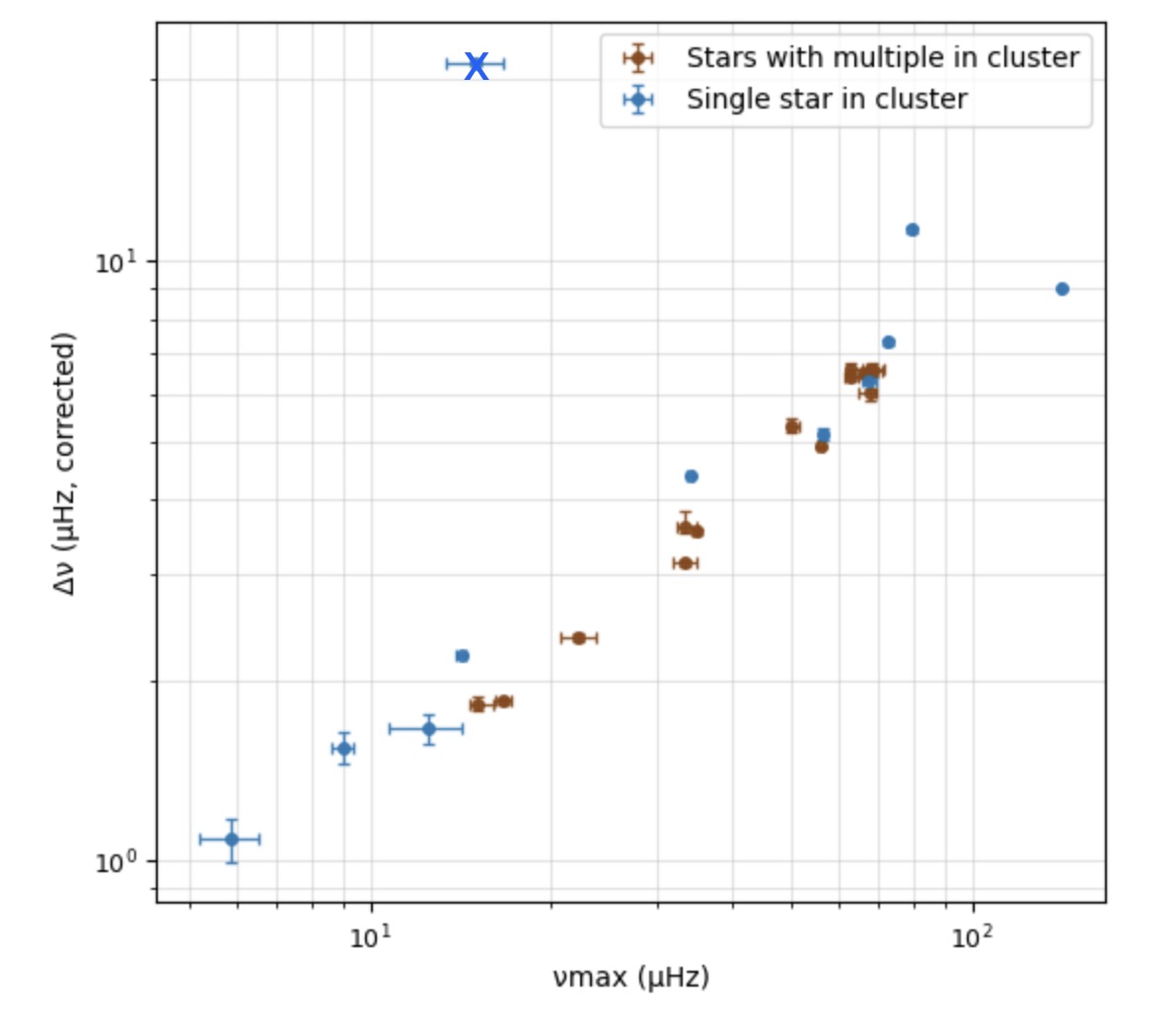}
    \caption{Corrected $\Delta\nu$ vs.\ \texorpdfstring{$\nu_{\max}$}{nu\_max}. With the exception of a single outlier (marked as a blue ``x''), our corrected $\Delta\nu$ values are strongly correlated with $\nu_{\max}$ in the expected way, indicating that the seismic parameters are reliable.}

  \label{fig:numaxdnu}
\end{figure}

As part of our analysis, we examine the $\Delta\nu$–$\nu_{\max}$ relation for the stars that passed the crowding, bleeding, and sector-quality filters described in \S\ref{subsec:crowding}. The retained sample follows the expected trend in $\Delta\nu$ versus $\nu_{\max}$, with outliers largely removed by these cuts; however, non-passing objects are still  in figures. This is displayed in Fig.~\ref{fig:numaxdnu} and confirms that the quality selection standards presented in this method preserve seismic scaling, reduce contamination, and identify true oscillating giants.

\subsection{\texorpdfstring{Log $g$ Comparison}{Log g Comparison}}

We validate surface gravity estimates by comparing spectroscopic $\log g$ with values inferred seismically via the $\nu_{\max}$ scaling relation, plotting $\log g_{\rm spec}$ against $\log g_{\rm seis}$ with uncertainties and a $1{:}1$ reference line. On the full set ($N=25$), we find a bias of $+0.108\,\mathrm{dex}$, median offset $+0.015\,\mathrm{dex}$, $\mathrm{RMSE}=0.236\,\mathrm{dex}$, robust spread $1.4826\times\mathrm{MAD}=0.124\,\mathrm{dex}$, and $R^{2}=0.744$, 
where $\mathrm{MAD}$ is the median absolute deviation. 
After removing two outliers using a MAD-$z$ cut (with MAD-$z$ defined as the absolute deviation from the median divided by $1.4826\times\mathrm{MAD}$), 
the remaining $N=23$ stars yield a bias of $+0.065\,\mathrm{dex}$, median offset $-0.000\,\mathrm{dex}$, 
$\mathrm{RMSE}=0.167\,\mathrm{dex}$, robust spread $1.4826\times\mathrm{MAD}=0.112\,\mathrm{dex}$, 
and $R^{2}=0.828$, indicating consistency between spectroscopic and seismic $\log g$.

\begin{figure}[htpb]
  \centering
  \includegraphics[width=\columnwidth]{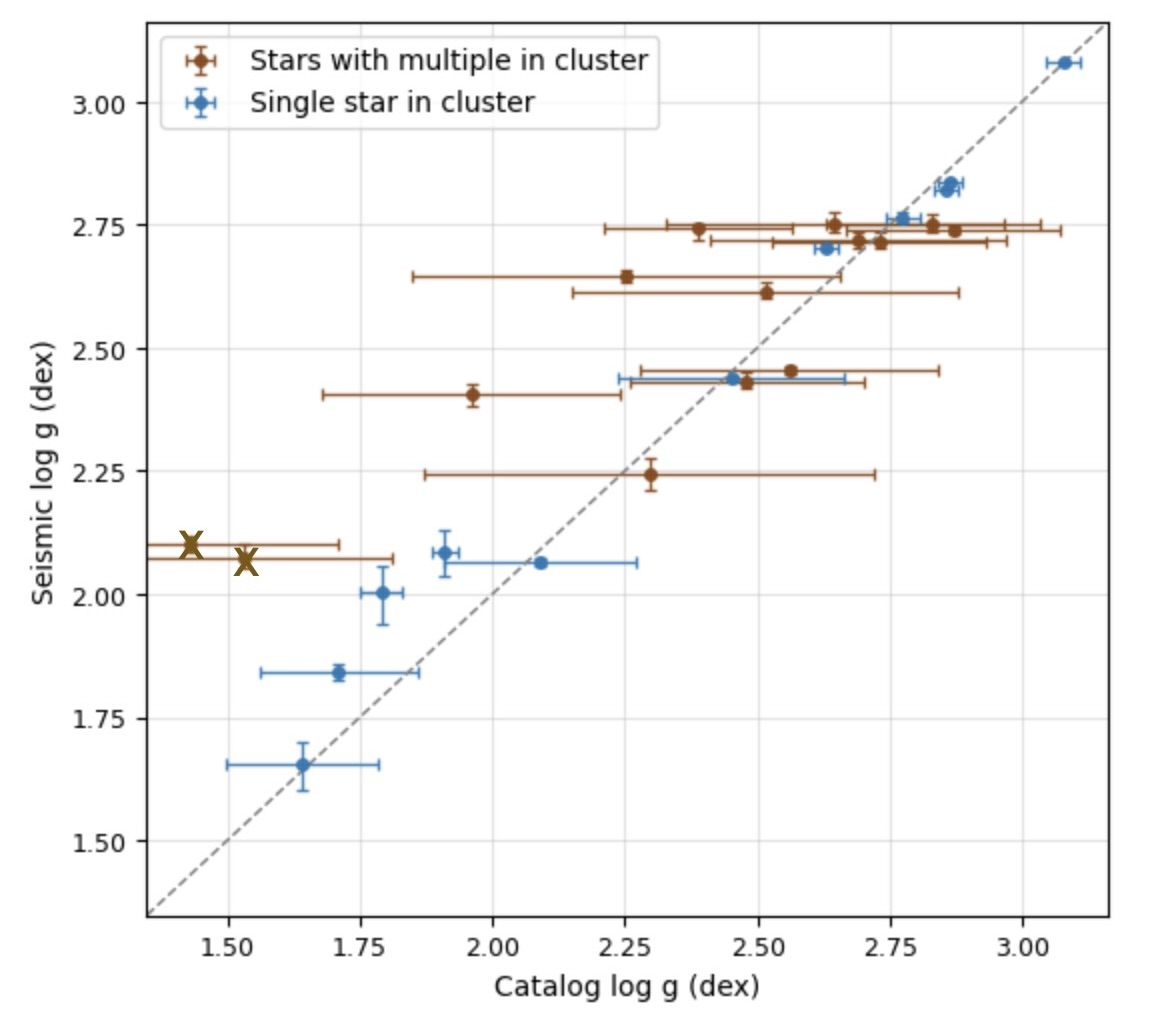}
  \caption{Seismic $\log g$ versus spectroscopic catalog $\log g$. The dashed line shows the 1:1 relation. With the exception of two outliers (marked as blue ``x'' symbols), our $\log g$ values inferred from the seismic $\nu_{\max}$ are strongly correlated with the spectroscopic $\log g$ values reported in the literature. We show only the $N=23$ stars with literature spectroscopic $\log g$ measurements; the full set of 25 seismic $\log g$ values is given in Table \ref{tab:coldefs}.}

  \label{fig:logg}
\end{figure}

\subsection{Radius Comparison}
Using the method outlined in \S\ref{subsec:mrval}, stellar radii are compared between seismic scaling-relation estimates and \textit{Gaia}-based determinations. For each star, we use $\Delta\nu$, $\nu_{\max}$, and $T_{\rm eff}$ to compute $R_{\rm seis}$ and compare these values against the \textit{FLAME} $R_{\rm \textit{Gaia}}$ derived from \textit{Gaia} parallaxes and bolometric corrections \citep{GaiaDR3_FLAME}. We quantify the agreement using fractional differences and dispersion, and present scatter plots with $1{:}1$ reference lines to assess systematic offsets. On the full member set ($N=13$), we find a bias of $+0.396\,R_{\odot}$, median offset $+0.165\,R_{\odot}$, $\mathrm{RMSE}=5.165\,R_{\odot}$, robust spread $1.4826\times\mathrm{MAD}=0.709\,R_{\odot}$, and $R^{2}=-0.132$. After removing four outliers with a MAD-$z$ cut, the remaining $N=9$ targets yield a bias of $+0.082\,R_{\odot}$, median offset $+0.005\,R_{\odot}$, $\mathrm{RMSE}=0.428\,R_{\odot}$, robust spread $1.4826\times\mathrm{MAD}=0.469\,R_{\odot}$, and $R^{2}=0.992$, indicating consistency between seismic and \textit{Gaia}-based radii and validating our choice to use $f_{\nu_{\max}}$ = 1.

\begin{figure}[htpb]
  \centering
  \includegraphics[width=\columnwidth]{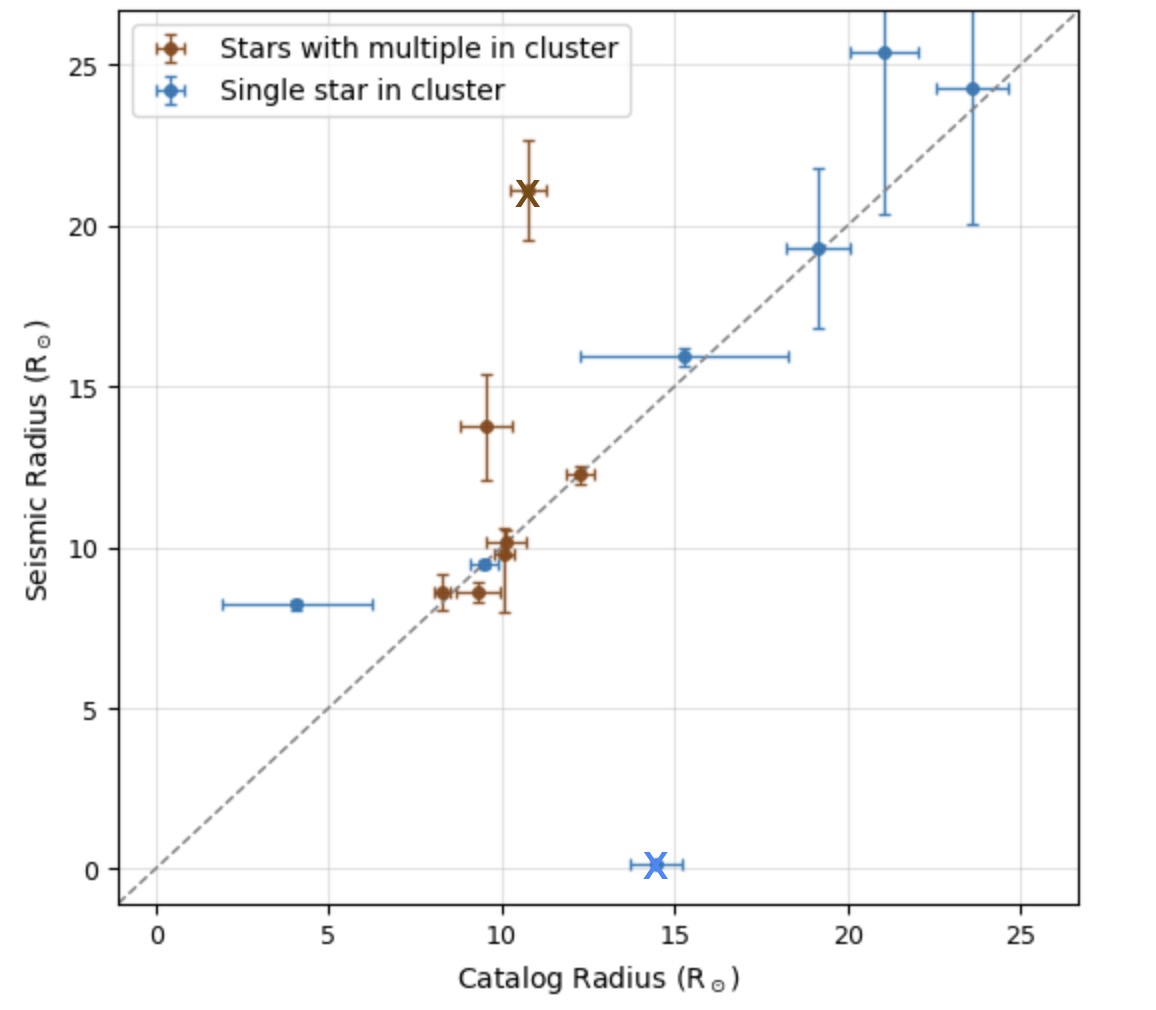}
  \caption{Seismic radius versus catalog radius with the 1:1 line shown as a dashed line. Error bars on the seismic radii reflect the uncertainties from our stellar age modeling, while error bars on the catalog radii use the adopted catalog radius uncertainties (including our inflated values, as described in ~\S\ref{subsec:mrval}).}

  \label{fig:rad}
\end{figure}

\subsection{Mass Comparison}

\begin{figure}[htpb]
  \centering
  \includegraphics[width=\columnwidth]{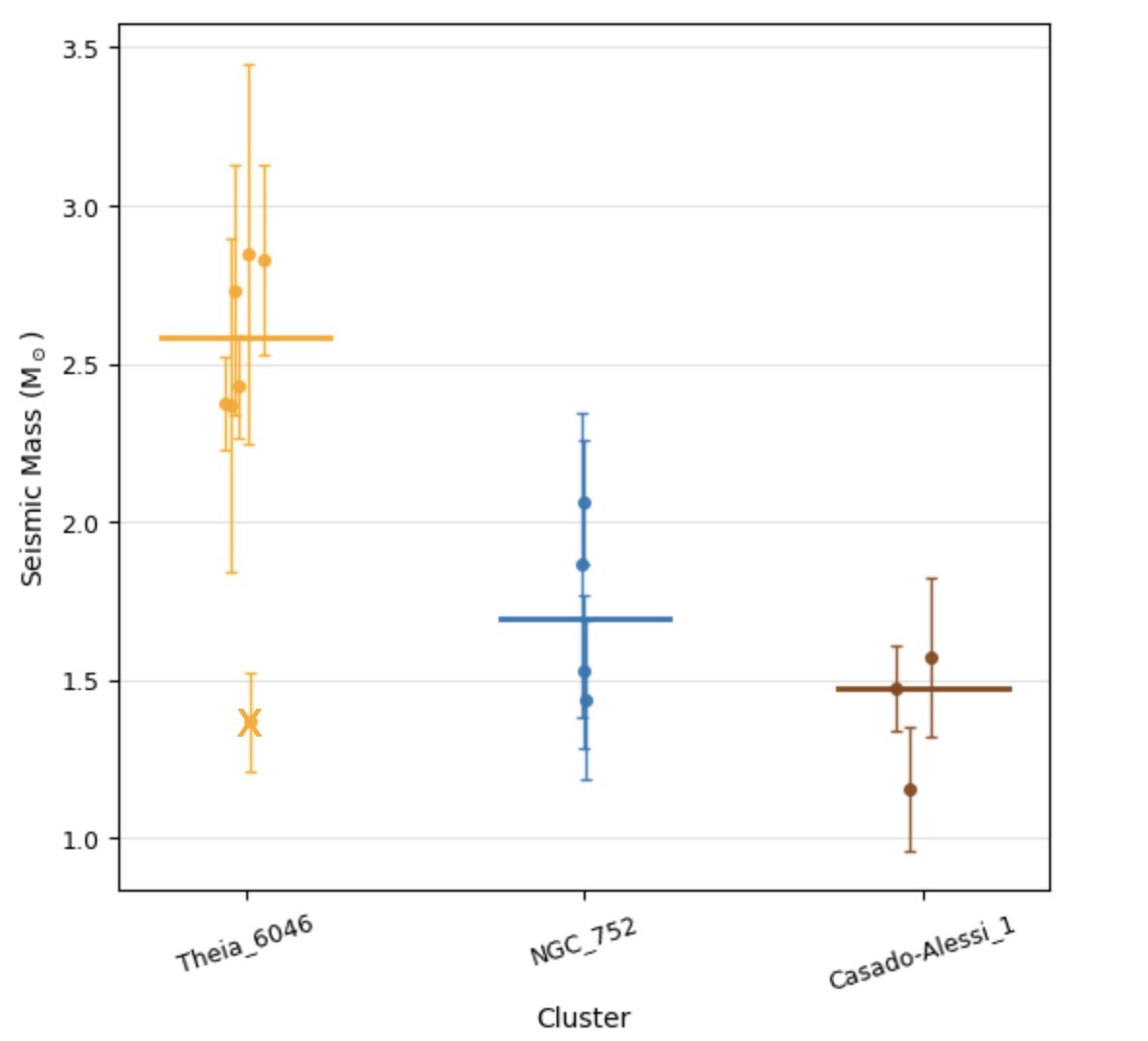}
  \caption{Seismic mass distributions by cluster with cluster medians (horizontal bars).}
  \label{fig:mass}
\end{figure}

Using the method described in \S\ref{subsec:mval}, we present the seismic masses for all clusters in our sample. We report on the internal coherence of each cluster’s mass distribution. In Figure \ref{fig:mass}, each cluster forms a generally tight cluster around a single mass. Each mass in our sample is derived independently from its own seismic observables and this type of coherence is expected for members of a coeval population. 

\subsection{Stellar Age Validation}
As stated in \S\ref{subsec:age}, we adopt four independent sets of stellar-evolution models (YREC, Dartmouth/DSEP, GARSTEC, and MIST) and follow the multi-model strategy suggested by \citet{Pinsonneault2025}. We use the dispersion among the four models to quantify the model-dependent component of the age uncertainty. For red clump stars, we assume no mass loss in the models when estimating ages, so the inferred ages correspond to stars that retain their initial masses, a reasonable limiting case for young clusters where mass loss is expected to be small \citep{Pinsonneault2025}.

Although our stars reside in clusters, not all of these clusters have a single well-determined age in the literature. For the 12 clusters with seismically inferred ages in this work, we compile the published ages and their spread in Table~\ref{tab:age_subset}. Our model-based cluster ages are then compared to the mean literature ages, with error bars reflecting the literature scatter, in Figure~\ref{fig:star_vs_cluster_age}. For all clusters, Figure~\ref{fig:ages_boxplot_by_cluster} shows the distribution of stellar ages within each cluster relative to the corresponding literature value. In general, our results are consistent with literature values, although in some cases, our data show a preference for one group of ages over another. The level of internal consistency indicates that the method yields reliable cluster ages and supports its use as a basis for deriving ages of stars outside of clusters, where such calibrated seismic ages are particularly valuable for galactic-archaeology applications.

\begin{table*}
\caption{Compiled ages of the star clusters. \label{tab:cluster_ages}}
\centering
\begin{tabular}{l c c c c c}
\hline
\hline
Cluster & 
\parbox{2.5cm}{\centering Number of lit.\ age sources} & 
\parbox{2.3cm}{\centering Mean lit.\ age (Gyr)} & 
\parbox{2.3cm}{\centering Cantat-Gaudin (2020) age (Gyr)} & 
\parbox{2.8cm}{\centering Hunt \& Reffert (2023) age (Gyr)} & 
\parbox{2.3cm}{\centering Seismic age (Gyr)} \\
\hline
IC\,348           & 8 & 0.005 $\pm$ 0.003 & 0.012 & 0.006 & 1.189 $\pm$ 0.309 \\
Unknown\,1        & 0 & --                & --    & --    & 0.734 $\pm$ 0.280 \\
Unknown\,2        & 0 & --                & --    & --    & 0.221 $\pm$ 0.237 \\
LISC\_3534        & 1 & 0.149             & --    & 0.149 & 0.118 $\pm$ 0.752 \\
COIN-\textit{Gaia}\_30     & 2 & 0.206 $\pm$ 0.052 & 0.257 & 0.154 & 4.251 $\pm$ 3.274 \\
UPK\,287          & 2 & 0.222 $\pm$ 0.060 & 0.257 & 0.154 & 0.667 $\pm$ 0.146 \\
HSC\_1052         & 1 & 0.068             & 0.068 & --    & 0.108 $\pm$ 0.059 \\
Unknown\,3        & 0 & --                & --    & --    & 6.405 $\pm$ 2.643 \\
UPK\,237          & 2 & 0.103 $\pm$ 0.048 & 0.151 & 0.055 & 0.262 $\pm$ 0.124 \\
Casado Alessi-1   & 4 & 1.111 $\pm$ 0.337 & 1.445 & 0.749 & 1.706 $\pm$ 0.583 \\
NGC\,752          & 4 & 1.299 $\pm$ 0.166 & 1.175 & 1.421 & 1.433 $\pm$ 0.366 \\
Theia\,6046       & 1 & 3.807             & --    & 3.807 & 0.525 $\pm$ 0.082 \\
\hline
\end{tabular}

\vspace{0.2cm}
\small
\textbf{References:}
(1) \citet{Strom1974IC348};
(2) \citet{Herbig1998IC348};
(3) \citet{TrullolsJordi1997IC348};
(4) \citet{Haisch2001IC348};
(5) \citet{Lada2006IC348};
(6) \citet{Luhman1999IC348};
(7) \citet{Nikoghosyan2015IC348};
(10) \citet{HuntReffert2023GaiaClusters};
(12) \citet{Carrera2019M67Halo};
(13) \citet{Sarajedini2009M67Age};
(14) \citet{Barnes2016M67Gyro};
(15) \citet{Geller2021M67Binaries};
(16) \citet{Carraro1994OldOpenClusters};
(17) \citet{Hassan1972NGC1662};
(18) \citet{PenaPeniche1994NGC1662};
(19) \citet{MaitzenHensberge1981NGC1662};
(20) \citet{Reyes2025}
\end{table*}

\subsubsection{Casado Alessi-1}
As shown in Figure \ref{fig:ages_boxplot_by_cluster}, the interpolated ages for members of Casado Alessi-1 are not fully consistent with the literature range presented in past works. We were able to locate four age estimates for the cluster, 0.78, 0.8, 1.445, and 1.45 Gyr. Across our three estimations, the mean ages for Casado Alessi-1 are $1.63$, $1.94$, and $1.55$\,Gyr; in each case, the quoted $1\sigma$ error bars extend down into the 0.8--1.45\,Gyr literature range, but the central values remain systematically higher. To explore this tension, we plot MIST isochrones of various ages on top of our catalog members of Casado Alessi-1, removing any targets that are outside of a $2\sigma$ window in parallax and proper motion around the adopted cluster means (or with $\mathrm{RUWE} > 1.4$) and may not be actual cluster members (Figure \ref{casadoiso}). A discrepancy may arise from different membership selections or fitting choices in past or current work. Given our seismic uncertainties (Table \ref{tab:age_subset}) and analysis of isochrones compared to the cluster as a whole, we suspect that the age of this cluster may have been previously underestimated, but more detailed study is likely needed to solidify this age. We include Table \ref{tab:age_subset} specifically to demonstrate that large per-star age uncertainties and grid-to-grid differences are common which makes the prevalence of error in cluster age determinations explicit. The full isochrone-fitting workflow
for Casado Alessi-1 is available on
\href{https://github.com/carlimankowski/mankowski2025-open-cluster-asteroseismology/blob/main/isochrones.ipynb}{GitHub}.

\begin{deluxetable*}{rrrrrrrrrr}
\tablecaption{Ages and model comparisons for three Casado Alessi-1 giants.\label{tab:age_subset}}
\tablewidth{0pt}
\tablehead{
\colhead{TICID} & \colhead{age\_gyr} & \colhead{age\_std} &
\colhead{age\_gyr\_$-$1$\sigma$} & \colhead{age\_gyr\_$+$1$\sigma$} &
\colhead{age\_MIST} & \colhead{age\_Dartmouth} &
\colhead{age\_GARSTEC} & \colhead{age\_YREC}
}
\startdata
306552813  & 1.627 & 0.811 & 2.254 & 1.127 & 1.843 & 2.746 & 2.469 & 2.299 \\
306955660  & 1.944 & 0.905 & 2.748 & 1.261 & 2.169 & 2.950 & 2.479 & 2.301 \\
306187638  & 1.547 & 0.780 & 2.203 & 1.150 & 1.764 & 2.689 & 2.465 & 2.297 \\
\enddata
\end{deluxetable*}

\subsubsection{NGC 752}
As shown in Figure \ref{fig:ages_boxplot_by_cluster}, the interpolated ages
for members of NGC\,752 are broadly consistent with the literature range
reported in past work.  Published ages for this nearby open
cluster fall in the 1.3--1.6\,Gyr range
\citep{CantatGaudin2020,Hunt2023}.
Our seismically derived ages for NGC\,752 lie within this interval; in each
case, the quoted $1\sigma$ error bars overlap the canonical literature range
and the central values do not show any strong offset.  

To further assess this agreement, we repeat the isochrone-based test used for
Casado Alessi-1.  In Figure \ref{ngciso} we plot
\textit{Gaia} DR3 color--magnitude data for likely members of NGC\,752,
removing stars that lie outside a $2\sigma$ window in parallax and proper
motion around the adopted cluster means (or with $\mathrm{RUWE} > 1.4$), and
marking the rejected targets with red crosses.  We overplot MIST isochrones
with ages of 1.30, 1.40, 1.50, and 1.60 Gyr.  The 1.3--1.5\,Gyr isochrones provide a
visibly good match to the main sequence, turnoff, and base of the red giant
branch, reinforcing the conclusion that our seismic ages for NGC\,752 are
consistent with the literature values.  The full isochrone-fitting workflow
for NGC\,752 is available on
\href{https://github.com/carlimankowski/mankowski2025-open-cluster-asteroseismology/blob/main/isochrones.ipynb}{GitHub}.

\subsubsection{Theia 6046} \label{subtheia}
As shown in Figure \ref{fig:ages_boxplot_by_cluster}, the interpolated ages for
members of Theia\,6046 differ from the value reported in large catalog work.
Hunt \& Reffert (\citeyear{Hunt2023}) give a cluster age of
$\log t \approx 9.58$ (roughly $3.8$\,Gyr), whereas our seismic
estimation yields ages on the order of $\sim$Gyr, substantially younger than
this catalog value. 

To explore this discrepancy, we repeat the isochrone-based test used for
Casado Alessi-1 and NGC 752.  In Figure \ref{theiaiso} we plot
\textit{Gaia} DR3 color--magnitude data for likely members of Theia\,6046,
removing stars that lie outside a $2\sigma$ window in parallax and proper
motion around the adopted cluster means (or with $\mathrm{RUWE} > 1.4$), and
marking the rejected targets with red crosses.  We overplot MIST isochrones
with ages of 0.3, 0.5, 0.7, 1.0, 1.5, and 3.7\,Gyr.  The younger ($\lesssim 1$\,Gyr) isochrones provide a
visibly better match to the main sequence and turnoff than the older
$\sim 3.7$\,Gyr track, suggesting that the true cluster age is likely younger
than the catalog value.  As with Casado Alessi-1, details of the membership
selection and fitting methodology likely contribute to the discrepancy, but more detailed study is likely needed to solidify this age. The full isochrone-fitting workflow for Theia 6046 is available on
\href{https://github.com/carlimankowski/mankowski2025-open-cluster-asteroseismology/blob/main/isochrones.ipynb}{GitHub}.

\begin{figure}
    \centering
    \includegraphics[width=\linewidth]{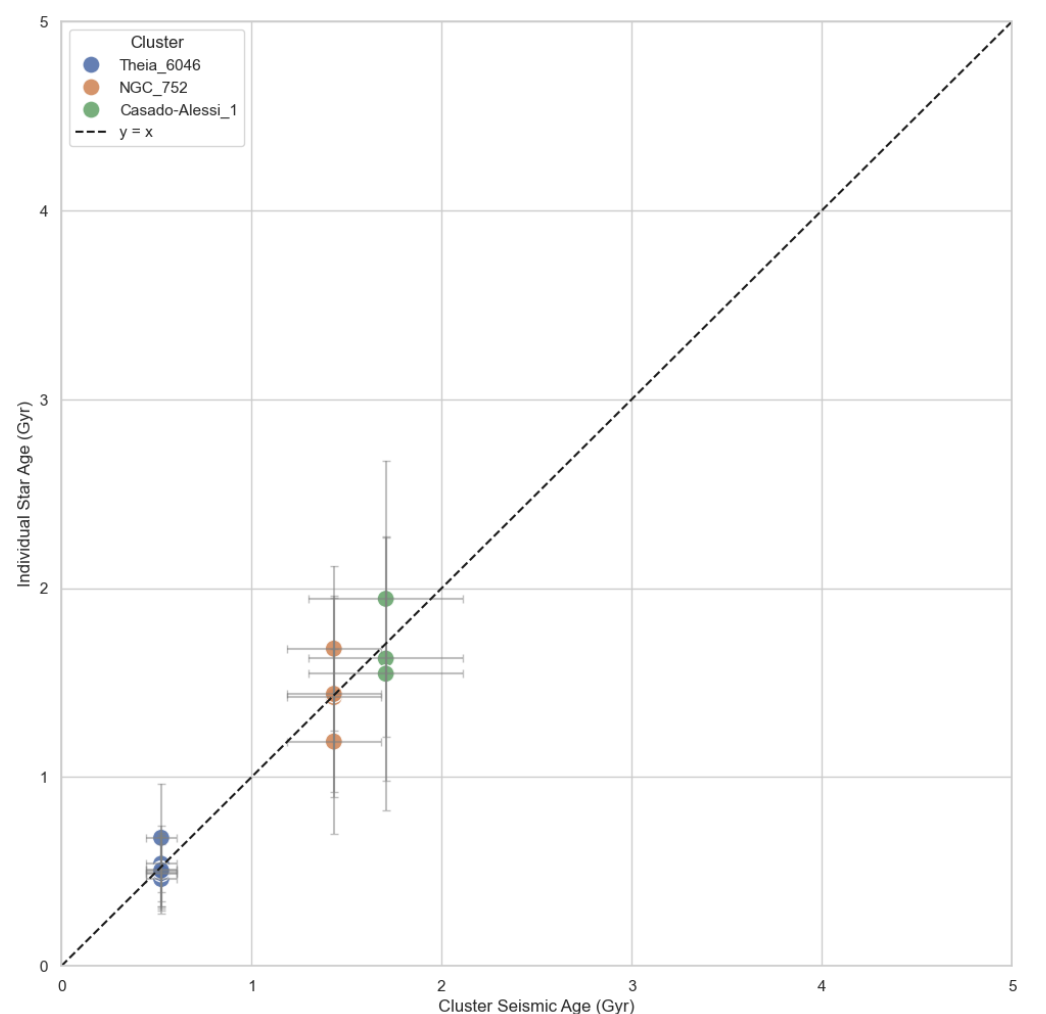}
    \caption{Individual star ages as a function of the corresponding cluster
    average age. Points are colored by cluster, with vertical error bars
    showing the uncertainty on each star’s fitted age.}
    \label{fig:star_vs_cluster_age}
\end{figure}

\begin{figure}[htbp]
    \centering
    \includegraphics[width=\linewidth]{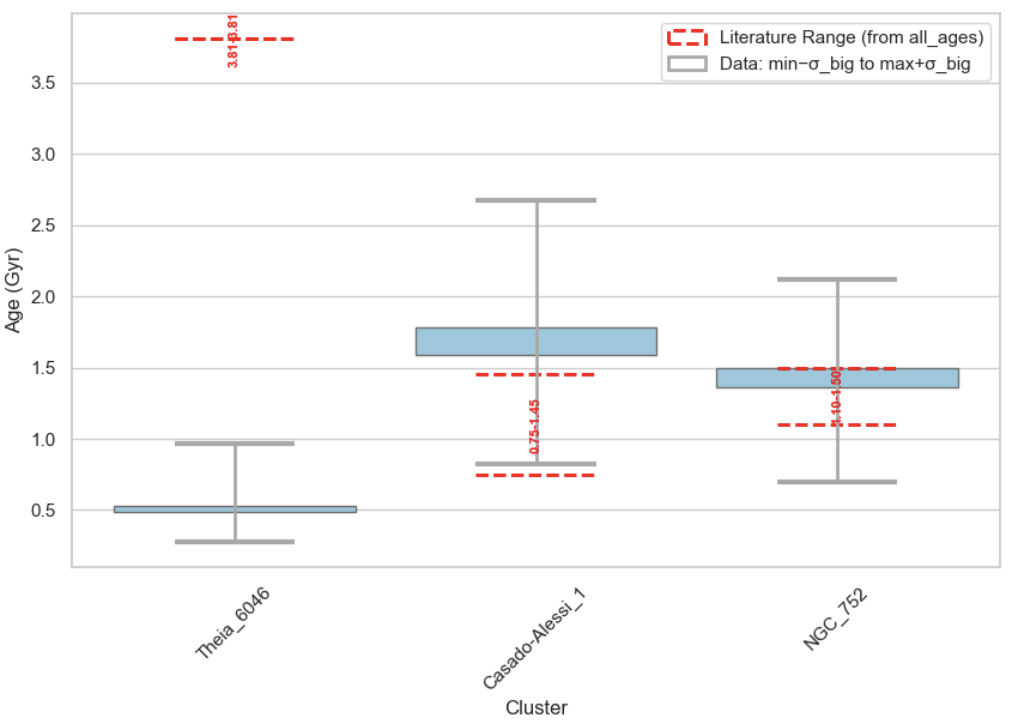}
    \caption{Boxplot of fitted ages for the clusters. Boxes represent the minimum-maximum age range in each cluster, whiskers show the full propagated uncertainty (age +/- error), and the bold line denotes the mean age. Red dashed lines indicate literature age estimates.
}
    \label{fig:ages_boxplot_by_cluster}
\end{figure}

\begin{figure*}[t]
    \centering
    \begin{subfigure}[t]{0.32\textwidth}
        \centering
        \includegraphics[width=\textwidth]{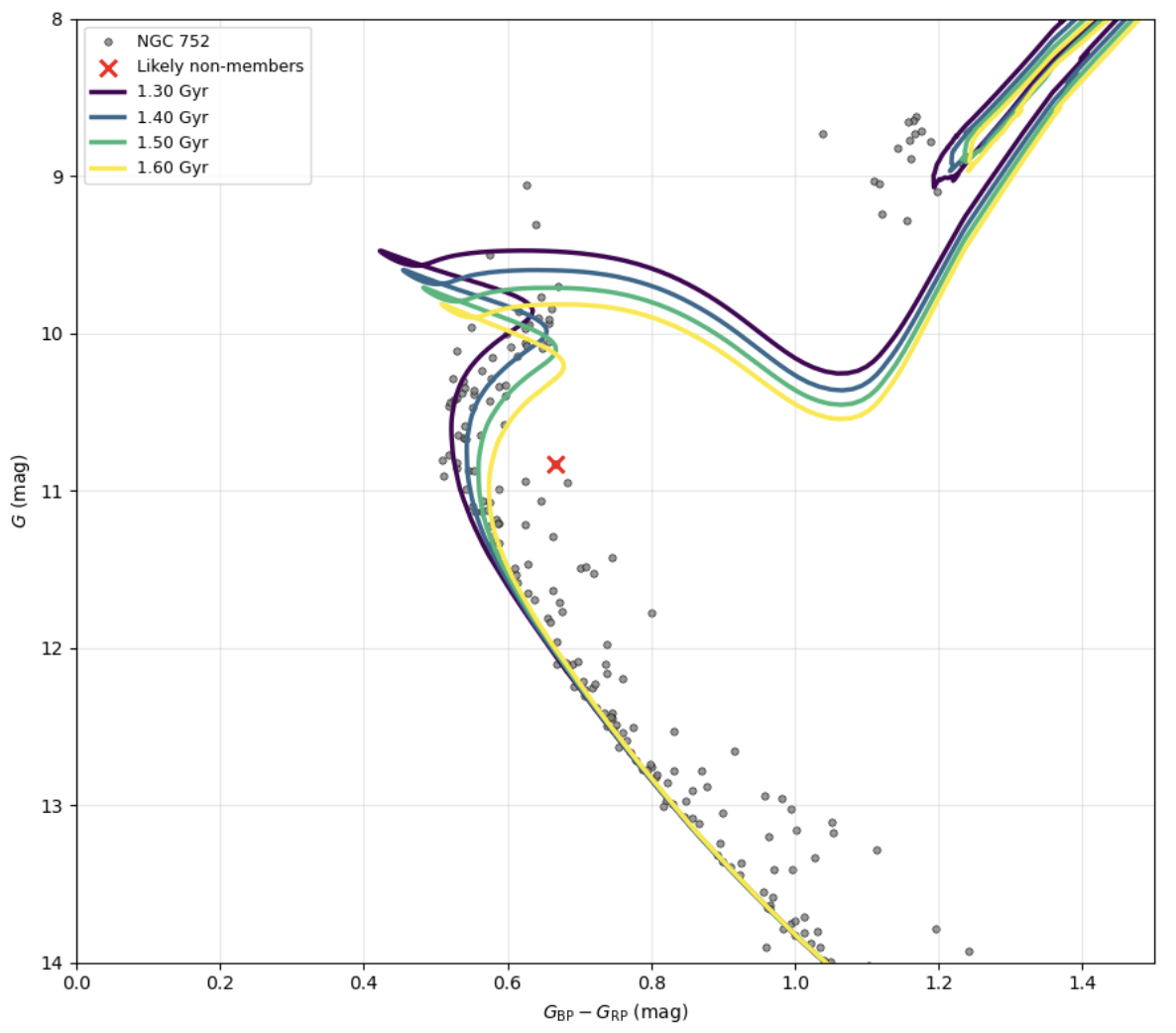}
        \caption{NGC 752}
        \label{ngciso}
    \end{subfigure}
    \hfill
    \begin{subfigure}[t]{0.32\textwidth}
        \centering
        \includegraphics[width=\textwidth]{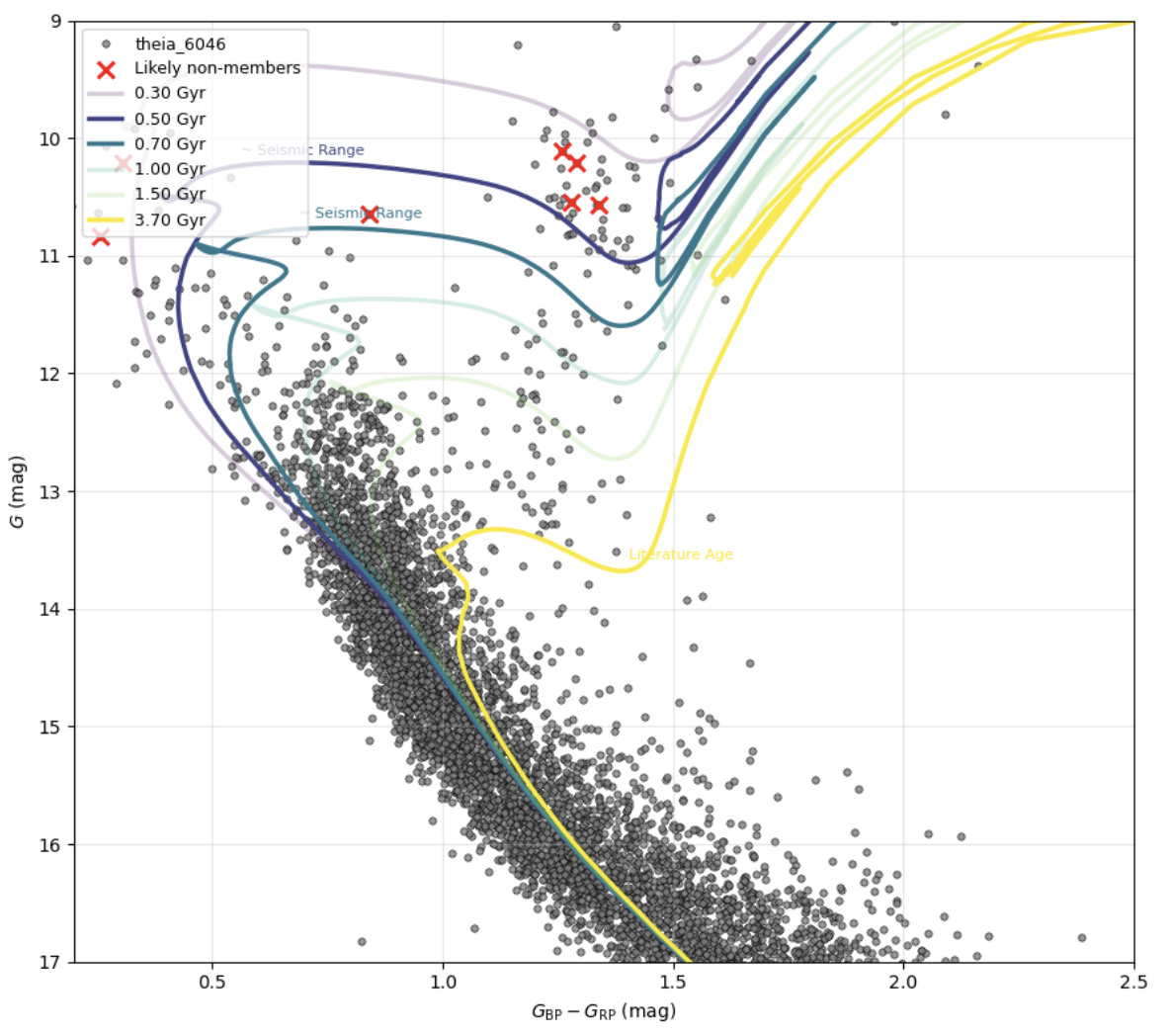}
        \caption{Theia 6046}
        \label{theiaiso}
    \end{subfigure}
    \hfill
    \begin{subfigure}[t]{0.32\textwidth}
        \centering
        \includegraphics[width=\textwidth]{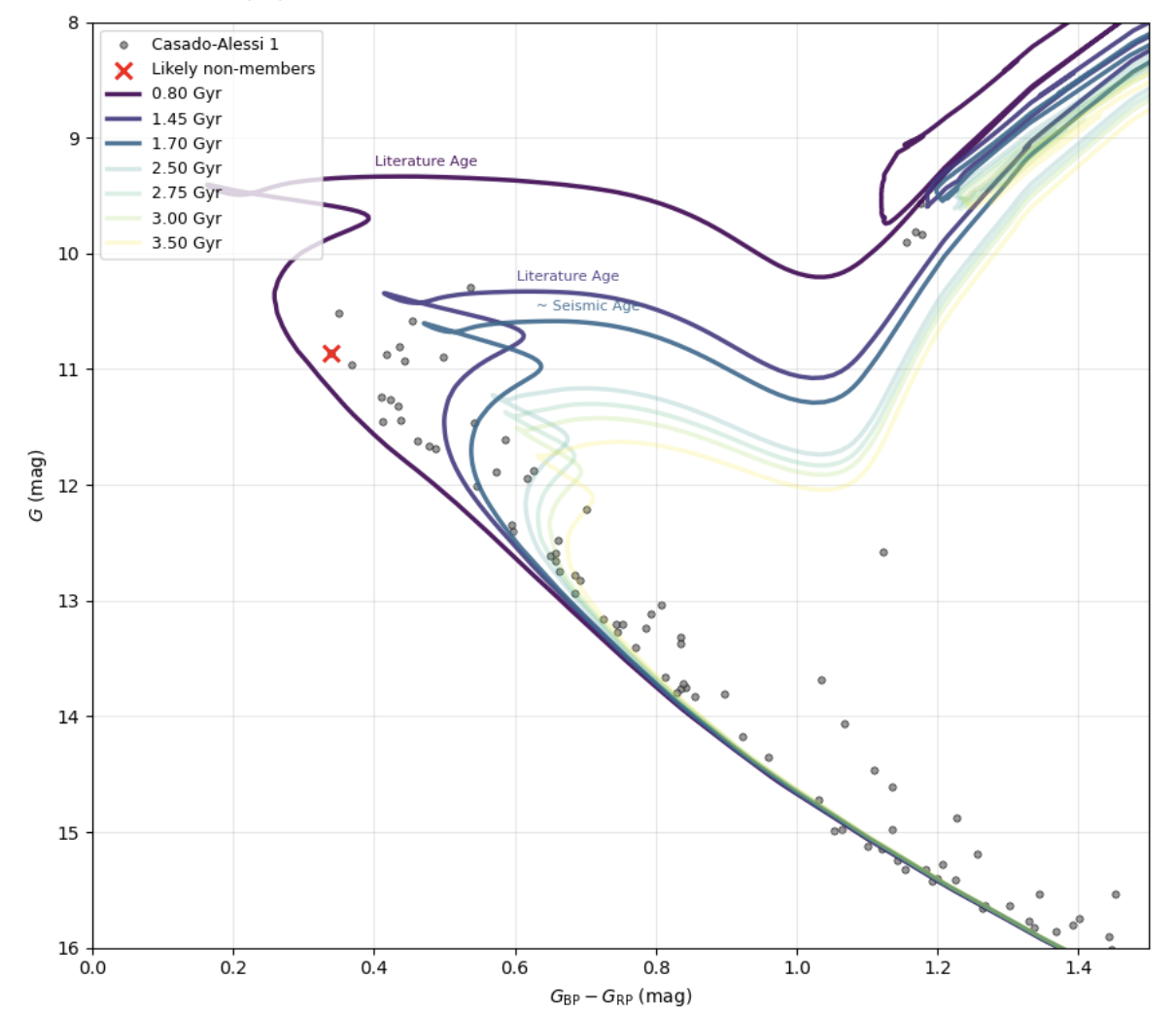}
        \caption{Casado Alessi 1}
        \label{casadoiso}
    \end{subfigure}
    \caption{Color--magnitude diagrams for the open clusters NGC 752 (left), Theia 6046 (middle), and Casado Alessi 1 (right) using \textit{Gaia} DR3 photometry. Grey points show likely cluster members; likely non-members are marked as red crosses. Colored curves are MIST isochrones at various ages plotted using the reported extinction $E(B-V)$ values from \citep{CantatGaudin2020} and respective distance moduli from \citep{Hunt2023}.}
    \label{fig:cluster_cmds}
\end{figure*}

\begin{figure*}
    \centering
    \begin{minipage}{0.9\linewidth}
        \centering
        \includegraphics[width=\linewidth]{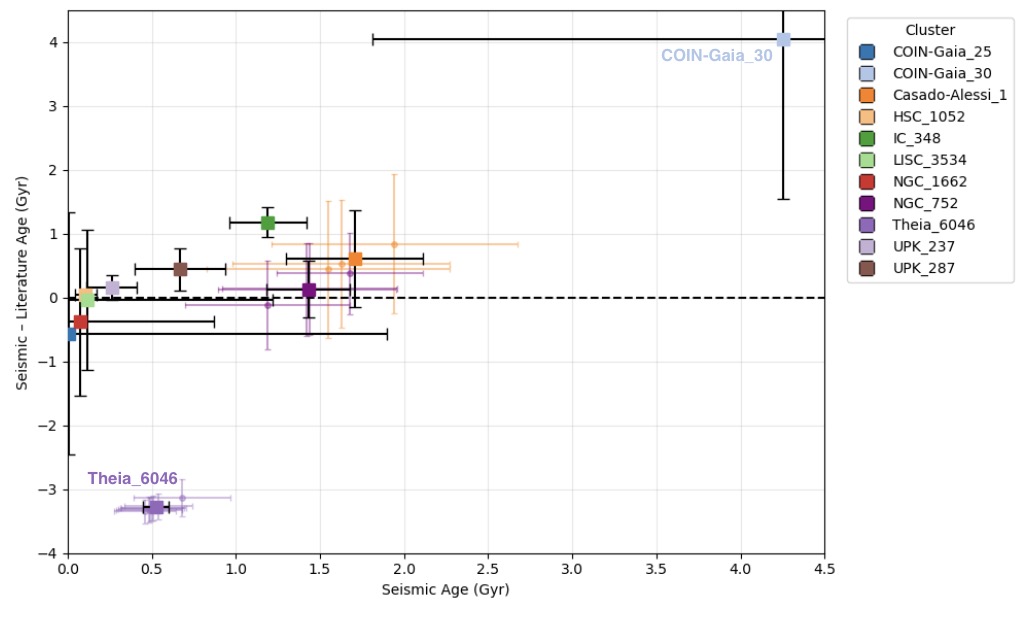}
        \vspace{2pt}
        \text{(a) All literature ages}
    \end{minipage}\hfill
    \begin{minipage}{0.9\linewidth}
        \centering
        \includegraphics[width=\linewidth]{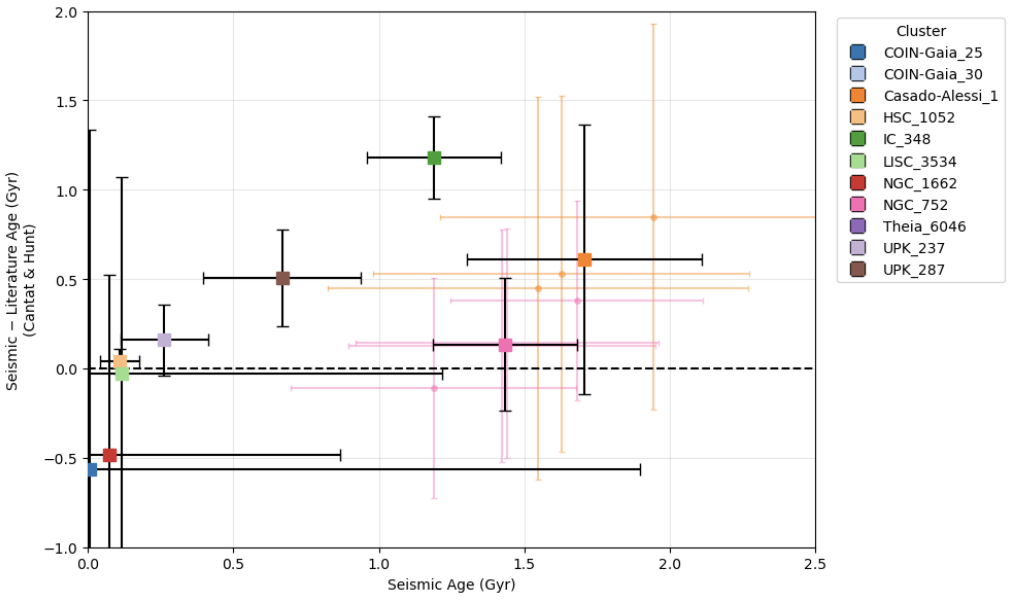}
        \vspace{2pt}
        \text{(b) Cantat-Gaudin (2020) and Hunt \& Reffert (2023) only}
    \end{minipage}
    \caption{
        Residual comparison between seismic ages and literature ages for all clusters in the sample. Cluster means are plotted as solid colors with black error bars. Clusters with more than one member have all members plotted behind it.
        In panel (a), the x-axis shows the seismic ages (Gyr) with their $1\sigma$ uncertainties, and the y-axis shows 
        the difference between the seismic age and the central literature age. 
        Panel (b) shows the same residuals, but using only the Cantat-Gaudin (2020) and Hunt \& Reffert (2023) ages as the literature references. Panel (b) discludes Theia 6046 (discussed in ~\S\ref{subtheia}) and outlier COIN-\textit{Gaia} 30.
    }
    \label{fig:seismic_literature_residuals}
\end{figure*}

\section{Conclusion and Future Work} 
We conclude that our method expands the available targets for asteroseismic analysis in star clusters using \textit{TESS} light curves.  We assembled a cluster-star catalog and workflow that integrates \textit{Gaia} astrometry with \textit{TESS} photometry to present red giants in clusters that are suitable for asteroseismic analysis. Our method (i) enforces distance, magnitude, and crowding/bleeding quality cuts, (ii) retrieves and preprocesses light curves, (iii) measures $\nu_{\max}$ and $\Delta\nu$ using PySYD, (iv) applies community and model-based corrections, (v) propagates uncertainties, (vi) validates radii, masses, and surface gravities across independent diagnostics, and (vii) compares our ages inferred from the asteroseismic analysis of red giants to the ages inferred in the literature from cluster CMDs. Our resulting catalogs, tables, cutouts, and notebooks provide a framework for expanding asteroseismology in clusters.

Our methodology is generally consistent with clusters having homogeneous ages. We show in Figure \ref{fig:star_vs_cluster_age} that our independent ages for each star in the cluster are consistent in a way that allows us to combine estimates and infer cluster ages to a high degree of precision. We also show in Figure \ref{fig:ages_boxplot_by_cluster} that our ages are generally consistent with the literature ages for our clusters. We show in Figure \ref{fig:cluster_cmds} that, across NGC\,752, Casado\,Alessi-1, and Theia\,6046, the CMD-based isochrone fits can favor the catalog ages, the seismic ages, or strongly contradict the older catalog value in favor of a younger seismic solution. More generally, this suggests that asteroseismology can be a competitive way to infer cluster age, but that in many cases more detailed analysis of each cluster can be used to improve age estimates \citep{Reyes2025}. We view this paper primarily as a demonstration of the possibility of using \textit{TESS}, even with its large pixel size and other limitations, to infer reliable cluster ages. We also hope that this analysis can serve as an initial step towards more detailed analysis of both individual stars and clusters in asteroseismic studies.
We expect that additional spectroscopy \citep{Otto2025}, more advanced light-curve processing techniques \citep{Pope2016}, and individual mode analyses in clusters \citep{Lindsay2025} would further tighten the constraints reported here. More generally, our work is intended as a contribution to a longer-term, community-wide effort to obtain robust asteroseismic mass and age estimates by combining stellar-evolution models and isochrones, detailed mode-by-mode asteroseismic analysis, comparisons to eclipsing binaries with dynamical masses, and the use of wide binaries and star clusters as coeval benchmarks \citep{ElBadry2021,GodoyRiveraChaname2018,Tayar2025,theodoridis2025_asteroseismically}. Within this broader framework, the cluster-calibrated ages presented here help to cement asteroseismology as a precise and accurate technique for interpreting both exoplanetary systems and large galactic-archaeology samples such as APOGEE.

Clusters have long been our benchmarks for inferring stellar ages and calibrating other techniques. Our results suggest that ages inferred from asteroseismology can be reliable in clusters, and that more clusters are available to enhance calibration efforts. Looking ahead, space missions such as ESA’s \textit{PLATO} \citep{PLATOmission} and NASA’s \textit{Roman} Space Telescope \citep{Roman} will offer the possibility of precise ages for hundreds of thousands of targets and extend this approach to even larger distances. To make full use of the precision of asteroseismology on such a broad scale, it is necessary to validate its accuracy against key calibration systems such as star clusters and wide binaries, and to compare it with other techniques over a range of ages and metallicities \citep{Miglio2013_HRGiants,Miglio2021_HAYDN}. Using tools like \textit{Roman}, \textit{PLATO}, and other dedicated missions \citep{Miglio2021} in this way will allow us to unlock the detailed evolutionary history of the Milky Way in a fully multidimensional sense.

\section*{Acknowledgments}
This work was supported by the University of Florida College of Liberal Arts and Sciences. 
C. Mankowski received course credit for this research through the Department of Astronomy. 
We thank J. Tayar for supervision and weekly discussions guiding the analysis. 
We acknowledge helpful assistance with \texttt{PySYD} from A. Beyer. We thank D. Stello, N. Myers, M. Howell, Y. Noureddine, and C. Wagner for careful reading of the manuscript and for constructive feedback that improved this work. This paper includes data collected by the \textit{TESS} mission and obtained from the Mikulski Archive for Space Telescopes (MAST) at the Space Telescope Science Institute (STScI). The \textit{TESS} mission is funded by NASA’s Science Mission Directorate, and funding for \textit{TESS} is provided by the NASA Explorer Program. STScI is operated by the Association of Universities for Research in Astronomy, Inc., under NASA contract NAS~5–26555. Some of the data presented in this paper were obtained from MAST. Support for MAST for non-HST data is provided by the NASA Office of Space Science via grant NAG5-7584 and by other grants and contracts.

This work has made use of data from the European Space Agency (ESA) mission
{\it Gaia} (\url{https://www.cosmos.esa.int/gaia}), processed by the {\it Gaia}
Data Processing and Analysis Consortium (DPAC,
\url{https://www.cosmos.esa.int/web/gaia/dpac/consortium}). Funding for the DPAC
has been provided by national institutions, in particular the institutions
participating in the {\it Gaia} Multilateral Agreement.

\section*{Author Contributions}

C. Mankowski carried out all aspects of this project, including data compilation, target selection, light curve analysis, seismic parameter extraction, application of corrections, age fitting, figure generation, and manuscript preparation, with the exception of Section 2.5 of the Methods, which was written and performed by C. Martin. 
J. Tayar conceived the project through her original proposal, provided overall supervision, provided manuscript edits, and met with C. Mankowski weekly to discuss methodology, results, and interpretation.

\section*{Software and third party data repository citations} \label{sec:cite}

This research made use of \texttt{astropy} \citep{astropy2013,astropy2018,astropy2022}, 
\texttt{lightkurve} \citep{lightkurve2018}, 
\texttt{PySYD} \citep{cantosy2022}, 
\texttt{asfgrid} \citep{sharma2016}, 
\texttt{MIST} \citep{Choi2016MIST}, 
\texttt{Dartmouth} \citep{Dotter2008}, 
\texttt{YREC} \citep{demarque2008}, 
\texttt{GARSTEC} \citep{weiss2008}, 
and visualization and analysis packages including 
\texttt{numpy}, \texttt{pandas}, \texttt{scipy}, \texttt{matplotlib}, 
\texttt{\'{E}chelle} \citep{echelle2016}, 
\texttt{Bokeh} \citep{bokeh2018}, \texttt{Panel}, \texttt{astroquery}, 
and \texttt{ebf}. 
This work was partially carried out in the TIKE (Timeseries Integrated Knowledge Engine) environment provided by STScI/MAST.  

We made use of data from the Mikulski Archive for Space Telescopes (MAST) for \textit{TESS} observations, 
the ESA \textit{Gaia} Archive for DR3 data.

\bibliography{sample7}
\bibliographystyle{aasjournalv7}

\appendix

\setcounter{table}{0}
\renewcommand{\thetable}{A\arabic{table}}
{\Large\bfseries APPENDIX A: SAMPLE CATALOG PRESENTATION\par}\vspace{0.6em}

\startlongtable
\begin{deluxetable}{llll}
\tablecaption{Description of columns in the machine-readable stellar catalog.\label{tab:coldefs}}
\tabletypesize{\scriptsize}
\tablewidth{0pt}
\tablehead{
\colhead{Column} &
\colhead{Description} &
\colhead{Unit} &
\colhead{Example}
}
\startdata
TICID & \textit{TESS} Input Catalog identifier & \nodata & 306345133 \\
GaiaDR3 & \textit{Gaia} DR3 source identifier & \nodata & 2990380948461581056 \\
cluster\_name & Adopted cluster name & \nodata & Theia\_6046 \\
RA\_ICRS & Right ascension (ICRS, \textit{Gaia} DR3) & deg & 78.5951 \\
DE\_ICRS & Declination (ICRS, \textit{Gaia} DR3) & deg & -9.7912 \\
pmRA & Proper motion in right ascension (\textit{Gaia} DR3) & mas\,yr$^{-1}$ & 1.93 \\
pmDE & Proper motion in declination (\textit{Gaia} DR3) & mas\,yr$^{-1}$ & 0.58 \\
Plx & Parallax (\textit{Gaia} DR3) & mas & 1.28 \\
Gmag & \textit{Gaia} $G$ magnitude & mag & 10.33 \\
BPmag & \textit{Gaia} $G_{\rm BP}$ magnitude & mag & 10.98 \\
RPmag & \textit{Gaia} $G_{\rm RP}$ magnitude & mag & 9.56 \\
Classification & Manual photometric class & \nodata & Potential Red Giant \\
Within1kpc & Flag: within 1\,kpc (y/n) & \nodata & y \\
Mag$<13$/\textit{TESS} & Flag: \textit{TESS} magnitude $<13$ (y/n) & \nodata & y \\
Giants & Flag: giant candidate (y/n) & \nodata & y \\
crowdbleedsec & Flag: any \textit{TESS} sector with crowding/bleeding issues (y/n) & \nodata & y \\
Crowding & \textit{TESS} crowding metric proxy & \nodata & 2.5 \\
Bleeding & \textit{TESS} bleeding metric proxy & \nodata & 4.5 \\
\# of Available Sector(s) & Number of \textit{TESS} sectors with usable light curves & \nodata & 5 \\
teff\_gspphot & \textit{Gaia} GSP-Phot effective temperature & K & 4703.2 \\
logg\_gspphot & \textit{Gaia} GSP-Phot surface gravity & dex & 2.58 \\
mh\_gspphot & \textit{Gaia} GSP-Phot metallicity [M/H] & dex & 0.24 \\
teff\_gspspec & \textit{Gaia} GSP-Spec effective temperature & K & 4448 \\
logg\_gspspec & \textit{Gaia} GSP-Spec surface gravity & dex & 2.01 \\
mh\_gspspec & \textit{Gaia} GSP-Spec metallicity [M/H] & dex & 0.25 \\
cluster\_prob & Adopted cluster membership probability & \nodata & 0.386 \\
numax\_pysyd & PySYD $\nu_{\max}$ (raw) & $\mu$Hz & 22.2 \\
numax\_error\_upper & Upper uncertainty on PySYD $\nu_{\max}$ & $\mu$Hz & 1.49 \\
numax\_error\_lower & Lower uncertainty on PySYD $\nu_{\max}$ & $\mu$Hz & 1.49 \\
numax\_corr & $\nu_{\max}$ after applied corrections & $\mu$Hz & 22.22 \\
dnu\_pysyd & PySYD $\Delta\nu$ (raw) & $\mu$Hz & 2.37 \\
dnu\_error\_upper & Upper uncertainty on PySYD $\Delta\nu$ & $\mu$Hz & 0.0348 \\
dnu\_error\_lower & Lower uncertainty on PySYD $\Delta\nu$ & $\mu$Hz & 0.0348 \\
fdnu\_used & Adopted $f_{\Delta\nu}$ correction factor & \nodata & 0.993 \\
fdnu\_source & Source of $f_{\Delta\nu}$ correction & \nodata & asf \\
dnu\_corr & $\Delta\nu$ after applied corrections & $\mu$Hz & 2.35 \\
PySYD\_Quality & PySYD quality flag (good/ok/bad) & \nodata & good \\
snr & Signal-to-noise ratio of detected power excess & \nodata & 16.1 \\
evo\_state & Adopted evolutionary state (RGB/RC) & \nodata & RGB \\
cat\_mass & \textit{Gaia} \textit{FLAME} mass & $M_\odot$ & 3.00 \\
M\_cat\_error\_upper & Upper error on \textit{FLAME} mass & $M_\odot$ & 0.8 \\
M\_cat\_error\_lower & Lower error on \textit{FLAME} mass & $M_\odot$ & 0.4 \\
cat\_mass\_err\_stated & Catalog-stated fractional mass error & \nodata & 0.0593 \\
cat\_radius & \textit{Gaia} \textit{FLAME} radius & $R_\odot$ & 10.76 \\
R\_cat\_error\_upper & Upper error on \textit{FLAME} radius & $R_\odot$ & 0.50 \\
R\_cat\_error\_lower & Lower error on \textit{FLAME} radius & $R_\odot$ & 0.50 \\
cat\_radius\_err\_stated & Catalog-stated fractional radius error & \nodata & 0.249 \\
cat\_met & \textit{Gaia} metallicity [M/H] & dex & -0.05 \\
cat\_met\_err & Error on \textit{Gaia} metallicity & dex & 0.0096 \\
cat\_logg & \textit{Gaia} surface gravity & dex & 2.2964 \\
cat\_logg\_err\_upper & Upper error on \textit{Gaia} $\log g$ & dex & 0.425 \\
cat\_logg\_err\_lower & Lower error on \textit{Gaia} $\log g$ & dex & 0.425 \\
cat\_teff & \textit{Gaia} effective temperature & K & 4575.6 \\
cat\_teff\_error\_upper & Upper error on \textit{Gaia} $T_{\rm eff}$ & K & 70.9 \\
cat\_teff\_error\_lower & Lower error on \textit{Gaia} $T_{\rm eff}$ & K & 70.9 \\
logg\_seis & Seismic surface gravity from scaling relations & dex & 2.244 \\
logg\_seis\_upper & Upper error on seismic $\log g$ & dex & 0.0294 \\
logg\_seis\_lower & Lower error on seismic $\log g$ & dex & 0.0294 \\
R\_seis & Seismic radius from scaling relations & $R_\odot$ & 21.09 \\
R\_seis\_error\_upper & Upper error on seismic radius & $R_\odot$ & 1.56 \\
R\_seis\_error\_lower & Lower error on seismic radius & $R_\odot$ & 1.56 \\
M\_seis & Seismic mass from scaling relations & $M_\odot$ & 2.85 \\
M\_seis\_error\_upper & Upper error on seismic mass & $M_\odot$ & 0.60 \\
M\_seis\_error\_lower & Lower error on seismic mass & $M_\odot$ & 0.60 \\
cluster & Flag: accepted cluster member (y/n) & \nodata & y \\
Removed? & Flag: removed from final analysis sample (y/n) & \nodata & n \\
age\_gyr & Adopted cluster age for this star & Gyr & 0.487 \\
sigma\_seis & Seismic age uncertainty term & Gyr & -0.114 \\
sigma\_grid & Grid-based age scatter term & Gyr & 0.159 \\
sigma\_big & Total age uncertainty combining error terms & Gyr & 0.196 \\
age\_minus1sigma & Cluster age at $-1\sigma$ bound & Gyr & 0.595 \\
age\_std\_minus1sigma & Uncertainty on age\_minus1sigma & Gyr & 0.260 \\
age\_plus1sigma & Cluster age at $+1\sigma$ bound & Gyr & 0.367 \\
age\_std\_plus1sigma & Uncertainty on age\_plus1sigma & Gyr & 0.0813 \\
age\_MIST & Best-fitting age from MIST model & Gyr & 0.517 \\
age\_Dartmouth & Best-fitting age from Dartmouth model & Gyr & 0.809 \\
age\_GARSTEC & Best-fitting age from GARSTEC model & Gyr & 0.860 \\
age\_YREC & Best-fitting age from YREC model & Gyr & 0.829 \\
age\_cantat\_2020 & Literature $\log_{10}$(age/yr) from Cantat-Gaudin (2020) & dex & \nodata \\
age\_hunt\_2023 & Literature $\log_{10}$(age/yr) from Hunt \& Reffert (2023) & dex & 9.5806 \\
age\_cantat\_Gyr & Cantat-Gaudin (2020) age converted to Gyr & Gyr & \nodata \\
age\_hunt\_Gyr & Hunt \& Reffert (2023) age converted to Gyr & Gyr & 3.807 \\
all\_ages & List of all adopted literature ages & Gyr & [3.807] \\
age\_std\_lit & Standard deviation of ages in all\_ages & Gyr & 0.0 \\
cluster\_age\_final & Cluster age considering all members & Gyr & 0.525 \\
cluster\_age\_err & Error on cluster age & Gyr & 0.0788 \\
\enddata
\end{deluxetable}
\clearpage 

\begin{table*}[!t]
\centering
{\Large\bfseries APPENDIX B: SEISMIC RESULTS\par}\vspace{0.6em}

\small
\setlength{\tabcolsep}{6pt}
\renewcommand{\arraystretch}{1.15}
\caption{Asteroseismic parameters and cluster ages for the 14 target stars.}
\label{tab:asteroseismic_params}

\begin{tabular*}{\textwidth}{@{\extracolsep{\fill}} l l
                S[table-format=2.3]
                S[table-format=1.3]
                S[table-format=1.3]
                S[table-format=1.3]
                S[table-format=1.3]
                S[table-format=1.3]}
\toprule
TIC ID & Cluster &
\multicolumn{1}{c}{$\nu_{\mathrm{max,corr}}$} &
\multicolumn{1}{c}{$\Delta\nu_{\mathrm{corr}}$} &
\multicolumn{1}{c}{Age} &
\multicolumn{1}{c}{$\sigma_{\mathrm{big}}$} &
\multicolumn{1}{c}{$\tau_{\mathrm{cluster}}$} &
\multicolumn{1}{c}{$\sigma_{\tau}$} \\
& &
\multicolumn{1}{c}{($\mu$Hz)} &
\multicolumn{1}{c}{($\mu$Hz)} &
\multicolumn{1}{c}{(Gyr)} &
\multicolumn{1}{c}{(Gyr)} &
\multicolumn{1}{c}{(Gyr)} &
\multicolumn{1}{c}{(Gyr)} \\
\midrule
306345133 & Theia\,6046 & 22.216 & 2.354 & 0.487 & 0.196 & 0.525 & 0.079 \\
43902016  & Theia\,6046 & 15.071 & 1.831 & 0.541 & 0.200 & 0.525 & 0.079 \\
24297458  & Theia\,6046 & 56.009 & 4.910 & 0.511 & 0.194 & 0.525 & 0.079 \\
24666306  & Theia\,6046 & 33.294 & 3.141 & 0.460 & 0.185 & 0.525 & 0.079 \\
24444542  & Theia\,6046 & 16.582 & 1.848 & 0.492 & 0.189 & 0.525 & 0.079 \\
249064439 & Theia\,6046 & 62.854 & 6.404 & 0.671 & 0.288 & 0.525 & 0.079 \\
34472483  & Theia\,6046 & 34.764 & 3.542 & 0.505 & 0.189 & 0.525 & 0.079 \\
67569102  & NGC\,752    & 67.760 & 6.014 & 1.188 & 0.491 & 1.433 & 0.247 \\
67420118  & NGC\,752    & 68.416 & 6.613 & 1.423 & 0.528 & 1.433 & 0.247 \\
186970424 & NGC\,752    & 50.083 & 5.290 & 1.440 & 0.520 & 1.433 & 0.247 \\
67419922  & NGC\,752    & 33.348 & 3.617 & 1.680 & 0.435 & 1.433 & 0.247 \\
306552813 & Casado-Alessi\,1 & 66.163 & 6.500 & 1.627 & 0.647 & 1.706 & 0.405 \\
306955660 & Casado-Alessi\,1 & 62.896 & 6.624 & 1.944 & 0.732 & 1.706 & 0.405 \\
306187638 & Casado-Alessi\,1 & 68.064 & 6.530 & 1.547 & 0.723 & 1.706 & 0.405 \\
\bottomrule
\end{tabular*}
\end{table*}
\clearpage

\clearpage
\setcounter{figure}{0}
\renewcommand{\thefigure}{C\arabic{figure}}

\graphicspath{{./}{visuals/}}

\newcommand{\RowGap}{\vspace{1.1em}}

\newcommand{\TICRow}[4]{%
\par\noindent{\centering\footnotesize\textbf{TIC~#1}\par}\vspace{0.25em}
\begin{minipage}[t]{0.32\textwidth}\centering
{\scriptsize Power Spectrum}\\[-0.25em]
\includegraphics[width=0.98\linewidth]{#2}
\end{minipage}\hfill
\begin{minipage}[t]{0.32\textwidth}\centering
{\scriptsize $\Delta\nu$ diagnostic}\\[-0.25em]
\includegraphics[width=0.98\linewidth]{#3}
\end{minipage}\hfill
\begin{minipage}[t]{0.32\textwidth}\centering
{\scriptsize $\nu_{\max}$ overlay}\\[-0.25em]
\includegraphics[width=0.98\linewidth]{#4}
\end{minipage}
\RowGap
}

\begin{figure*}[!t]
\centering
{\Large\bfseries APPENDIX C: $\Delta\nu$ AND $\nu_{\text{max}}$ VISUALS\par}\vspace{0.6em}

\TICRow{306345133}{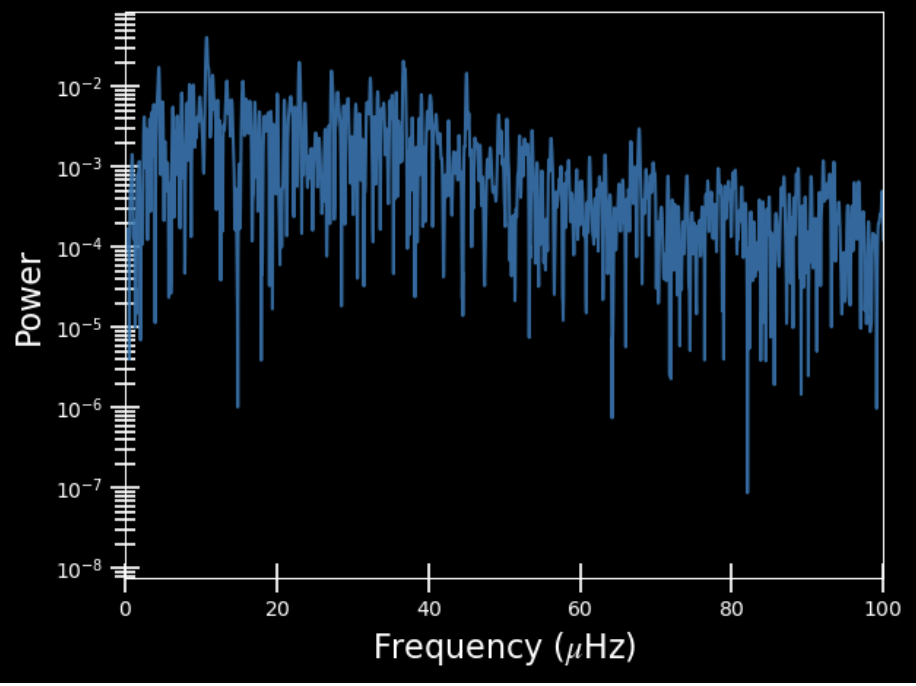}{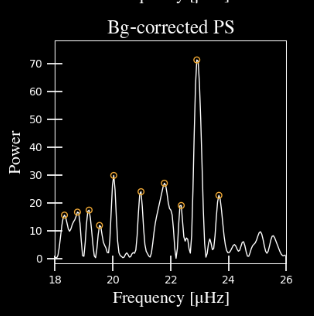}{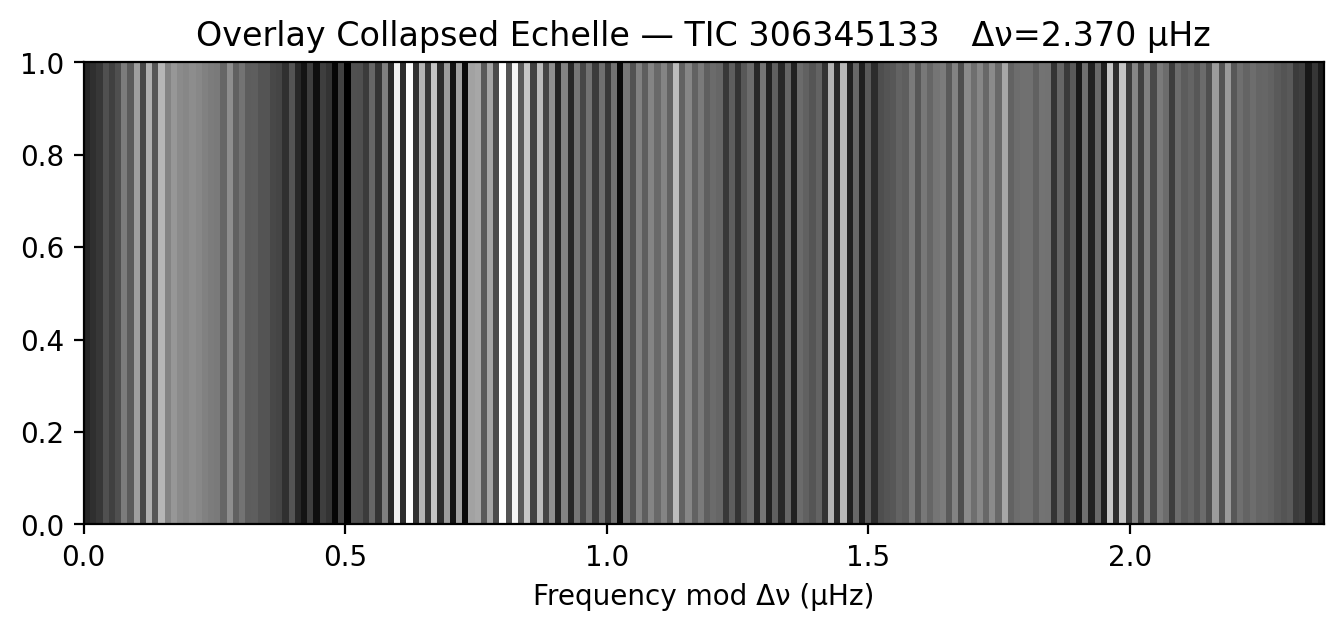}
\TICRow{43902016}{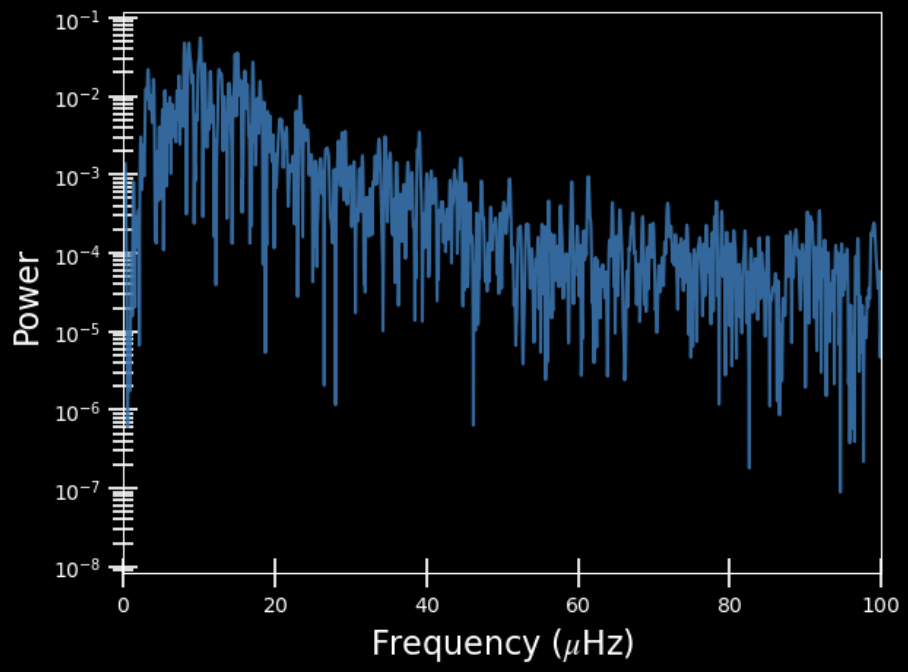}{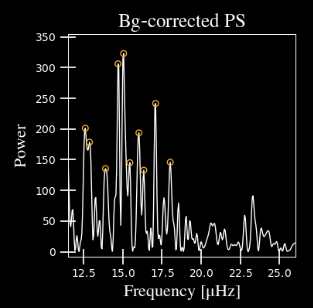}{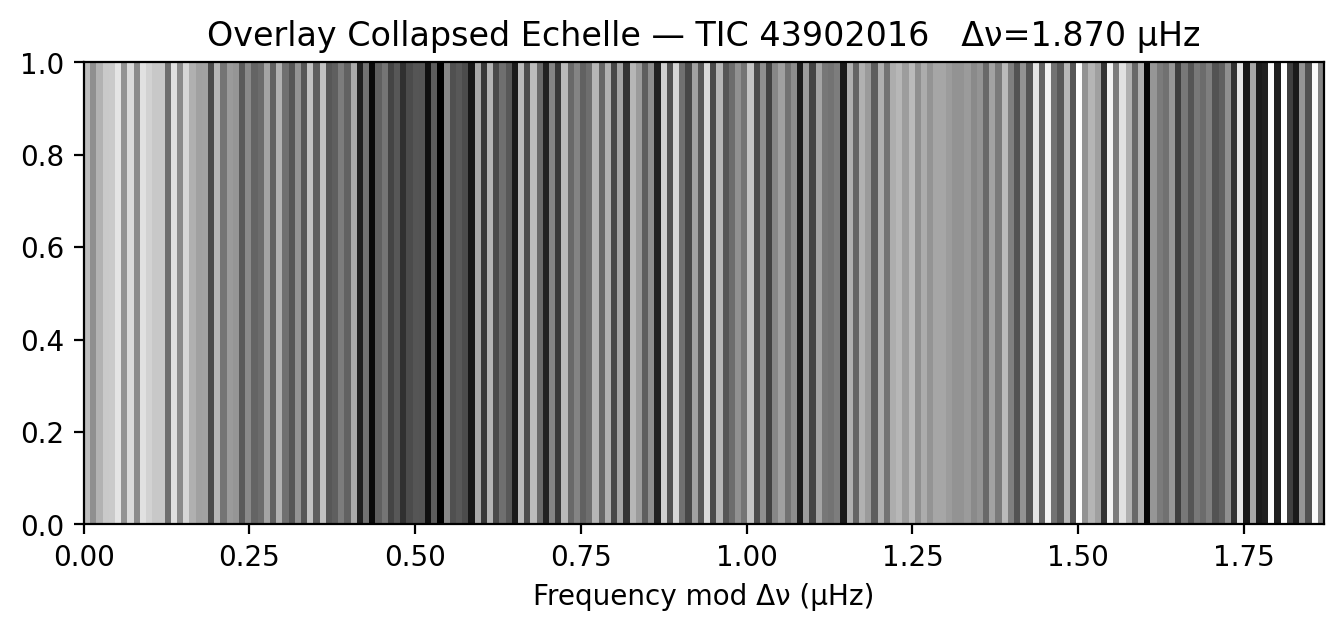}
\TICRow{24297458}{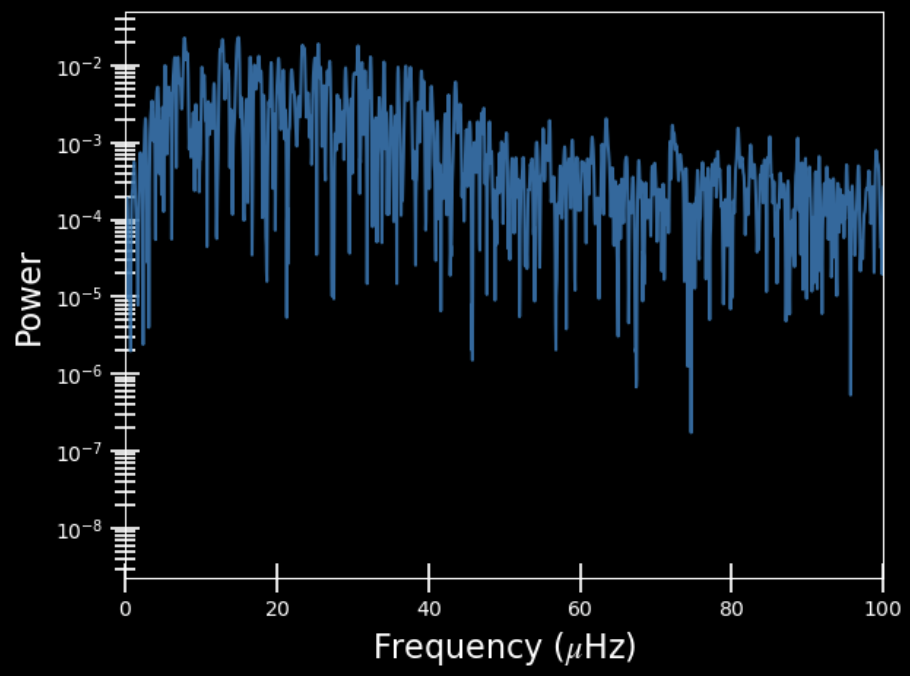}{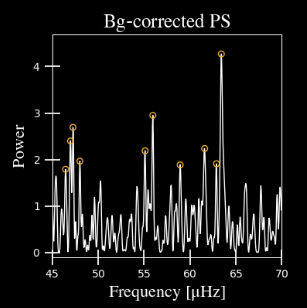}{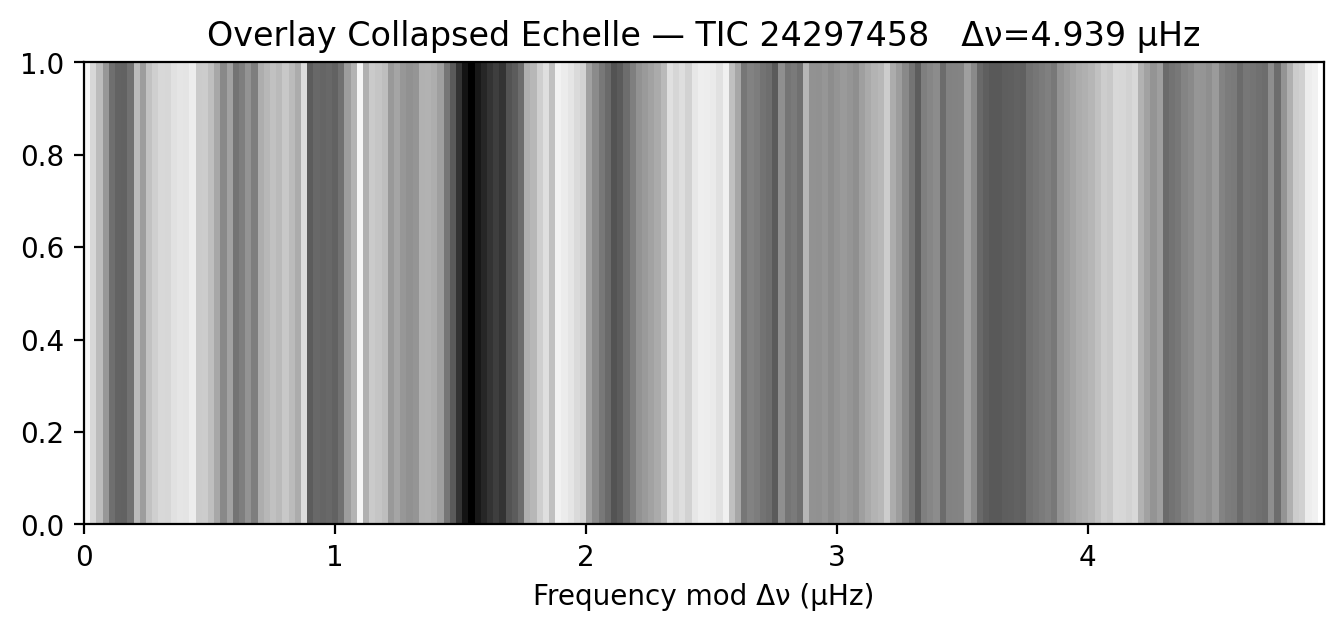}
\label{fig:appC:1}
\end{figure*}

\begin{figure*}[!t]
\centering
\TICRow{24666306}{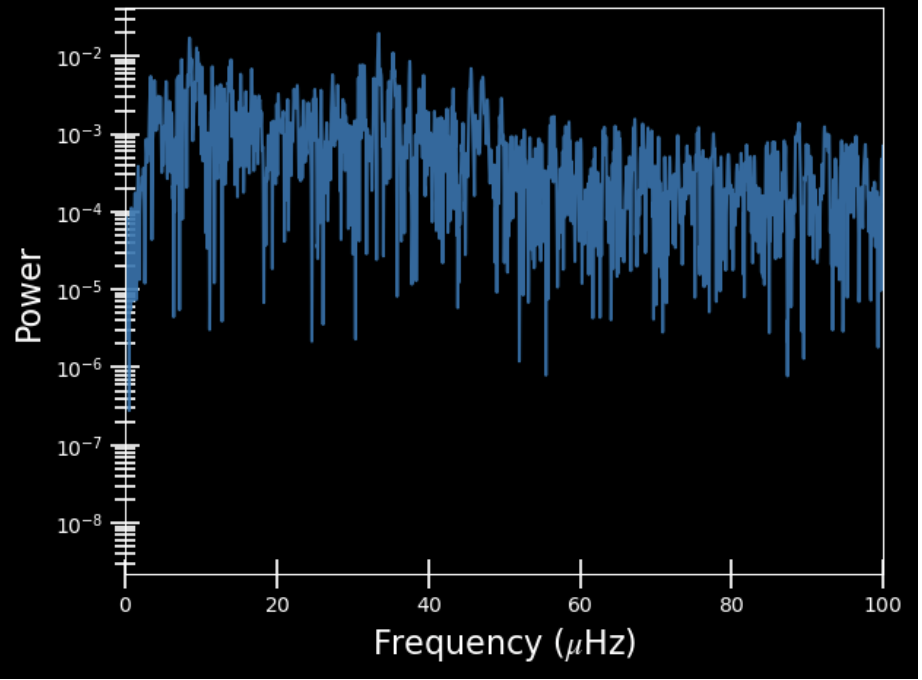}{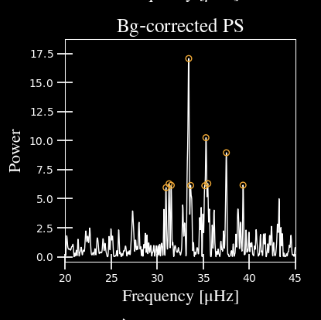}{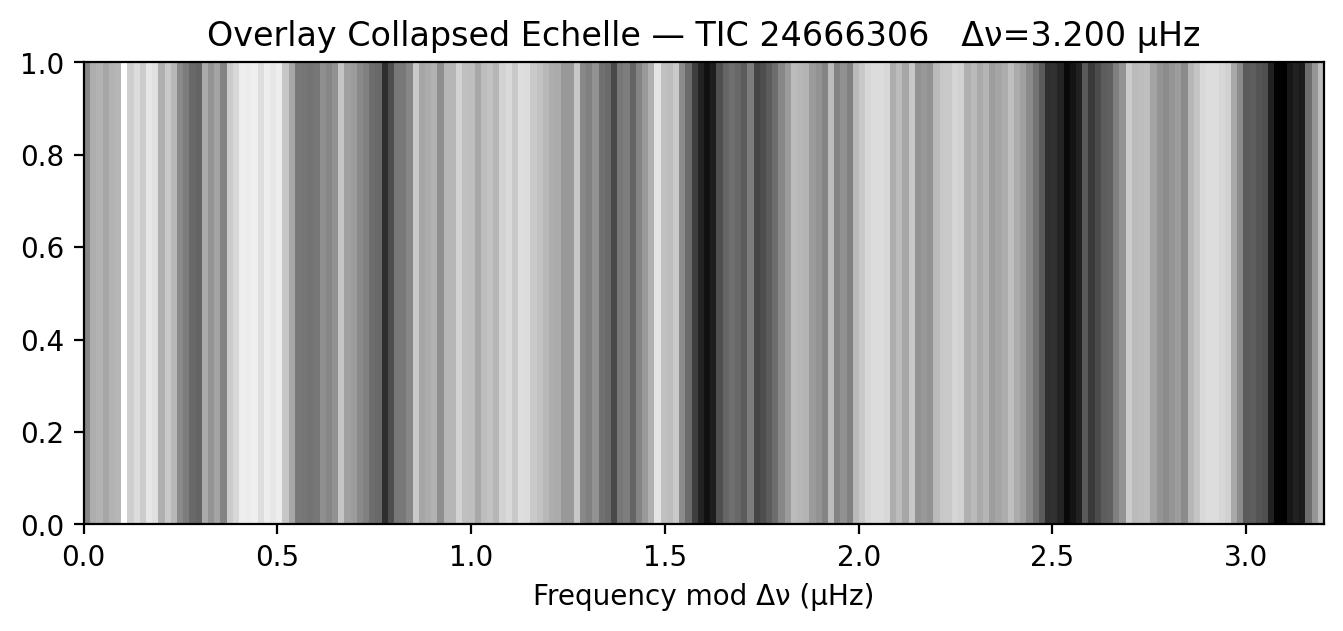}
\TICRow{24444542}{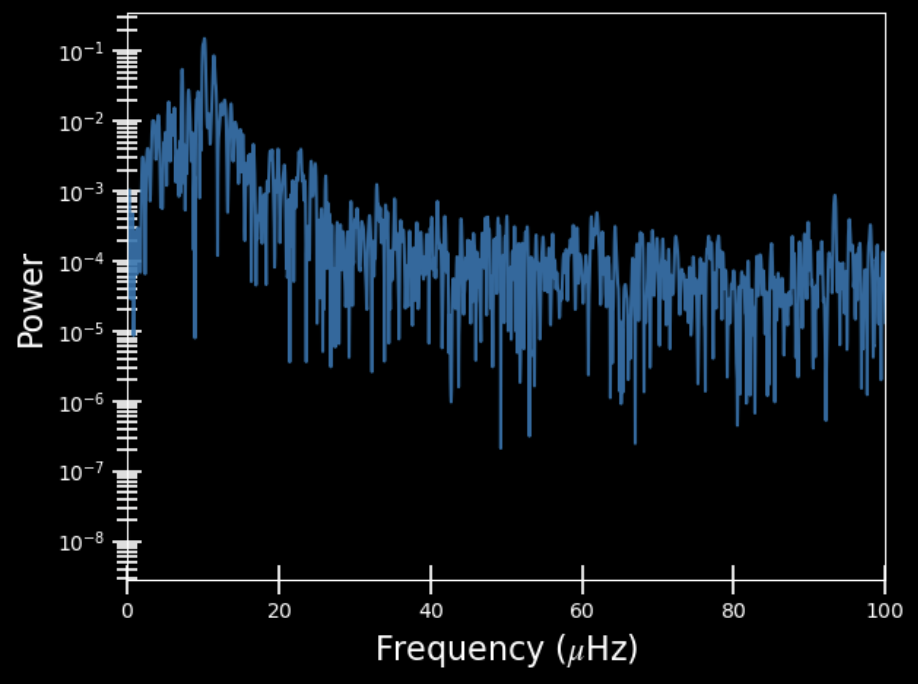}{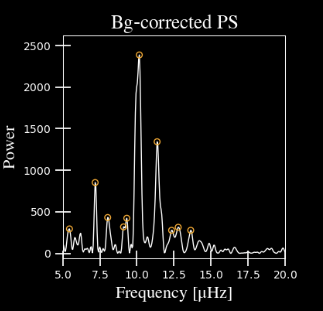}{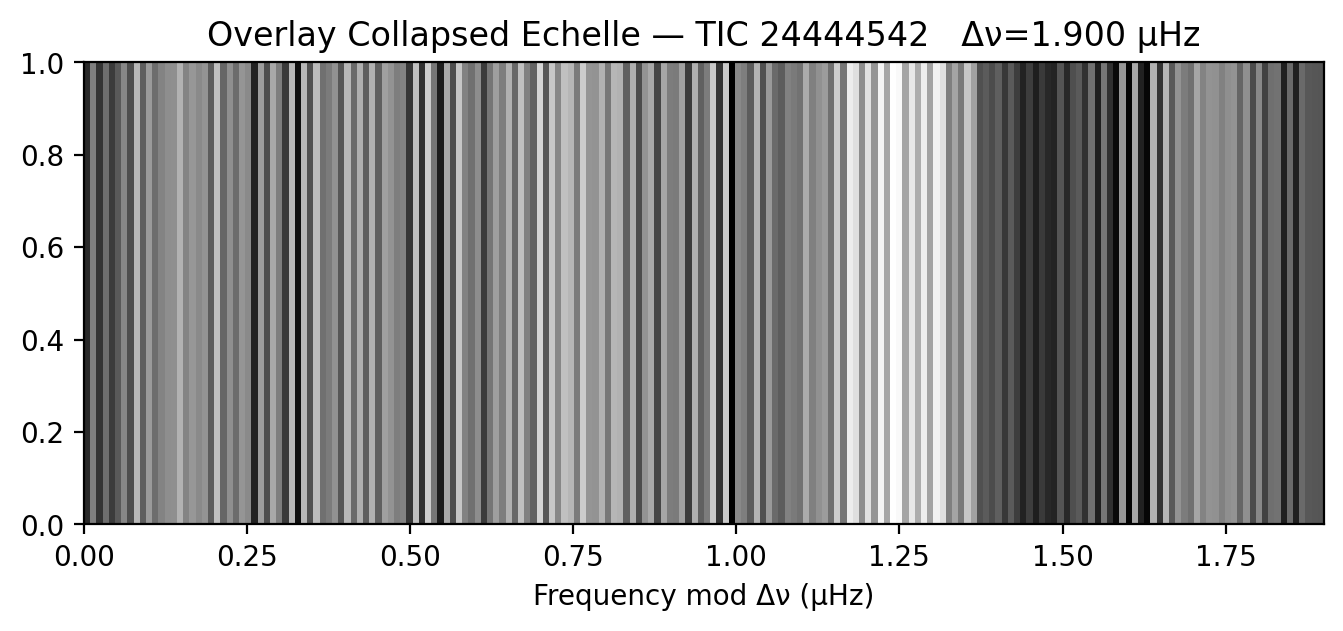}
\TICRow{34472483}{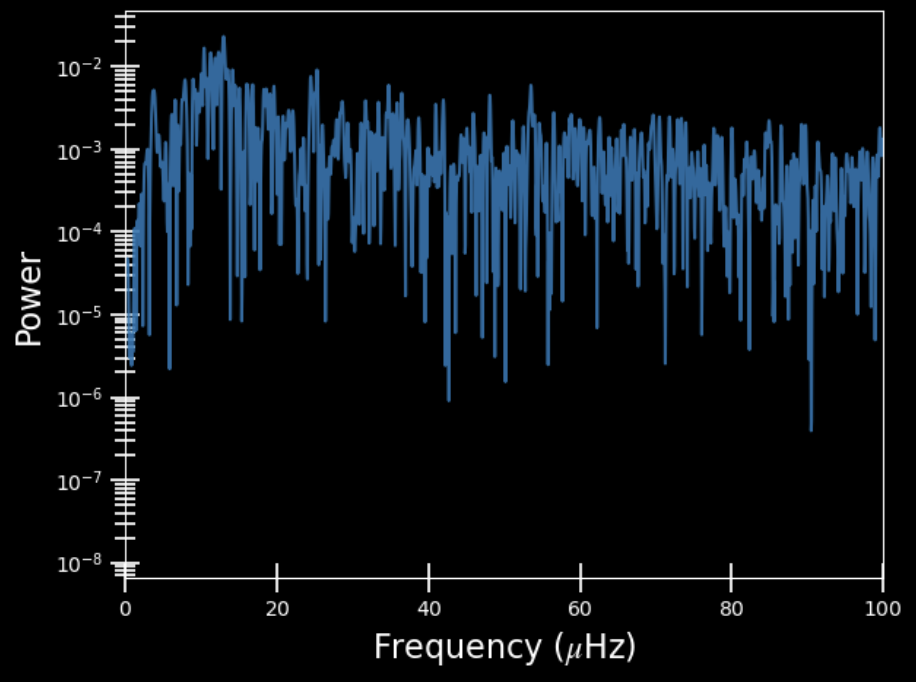}{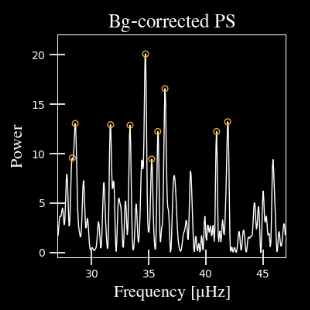}{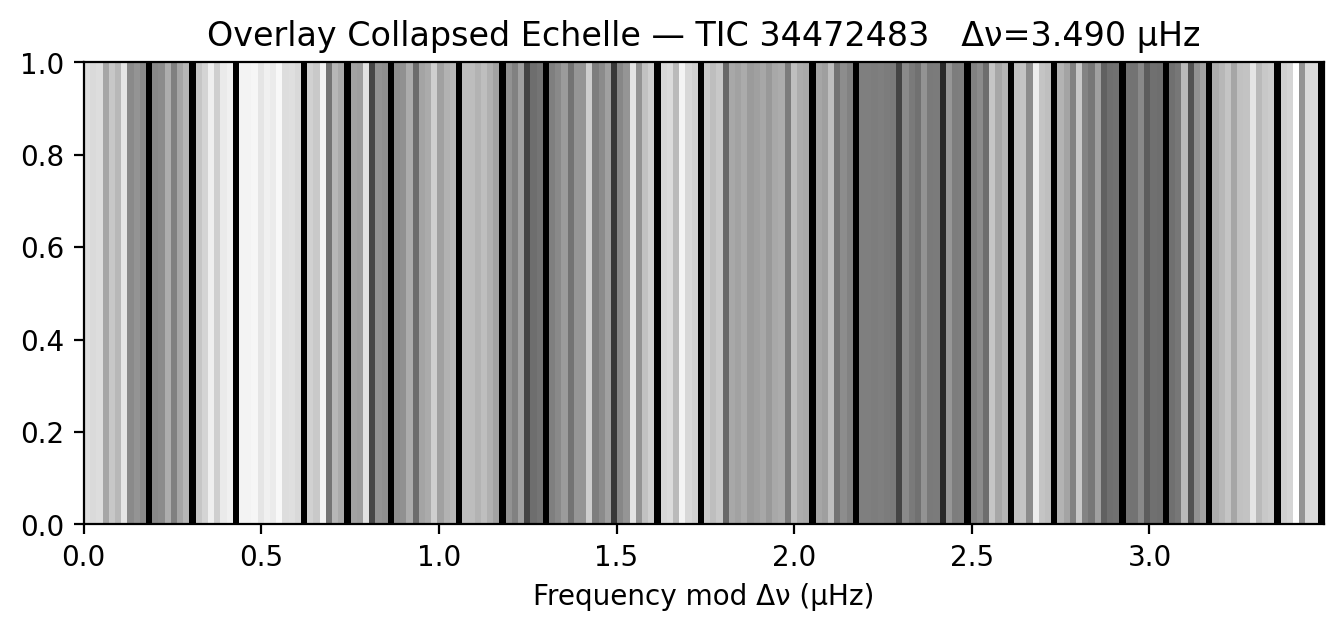}
\label{fig:appC:2}
\end{figure*}

\begin{figure*}[!t]
\centering
\TICRow{67569102}{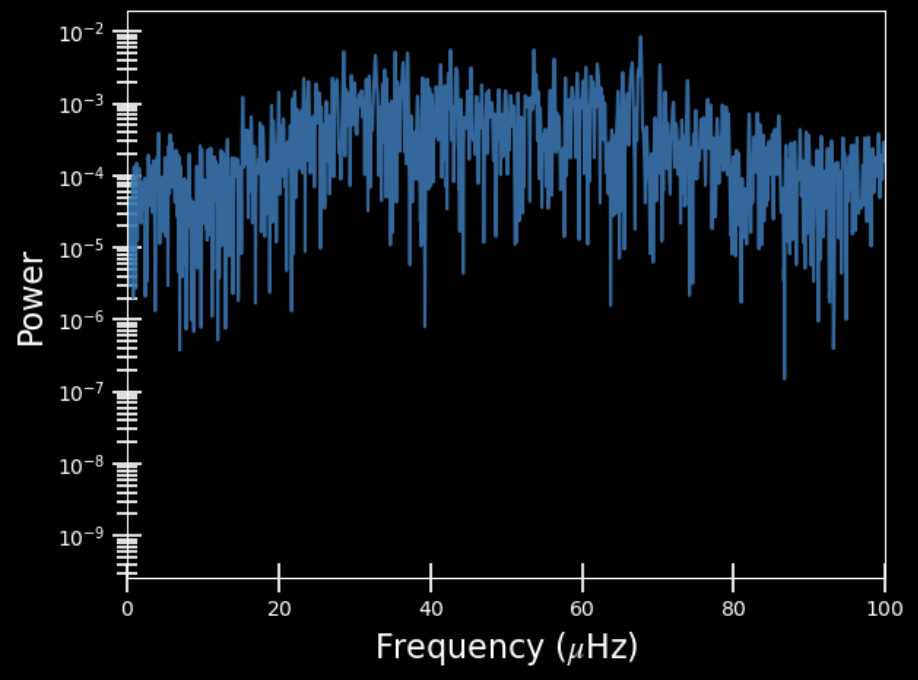}{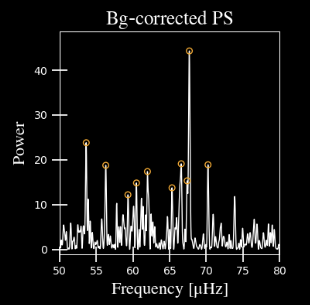}{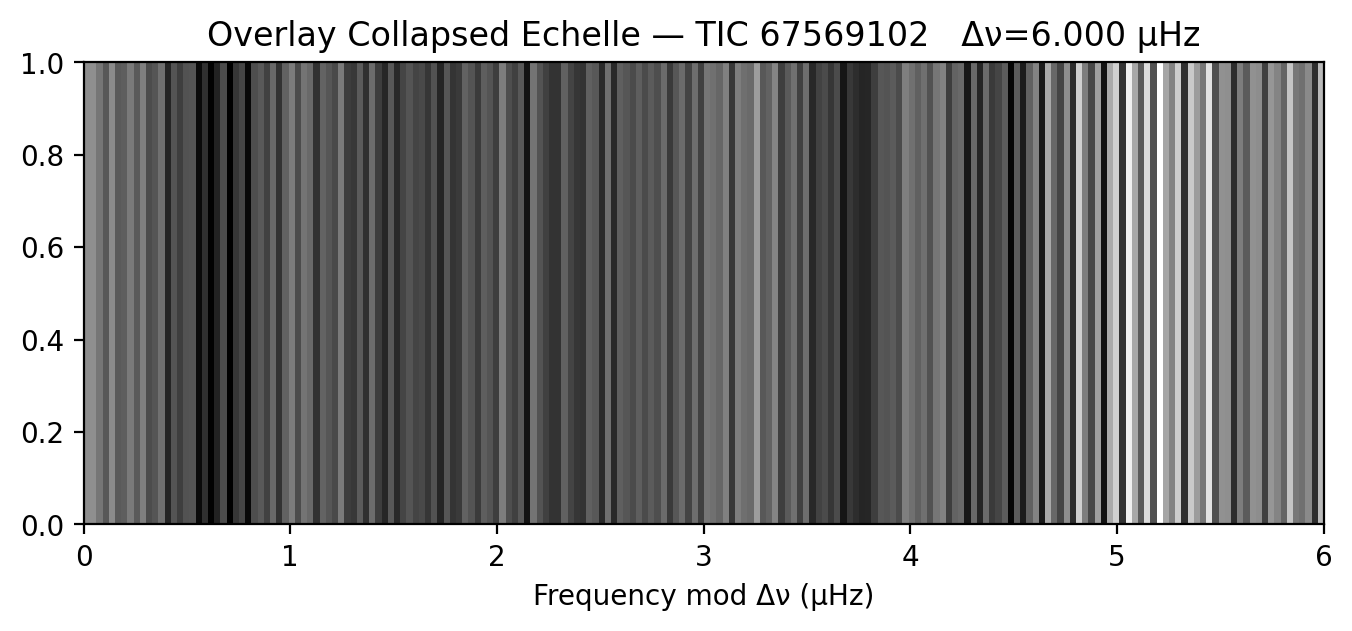}
\TICRow{67420118}{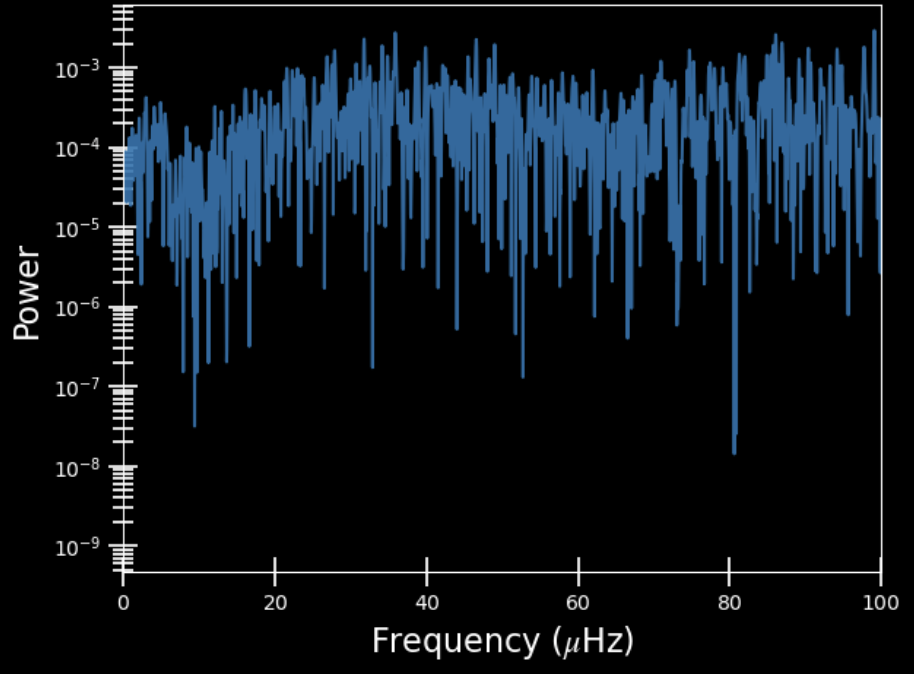}{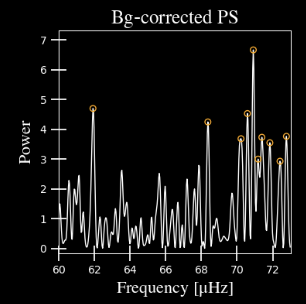}{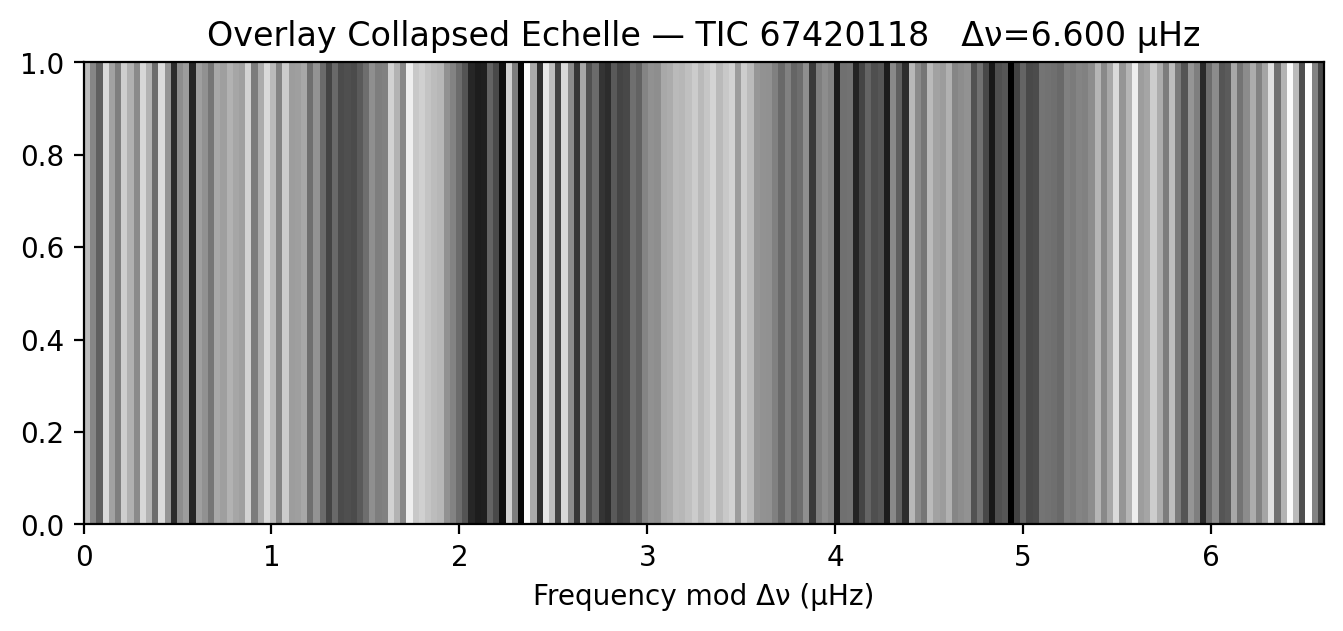}
\TICRow{67419922}{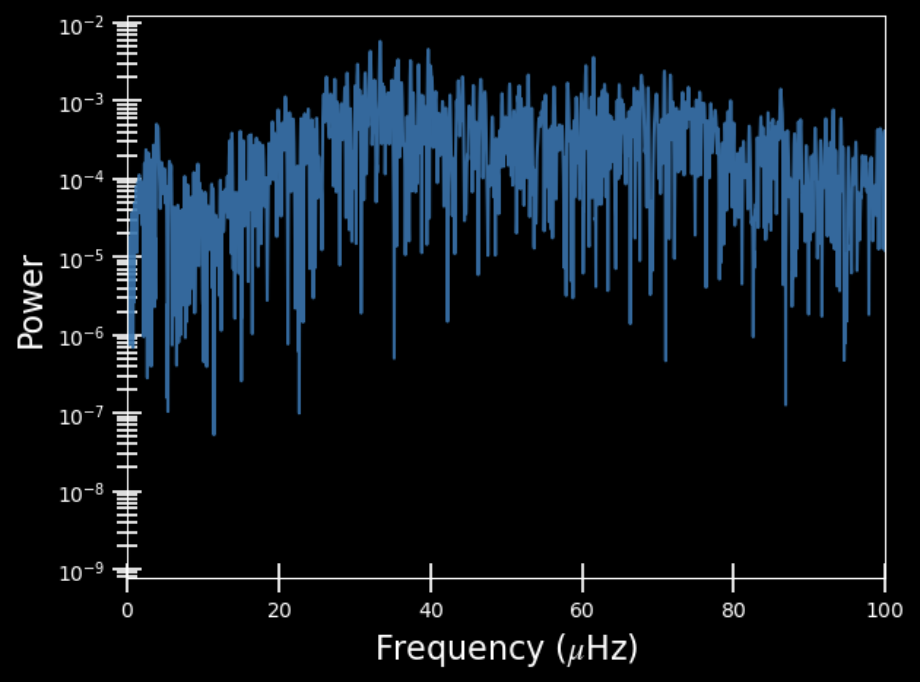}{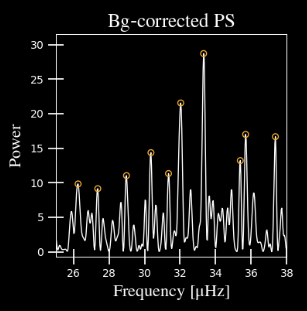}{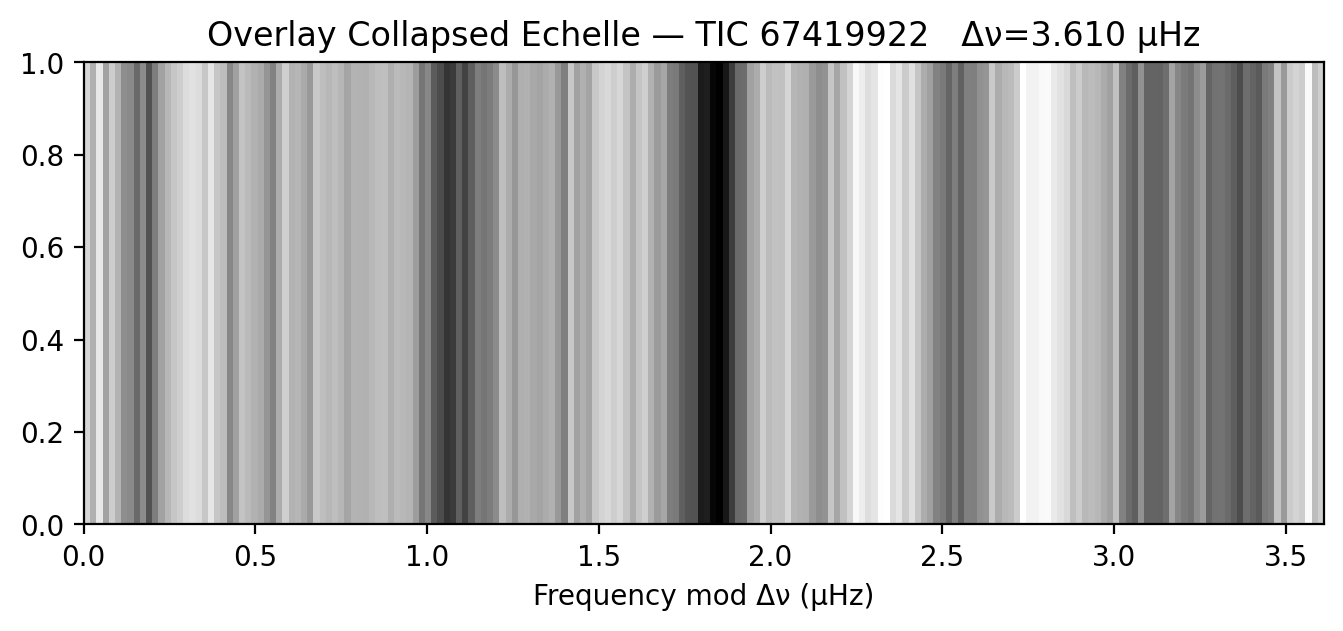}
\caption{Appendix C continued.}
\label{fig:appC:3}
\end{figure*}

\begin{figure*}[!t]
\centering
\TICRow{186970424}{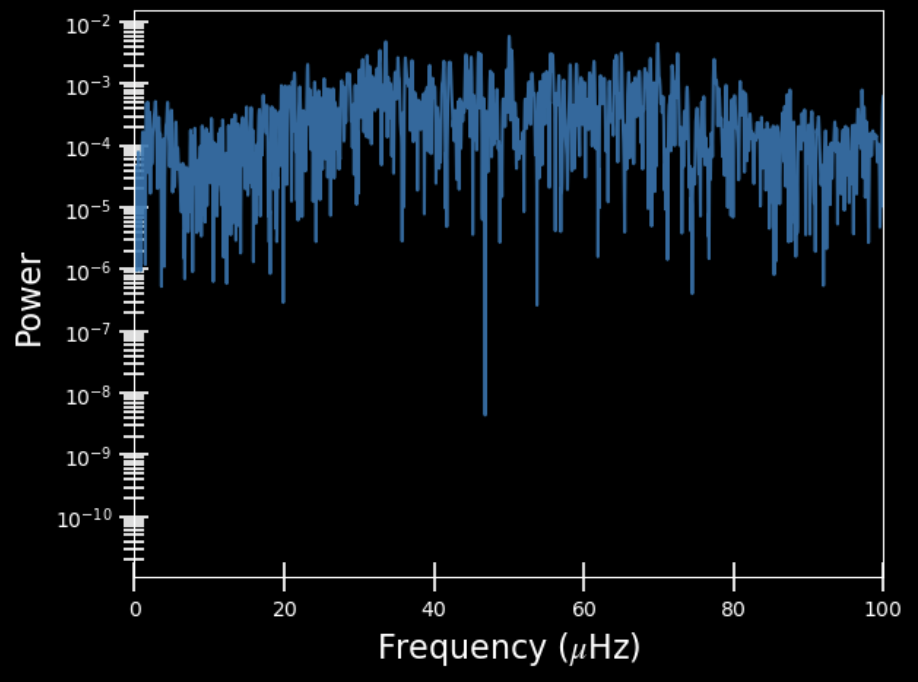}{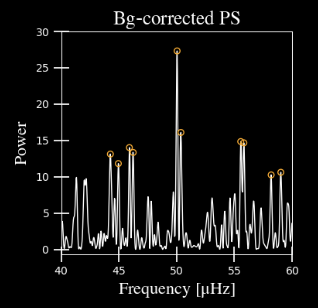}{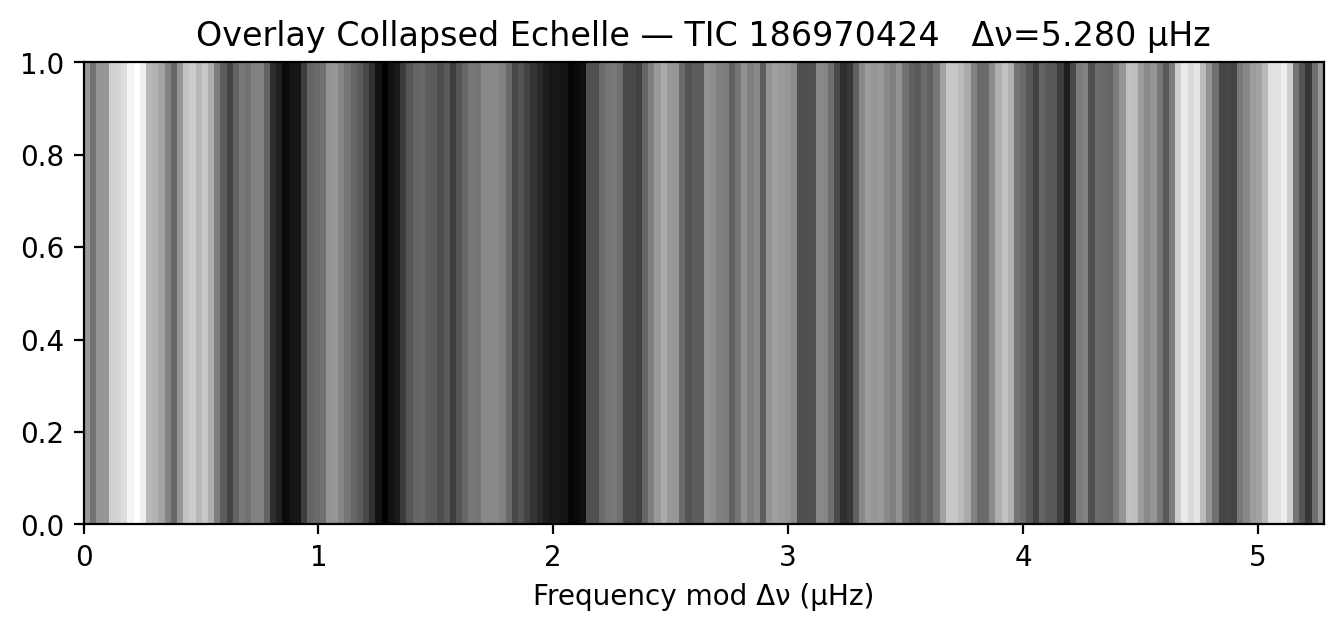}
\TICRow{306552813}{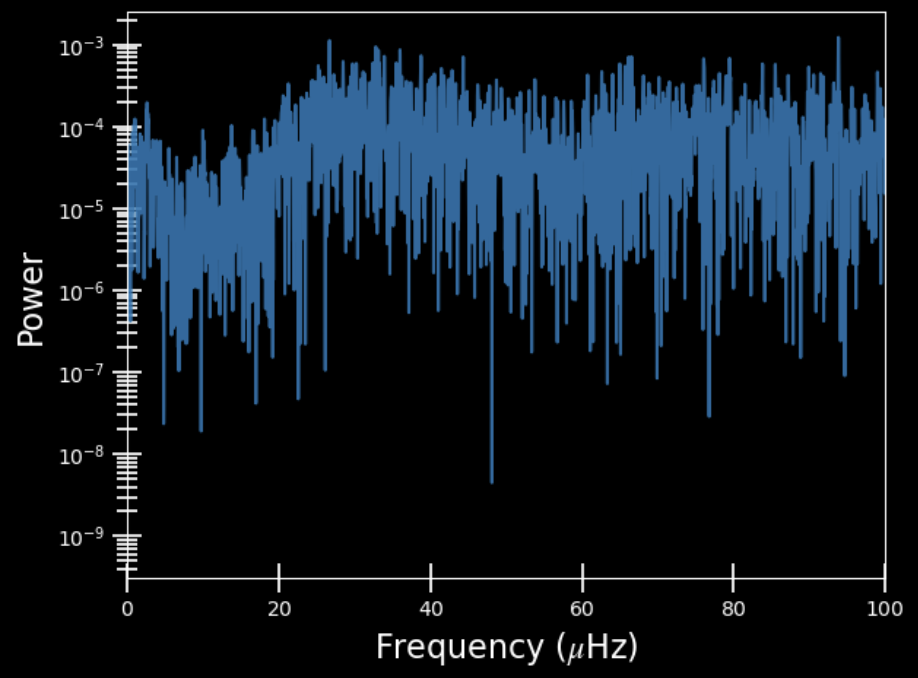}{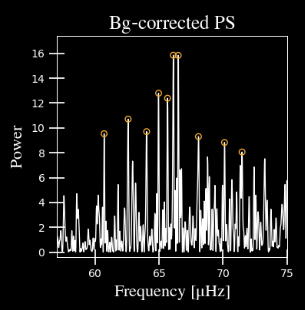}{overlay_TIC306552813.png}
\TICRow{306955660}{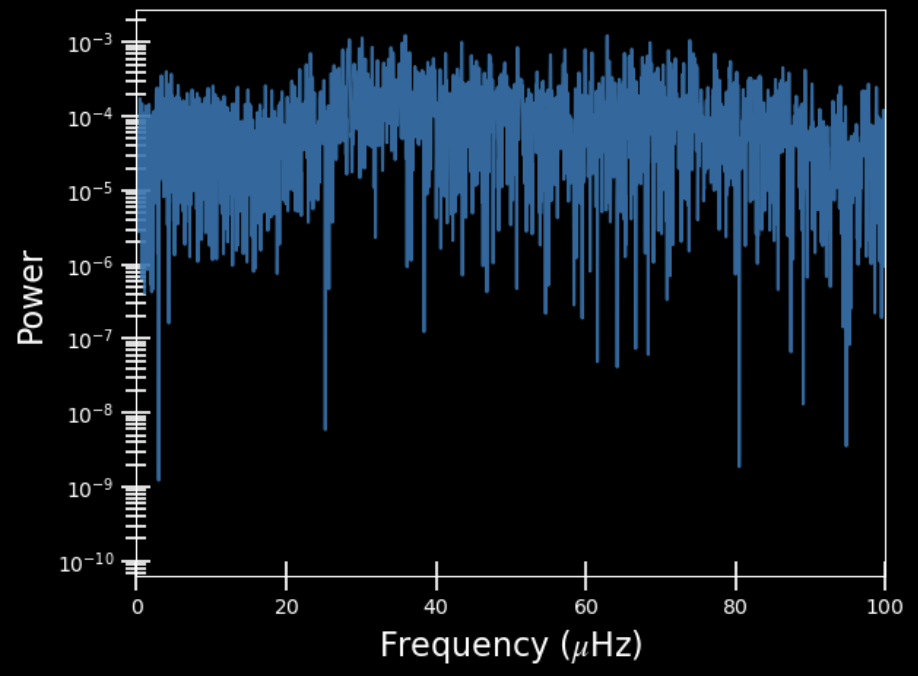}{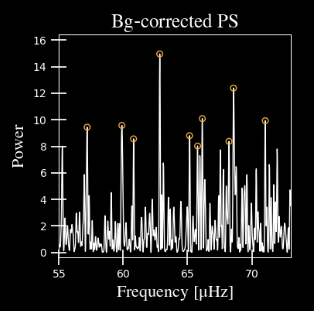}{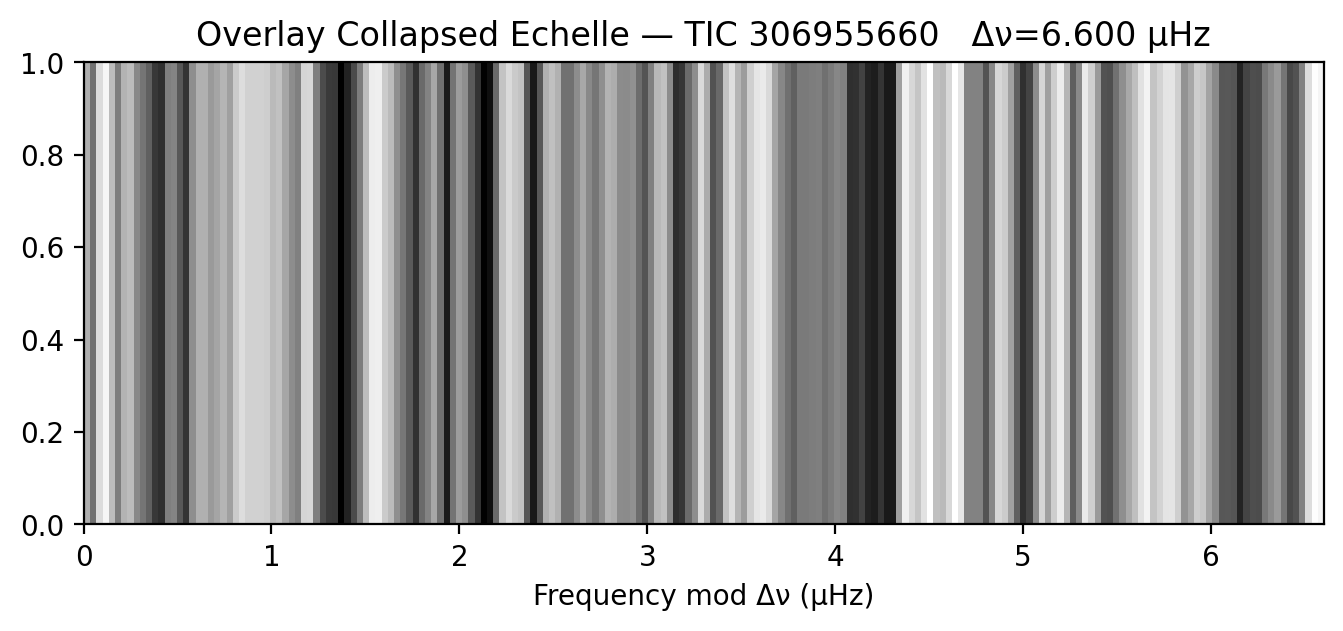}
\label{fig:appC:4}
\end{figure*}

\begin{figure*}[!t]
\centering
\TICRow{306187638}{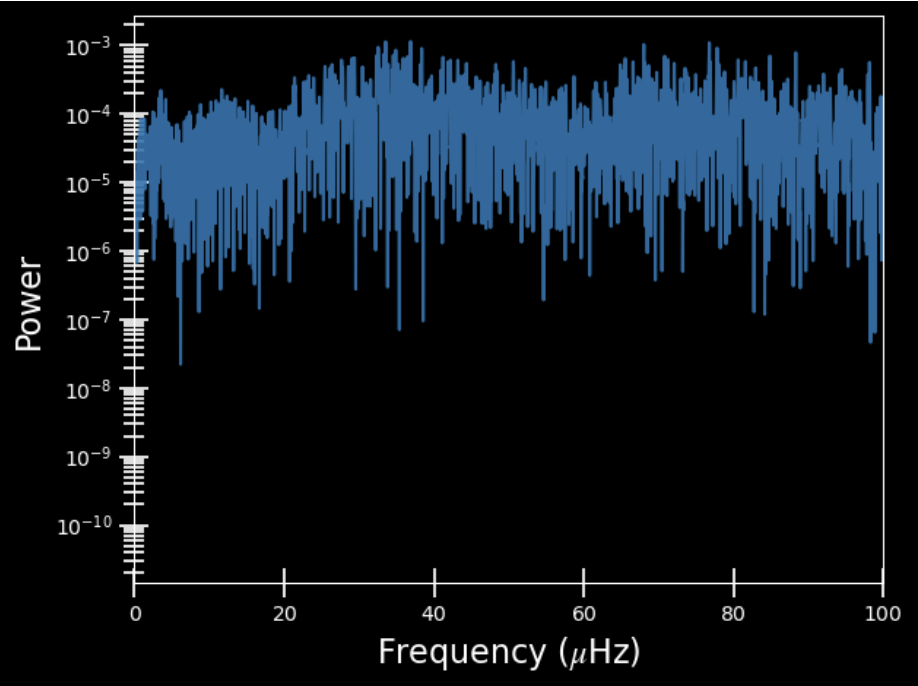}{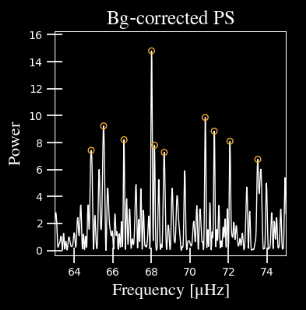}{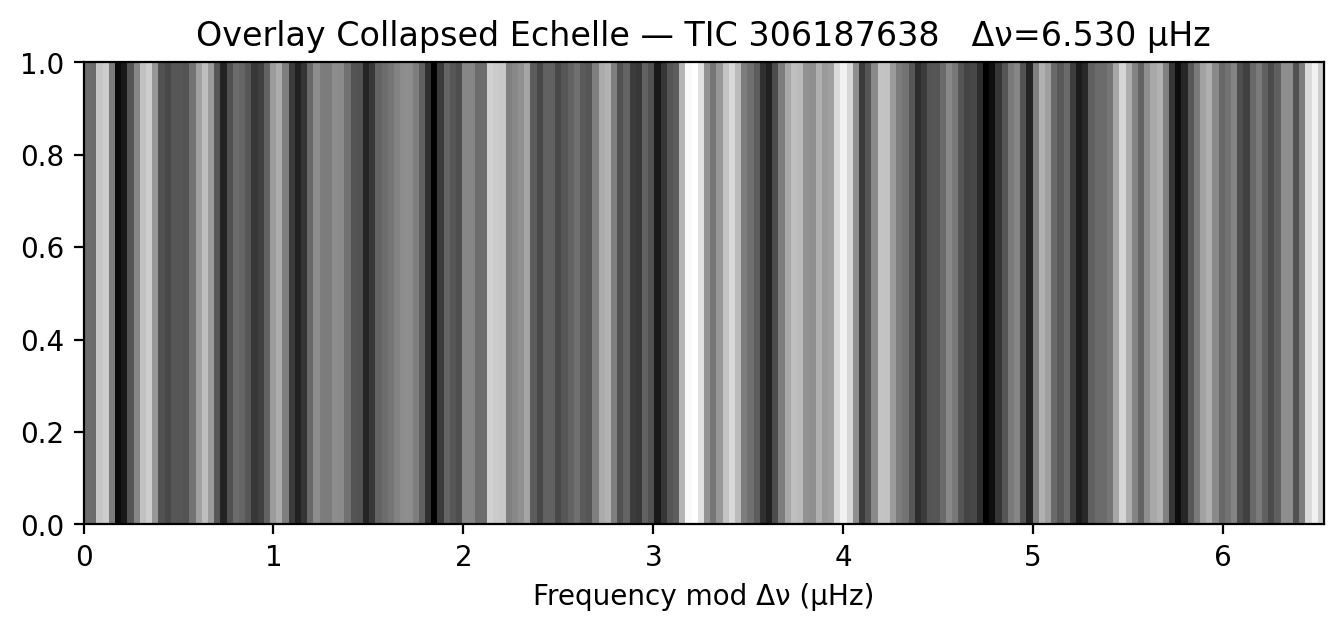}
\TICRow{26390802}{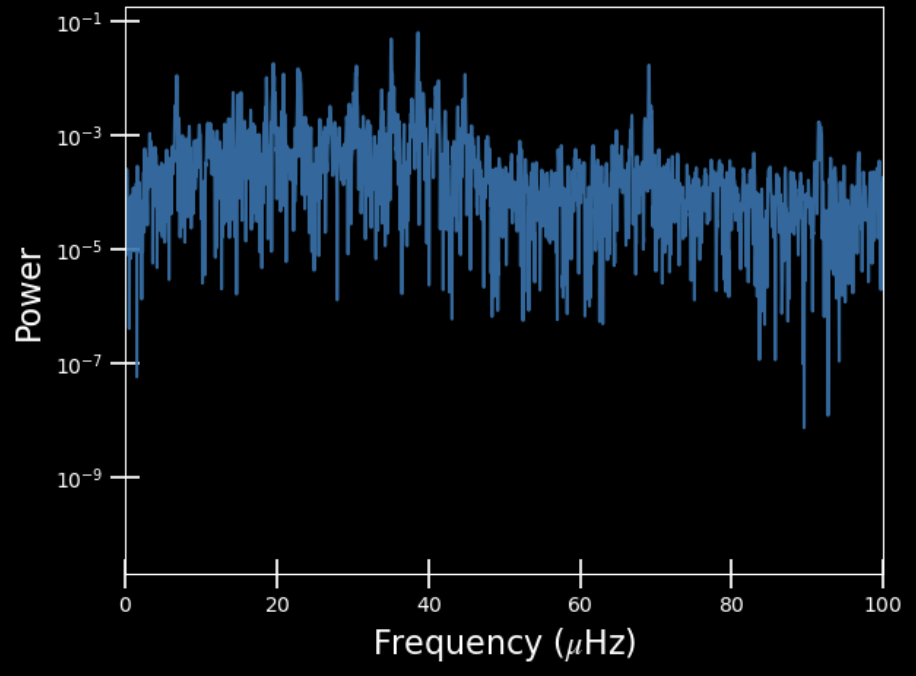}{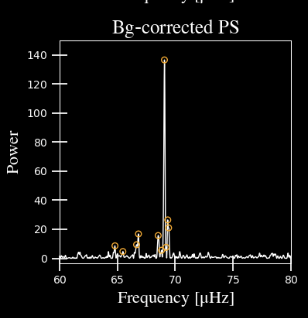}{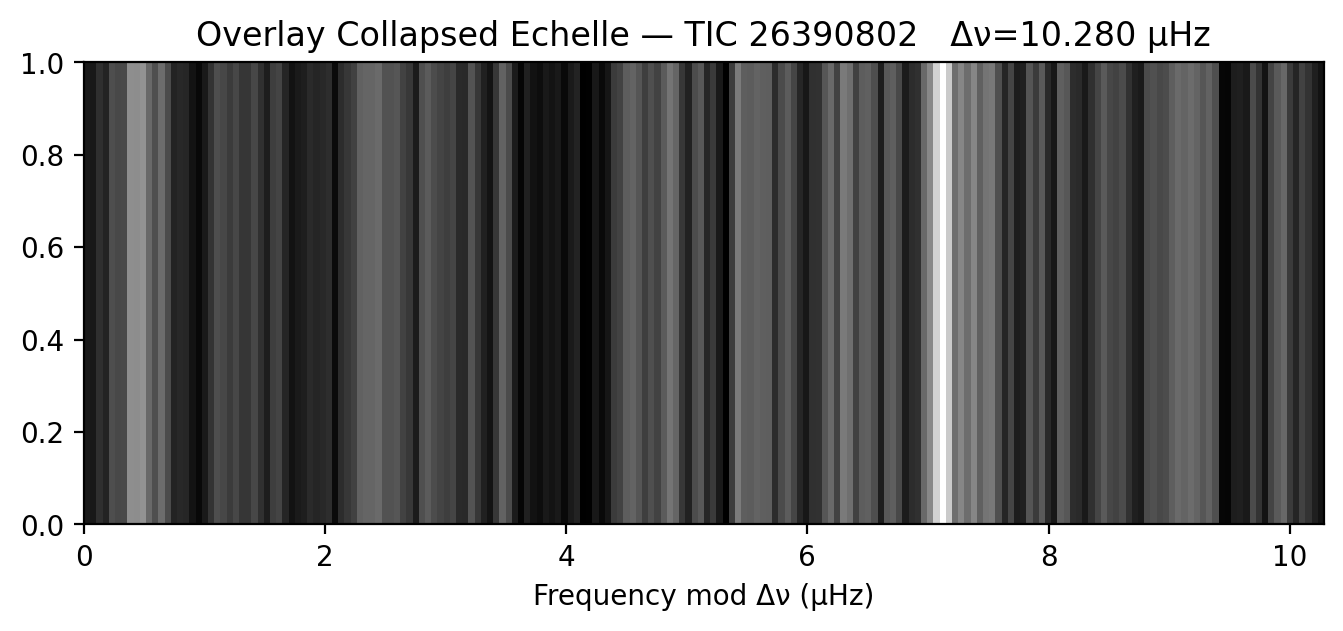}
\TICRow{306552754}{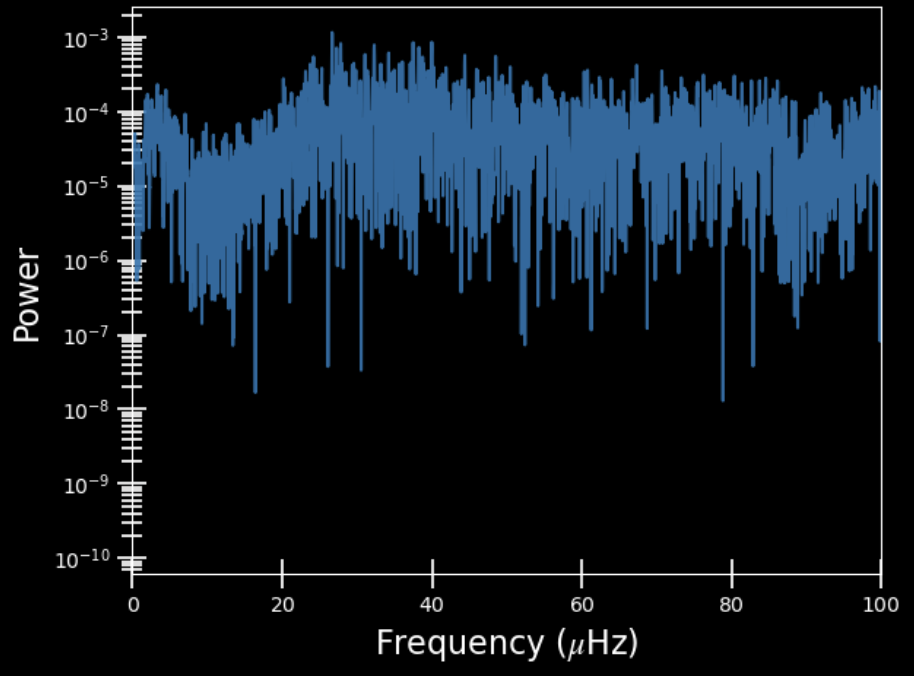}{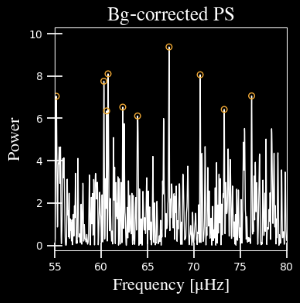}{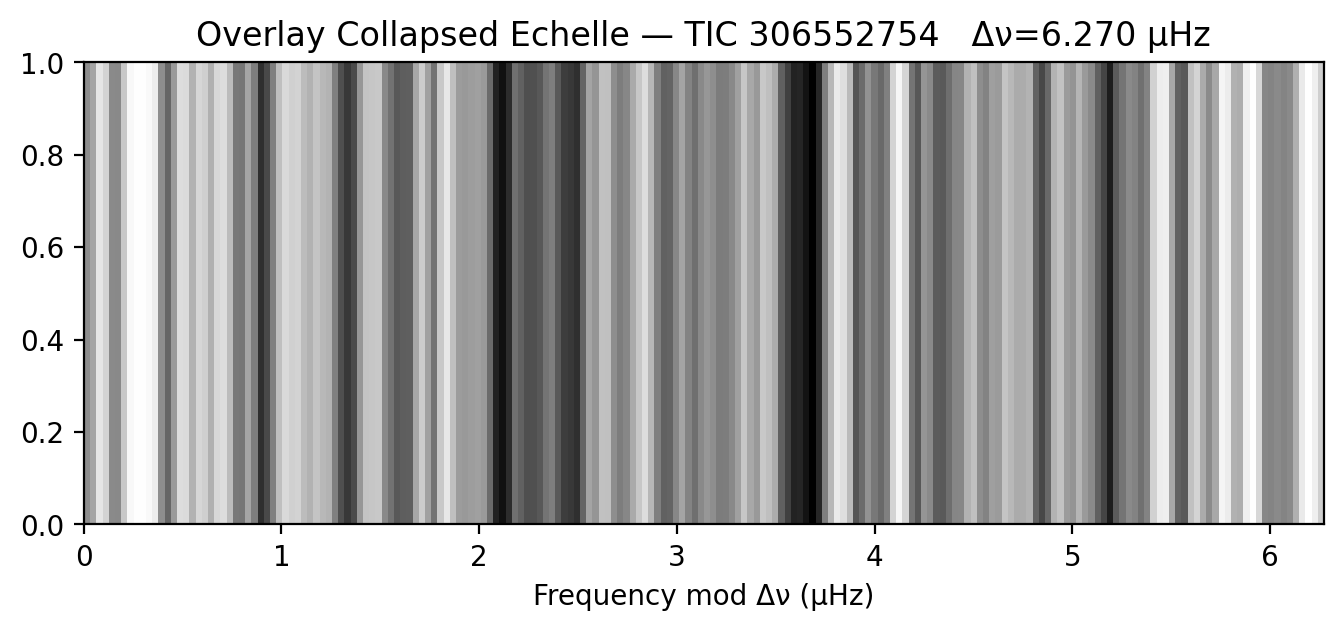}
\label{fig:appC:5}
\end{figure*}

\begin{figure*}[!t]
\centering
\TICRow{306552795}{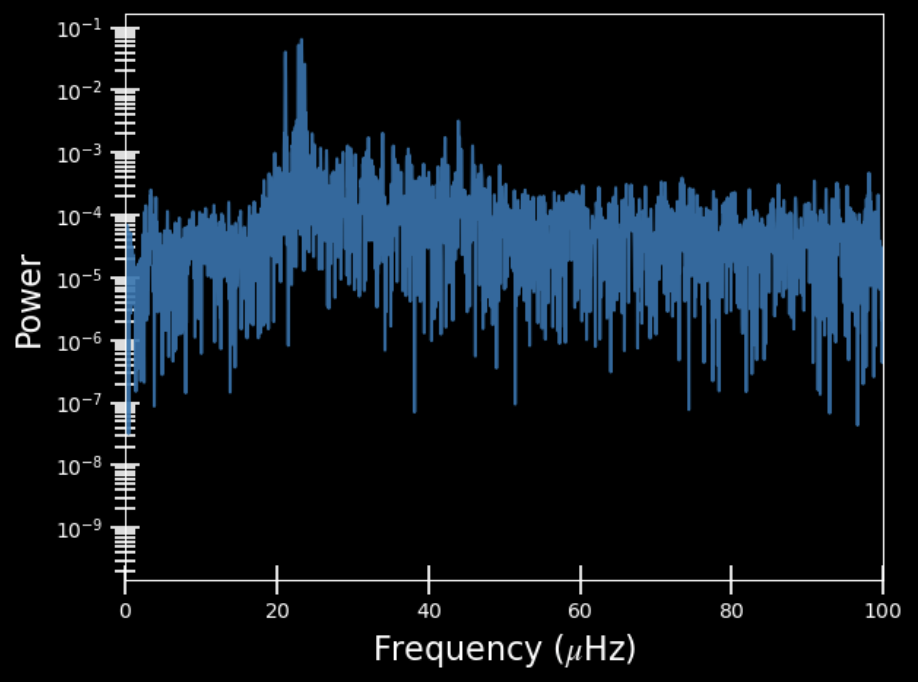}{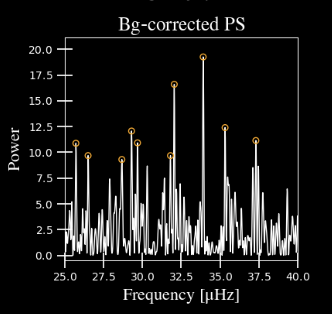}{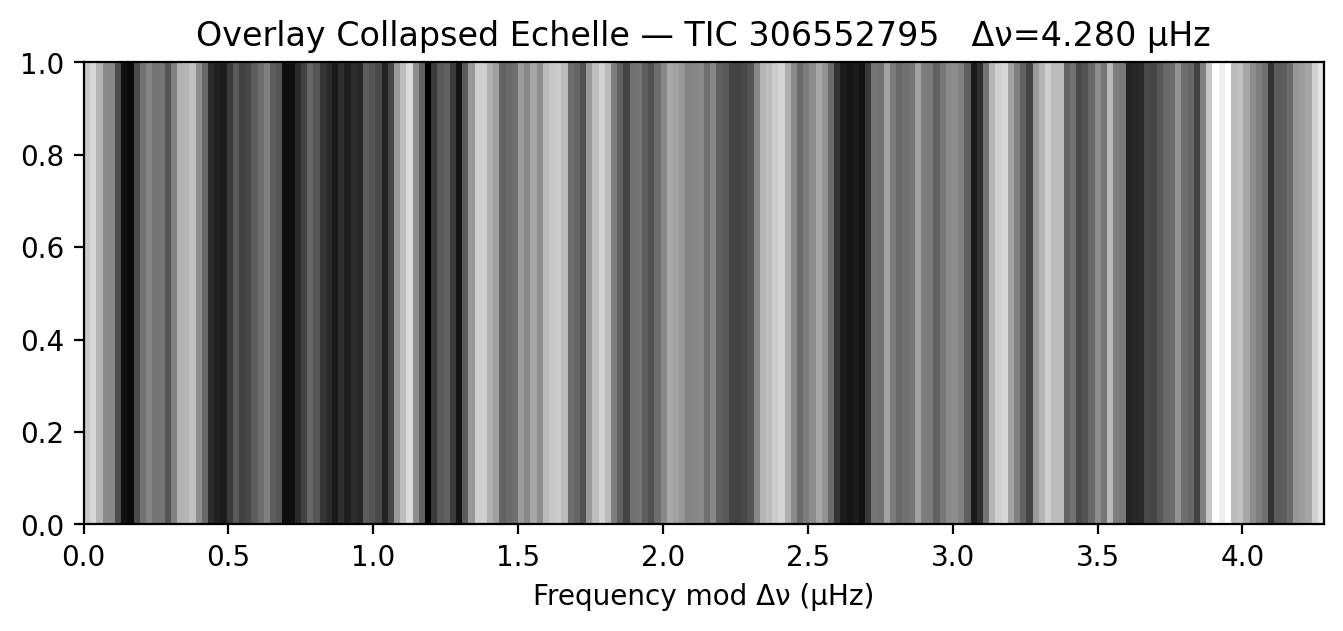}
\TICRow{417266022}{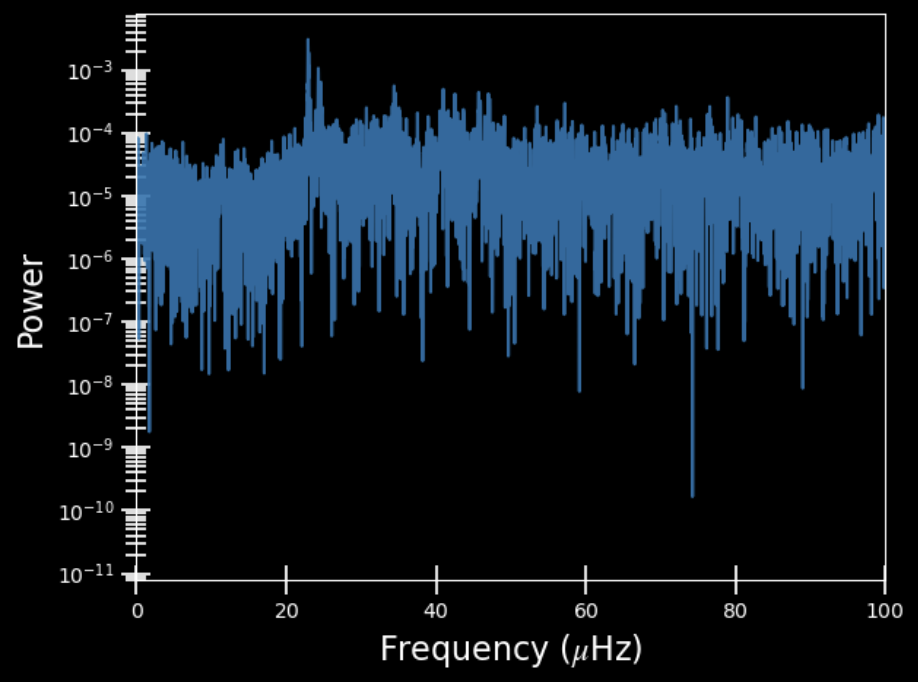}{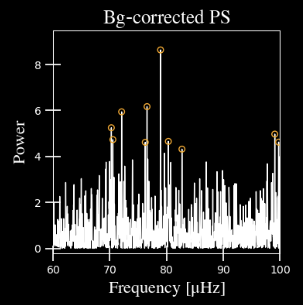}{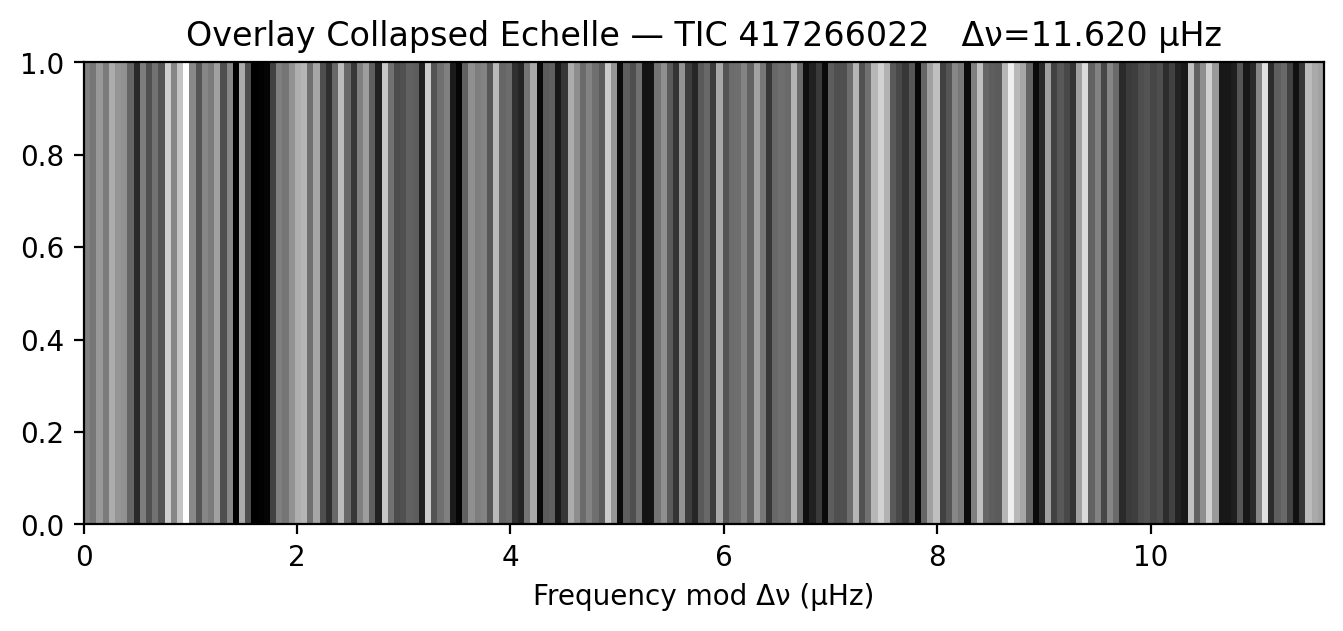}
\TICRow{417547686}{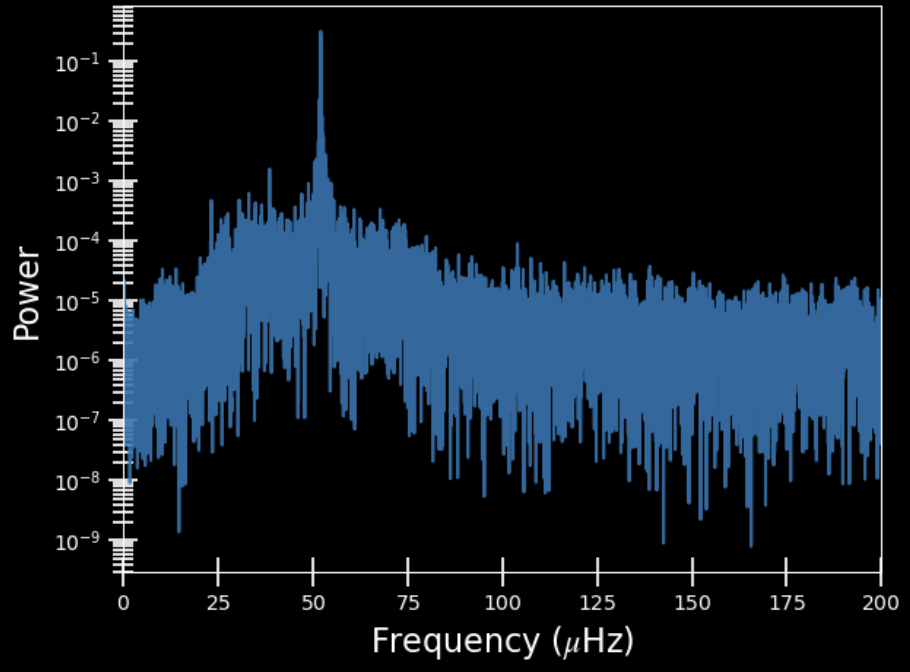}{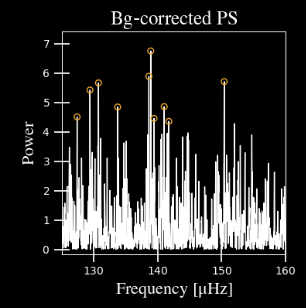}{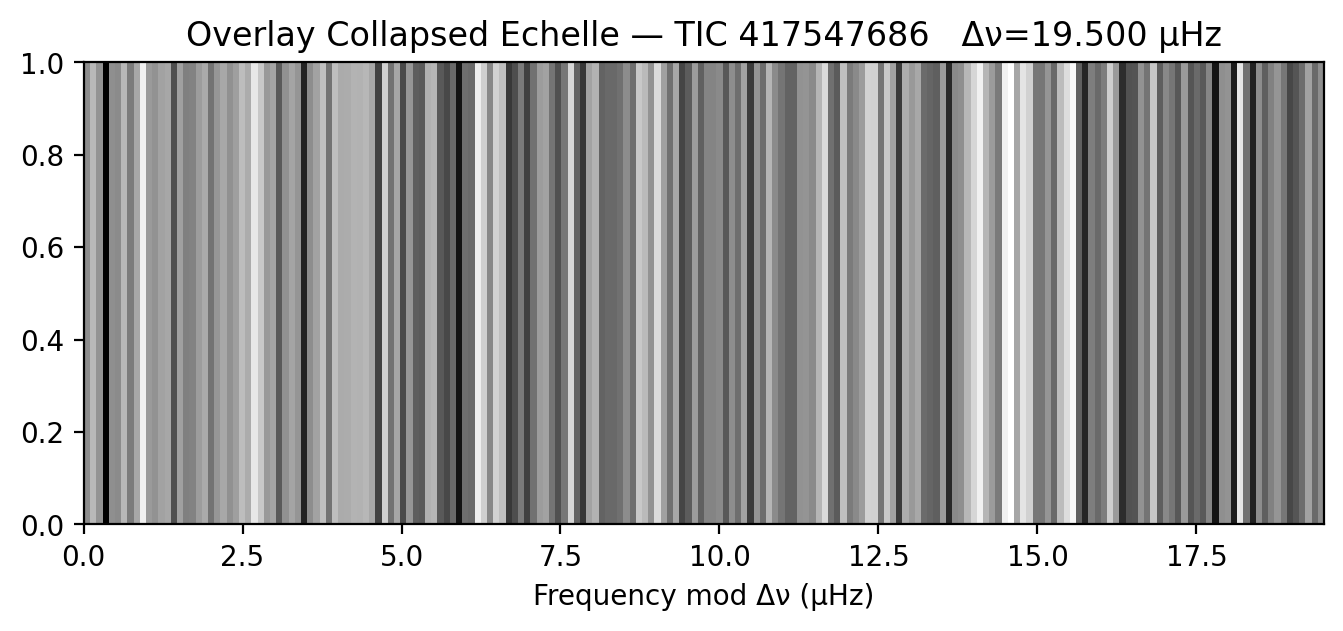}
\label{fig:appC:6}
\end{figure*}

\begin{figure*}[!t]
\centering
\TICRow{459055617}{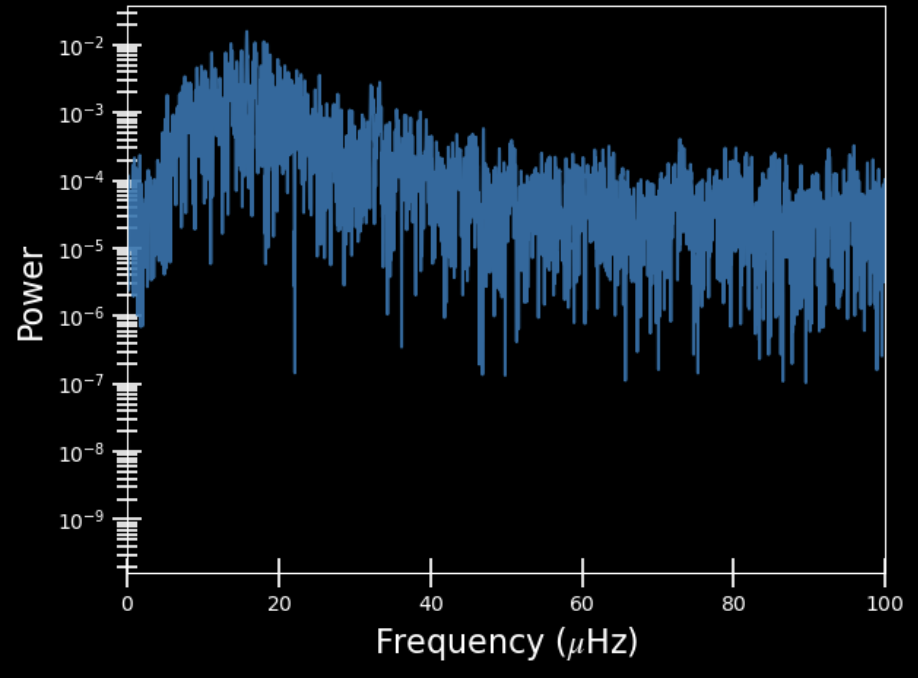}{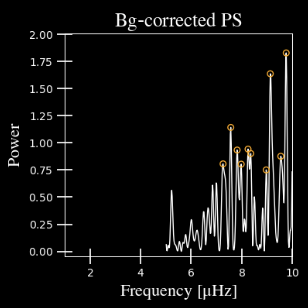}{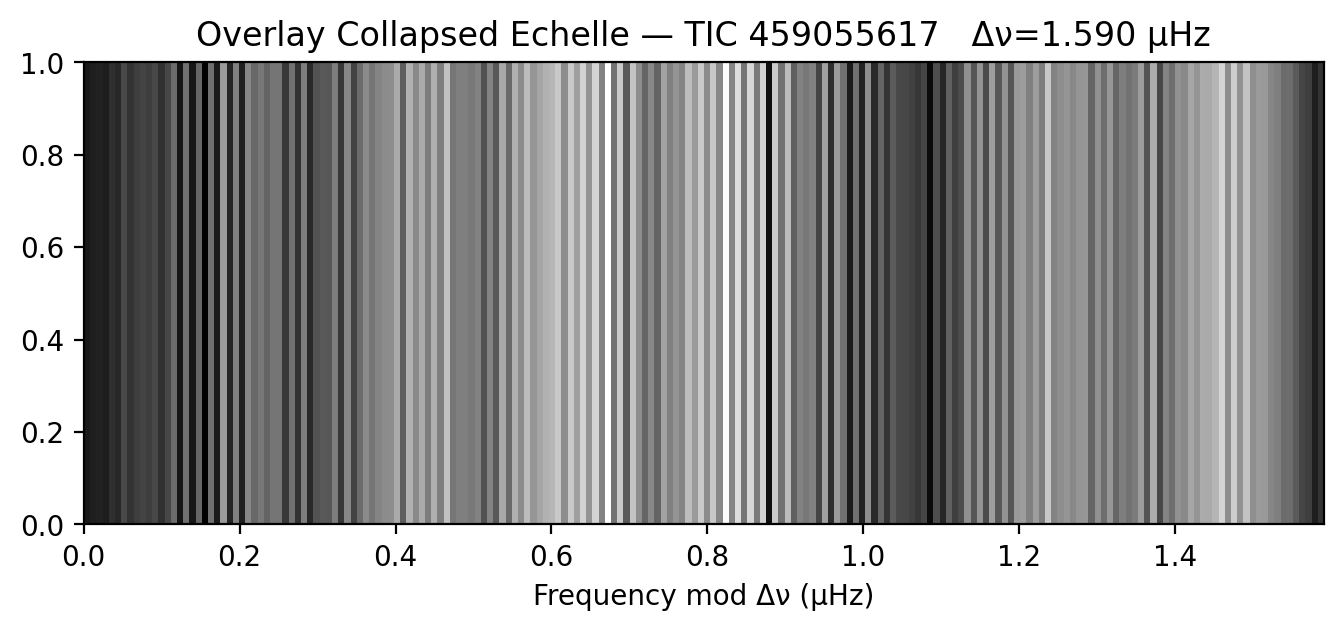}
\end{figure*}

\clearpage
\end{document}